\RequirePackage{ifpdf}
\ifpdf 
\documentclass[pdftex]{sigma}
\else
\documentclass{sigma}
\fi

\usepackage{dsfont}
\let\mathbb\mathds

\newcommand{\moyal}{Groenewold--Moyal}
\newcommand{\uim}{UV/IR mixing}
\newcommand{\nc}{non-commu\-ta\-tive}
\newcommand{\Nc}{Non-commu\-ta\-tive}
\newcommand{\nl}{non-local}
\newcommand{\nA}{non-Abelian}
\newcommand{\naiv}{na\"ive}
\newcommand{\np}{non-planar}

\newcommand{\secref}[1]{Section~\ref{#1}}		
\newcommand{\co}[2]{\big[#1,#2\big]}					
\newcommand{\aco}[2]{\big\{#1,#2\big\}}				
\newcommand{\starco}[2]{\big[ #1\stackrel{\star}{,}#2\big] }		
\newcommand{\staraco}[2]{\big\{ #1\stackrel{\star}{,}#2\big\} }	
\newcommand{\dif}[2]{\frac{\delta#1}{\delta#2}}
\newcommand{\var}[2]{\frac{\d #1}{\d #2}}				
\newcommand{\vvar}[3]{\frac{\d^2 #1}{\d #2\d #3}}			
\newcommand{\pa}{\partial}						
\newcommand{\diff}[2]{\frac{\pa #1}{\pa #2}}				
\newcommand{\ri}{\mathrm{i}}						
\newcommand{\re}{e}
\renewcommand{\k}{\tilde{k}}						
\newcommand{\p}{\tilde{p}}						
\newcommand{\Dt}{\widetilde{D}}						
\newcommand{\wtild}{\widetilde}						
\newcommand{\bc}{\bar{c}}						
\newcommand{\St}{S_{\textrm{tot}}}					
\newcommand{\Act}{S}
\newcommand{\id}{\mathds{1}}						
\newcommand{\R}{\mathds{R}}						
\newcommand{\N}{\mathds{N}}						
\newcommand{\C}{\mathds{C}}						
\newcommand{\cA}{\mathcal{A}}						
\renewcommand{\a}{\alpha}
\renewcommand{\b}{\beta}
\newcommand{\g}{\gamma}
\renewcommand{\d}{\delta}
\newcommand{\e}{\epsilon}
\newcommand{\vare}{\varepsilon}

\renewcommand{\th}{\theta}

\renewcommand{\l}{\lambda}
\newcommand{\m}{\mu}
\newcommand{\n}{\nu}

\renewcommand{\r}{\rho}
\newcommand{\s}{\sigma}

\newcommand{\ph}{\phi}

\renewcommand{\L}{\Lambda}
\renewcommand{\Xi}{\Xi}

\newcommand{\W}{\Omega}

\newcommand{\inv}[1]{\frac{1}{#1}}				
\newcommand{\tinv}[1]{\tfrac{1}{#1}}
\newcommand{\intk}{\int d^4k}						
\newcommand{\intx}{\int d^4x}						
\newcommand{\nn}{\nonumber}						
\newcommand{\F}{\widetilde{F}}

\newcommand{\wsq}{\widetilde{\square}}
\newcommand{\wprod}[2]{\ensuremath{#1\theta #2}}			
\newcommand{\ig}{\mathrm{i}g}
\newcommand{\bpsi}{\bar{\psi}}					
\newcommand{\bphi}{\bar{\phi}}					
\newcommand{\bB}{\bar{B}}					
\newcommand{\oldB}{\mathfrak{B}}			
\newcommand{\mth}{\theta}      
\newcommand{\sth}{\varepsilon} 
\newcommand{\cth}{\Theta}      
\newcommand{\st}{\bar{\sigma}}
\newcommand{\tri}{\triangle}

\newcommand{\hd}{\hat{\delta}}
\renewcommand{\eth}[1]{\frac{\e_{#1}}{\th}}
\newcommand{\xid}{\xi \cdot \partial}
\newcommand{\eid}{\vare \cdot \partial}

\newcommand{\bS}{{\bar S}}

\newcommand{\hAstar}{{{\hat A}^*}{}}
\newcommand{\BB}{{\mathcal B}}

\makeatletter
\def\fmslash{\@ifnextchar[{\fmsl@sh}{\fmsl@sh[0mu]}}
\def\fmsl@sh[#1]#2{%
  \mathchoice
    {\@fmsl@sh\displaystyle{#1}{#2}}%
    {\@fmsl@sh\textstyle{#1}{#2}}%
    {\@fmsl@sh\scriptstyle{#1}{#2}}%
    {\@fmsl@sh\scriptscriptstyle{#1}{#2}}}
\def\@fmsl@sh#1#2#3{\m@th\ooalign{$\hfil#1\mkern#2/\hfil$\crcr$#1#3$}}

\numberwithin{equation}{section}

\begin{document}

\allowdisplaybreaks

\renewcommand{\thefootnote}{$\star$}

\renewcommand{\PaperNumber}{062}

\FirstPageHeading

\ShortArticleName{Gauge Theories on Deformed Spaces}

\ArticleName{Gauge Theories on Deformed Spaces\footnote{This paper is a
contribution to the Special Issue ``Noncommutative Spaces and Fields''. The
full collection is available at
\href{http://www.emis.de/journals/SIGMA/noncommutative.html}{http://www.emis.de/journals/SIGMA/noncommutative.html}}}

\Author{Daniel N. BLASCHKE~$^\dag$, Erwin KRONBERGER~$^\ddag$, Ren\'e I.P. SEDMIK~$^\ddag$ \\ and Michael WOHLGENANNT~$^\ddag$}

\AuthorNameForHeading{D.N.~Blaschke, E.~Kronberger, R.I.P.~Sedmik and M.~Wohlgenannt}

\Address{$^\dag$~Faculty of Physics, University of Vienna, Boltzmanngasse 5 A-1090 Vienna, Austria}
\EmailD{\href{mailto:daniel.blaschke@univie.ac.at}{daniel.blaschke@univie.ac.at}}

\Address{$^\ddag$~Institute for Theoretical Physics, Vienna University of Technology,\\
\hphantom{$^\ddag$}~Wiedner Hauptstrasse 8-10, A-1040 Vienna, Austria}
\EmailD{\href{mailto:kronberger@hep.itp.tuwien.ac.a}{kronberger@hep.itp.tuwien.ac.at}, \href{mailto:sedmik@hep.itp.tuwien.ac.at}{sedmik@hep.itp.tuwien.ac.at},\\
\hspace*{17mm}\href{mailto:miw@hep.itp.tuwien.ac.at}{miw@hep.itp.tuwien.ac.at}}

\ArticleDates{Received April 13, 2010, in f\/inal form July 14, 2010;  Published online August 04, 2010}

\Abstract{The aim of this review is to present an overview over available models and approaches to {\nc} gauge theory. Our main focus thereby is on gauge models formulated on f\/lat {\moyal} spaces and renormalizability, but we will also review other deformations and try to point out common features. This review will by no means be complete and cover all approaches, it rather ref\/lects a highly biased selection.}


\Classification{81T13; 81T15; 81T75}

\tableofcontents

\renewcommand{\thefootnote}{\arabic{footnote}}
\setcounter{footnote}{0}

\section{Introduction}
\label{sec:intro}

Even in the early days of quantum mechanics and quantum f\/ield theory (QFT), continuous space-time and Lorentz symmetry were considered inappropriate to describe the small scale structure of the universe \cite{Schroedinger:1934}. Four dimensional QFT suf\/fers from infrared (IR) and ultraviolet (UV) divergences as well as from the divergence of the renormalized perturbation expansion. Despite the impressive agreement between theory and experiments and many attempts, these problems are not settled and remain a big challenge for theoretical physics. In \cite{Heisenberg:1938} the introduction of a~fundamental length is suggested to cure the UV divergences. H.~Snyder was the f\/irst to formulate these ideas mathematically \cite{Snyder:1946,Snyder:1947} and introduced {\nc} coordinates. Therefore a~position uncertainty arises naturally. But the success of the (commutative) renormalization program made people forget about these ideas for some time. Only when the quantization of gravity was considered thoroughly, it became clear that the usual concepts of space-time are inadequate and that space-time has to be quantized or {\nc}, in some way. This situation has been analyzed in detail by S.~Doplicher, K.~Fredenhagen and J.E.~Roberts in~\cite{Doplicher:1994b}. Measuring the distance between two particles, energy has to be deposited in that space-time region, proportional to the inverse distance. If the distance is of the order of the Planck length, the bailed energy curves space-time to such an extent that light will not be able to leave that region and generates a black hole. The limitations arising from the need to avoid the appearance of black holes during a measurement process lead to uncertainty relations between space-time coordinates. This already allows to catch a glimpse of the deep connection between gravity and {\nc} geometry, especially {\nc} gauge theory. We will provide some further comments on this later. At this point, one also has to mention the extensive work of A.~Connes~\cite{Connes:1994}, who wrote the f\/irst book on the underlying mathematical concepts of {\nc} spaces\footnote{Also noteworthy, is the attempt of formulating the Standard Model of particle physics using so-called spectral action principle and ideas based on {\nc} geometry~\cite{Connes:1994,Connes:2006a,Barrett:2006a}.}.

{\bf {\Nc} coordinates.}
In {\nc} quantum f\/ield theories, the coordinates themselves have to be considered as operators $\hat x^i$ (denoted by hats) on some Hilbert space~$\mathcal{H}$, satisfying an algebra def\/ined by commutation relations. In general, they have the form
\begin{gather}
\label{eq:intro_main:commutator}
[\hat x^i, \hat x^j] = \ri \cth^{ij}(\hat x) ,
\end{gather}
where $\cth^{ij}(\hat x)$ might be any function of the generators with $\cth^{ij} = - \cth^{ji}$ and satisfying the Jacobi identity. Most commonly, the commutation relations are chosen to be either constant, linear or quadratic in the generators. In the canonical case the relations are constant,
\begin{gather}
\label{eq:intro_main:canonicalR}
[\hat x^i, \hat x^j] = \ri \cth^{ij}= \textrm{const} .
\end{gather}
This case will be discussed in \secref{sec:covar}. The linear or Lie-algebra case
\begin{gather*}
[\hat x^i, \hat x^j] = \ri \l^{ij}_k \hat x^k ,
\end{gather*}
where $\lambda^{ij}_k\in \C$ are the structure constants, basically has been discussed in two dif\/ferent settings, namely fuzzy spaces \cite{Madore:1990,Madore:1991} and $\kappa$-deformation \cite{Lukierski:1991,Majid:1994,Wohlgenannt:2003}. Those approaches will keep us busy in~\secref{sec:noncanon:kappa} and~\secref{sec:noncanon:fuzzy}, respectively. The third commonly used choice is a quadratic commutation relation,
\begin{gather}
[\hat x^i, \hat x^j] = \left(\frac 1q \hat R^{ij}_{kl} - \delta^i_l \delta^j_k\right)  \hat x^k \hat x^l ,
\end{gather}
where $\hat R^{ij}_{kl}\in \C$ is the so-called $\hat R$-matrix, corresponding to quantum groups \cite{Reshetikhin:1990,Wess:1994}. We will brief\/ly comment on this case in~\secref{sec:noncanon:other}.

Independent of the explicit form of $\cth^{ij}$, the commutative algebra of functions on space-time has to be replaced by the {\nc} algebra $\widehat{\mathcal A}$ generated by the coordinates $\hat x^i$, subject to the ideal $\mathcal I$ of relations generated by the commutation relations,
\[
\widehat{\mathcal A} = \frac {\C \langle  \hat x^i \rangle } {\mathcal I} .
\]
However, there is an isomorphism mapping of the {\nc} function algebra $\widehat{\mathcal A}$ to the commutative one equipped with an additional {\nc} product $\star$, $\{\mathcal A, \star\}$. This isomorphism exists, if\/f the {\nc} algebra together with the chosen basis (ordering) satisf\/ies the so-called Poincar\'e--Birkhof\/f--Witt property, i.e.\ any monomial of order $n$ can be written as a sum of the basis monomials of order $n$ or smaller, by reordering and thereby using the algebra relations \eqref{eq:intro_main:commutator}. Let us choose, for example, the basis of normal ordered monomials:
\[
1,\  \hat x^i, \ \dots, \ (\hat x^{i_1})^{n_1}\cdots (\hat x^{i_m})^{n_m} , \ \dots, \qquad \textrm{where}\quad  i_a < i_b, \quad \textrm{for} \ \  a < b.
\]
We can map the basis monomials in $\mathcal A$ onto the respective normally ordered basis elements of $\widehat{\mathcal A}$
\begin{alignat*}{3}
& W :\  && \mathcal{A}   \to   \widehat{\mathcal A} , & \nonumber\\
&&& x^i   \mapsto   \hat x^i,& \nonumber\\
&&& x^i x^j   \mapsto   \hat x^i \hat  x^j \equiv\,
	  :\hat x^i \hat x^j:  , \quad \textrm{for} \ \ i<j  . & 
\end{alignat*}
The ordering is indicated by $:\, :$. $W$ is an isomorphism of vector spaces. In order to extend $W$ to an algebra isomorphism, we have to introduce a new non-commutative multiplication $\star$ in~$\mathcal A$. This star product is def\/ined by
\begin{gather}
\label{eq:intro_main:star_prod_def}
W(f\star g) := W(f) \cdot W(g) = \hat f\cdot \hat g,
\end{gather}
where $f, g \in \mathcal A$, $\hat f, \hat g \in \widehat{\mathcal A}$. Thus, an algebra isomorphism is established,
\[
(\mathcal A,\star) \cong (\widehat{\mathcal A},\cdot) .
\]
The information about the non-commutativity of $\widehat{\mathcal A}$ is encoded in the star product. If we chose a symmetrically ordered basis, we can use the Weyl-quantization map for $W$
\begin{gather}
\hat f=W(f)  =  \inv{(2\pi)^{D}} \int d^D k\, e^{\ri k_j \hat x^j}
	\tilde{f}(k)
,\qquad
\tilde f(k)   =   \int d^D x \, e^{-\ri k_j x^j} f(x),\label{eq:intro_main:invfour}
\end{gather}
where we have replaced the commutative coordinates by {\nc} ones 
in the inverse Fourier transformation~\eqref{eq:intro_main:invfour}. The exponential takes care of the symmetrical ordering. Using equation~\eqref{eq:intro_main:star_prod_def}, we get
\[
W(f \star g)=\inv{(2\pi)^D} \int d^D k \, d^D p \,
e^{\ri k_i\hat x^i}e^{\ri p_j \hat x^j} \tilde f (k) \tilde g (p).
\]
Because of the non-commutativity of the coordinates $\hat x^i$,
we have to apply the Baker--Campbell--Hausdorf\/f (BCH) formula
\[
e^A e^B = e^{A+B +\inv{2}[A,B] + \inv{12}[[A,B],B] -
\inv{12} [[A,B],A] + \cdots} .
\]
Clearly, we need  to specify $\cth^{ij}(\hat x)$ in order to calculate the star product explicitly, which we will do in the respective sections.

{\bf {\Nc} quantum f\/ield theory.}
Knowing about the structure of deformed spaces, we have to expose these ideas to the real world. We need to formulate models on them~-- f\/irst toy models, then more physical ones~-- and try to make testable predictions. In recent years, a lot of ef\/forts have been made to construct Quantum Field Theories on {\nc} spaces. For some earlier reviews, see e.g. \cite{Harvey:2001yn,Szabo:2001,Paniak:2003yf,Paniak:2003xm}.
In the present one, we will discuss new developments and emphasize renormalizability properties of the models under consideration. We will not discuss the transition from Euclidean to Minkowskian signature (or vice versa). This is still an open and undoubtedly very interesting question in {\nc} geometry. For a~reference, see e.g.~\cite{Bahns:2009, Grosse:2008c}. We will stay on either side and will not try to f\/ind a match for the theory on the other side.

Scalar theories on deformed spaces have f\/irst been studied in a {\naiv} approach, replacing the pointwise product by the {\moyal} product~\cite{Groenewold:1946,Moyal:1949}, corresponding to \eqref{eq:intro_main:canonicalR}. T.~Filk~\cite{Filk:1996} developed Feynman rules, and soon after S.~Minwalla, M.~van Raamsdonk and N.~Seiberg~\cite{Minwalla:1999} encountered a serious  problem when considering perturbative expansions. Two dif\/ferent kinds of contributions arise: The planar loop contributions show the standard singularities which can be handled by a renormalization procedure. The {\np} ones are f\/inite for generic momenta. However they become singular at exceptional momenta. The usual UV divergences are then ref\/lected by new singularities in the IR. This ef\/fect is referred to as ``{\uim}'' and is the most important feature for any {\nc} f\/ield theory. It also spoils the usual renorma\-li\-za\-tion procedure: Inserting many such {\np} loops to a higher order diagram generates singularities of arbitrary inverse power. Without imposing a special structure such as supersymmetry, the renormalizability seemed lost~\cite{Chepelev:2000}. Crucial progress was achieved when two dif\/ferent, independent approaches yielded a solution of this problem for the special case of a scalar four dimensional theory def\/ined on the Euclidean canonically deformed space. Consequently, the renormalizability to all orders in perturbation theory could be showed. Both models modify the theory in the IR by adding a new term. These modif\/ications alter the propagator and lead to a crucial damping behaviour in the IR.

First, H.~Grosse and R.~Wulkenhaar \cite{Grosse:2003,Grosse:2004b} took the {\uim} contributions properly into account through a modif\/ication of the free Lagrangian by adding an oscillator like term with parameter $\Omega$. This term modif\/ies the spectrum of the free Hamiltonian. The harmonic oscillator term was obtained as a result of the renormalization proof. Remarkably, the model fulf\/ills the so-called Langmann--Szabo duality~\cite{Langmann:2002} relating short and long distance behaviour. There are indications that even a constructive procedure might be possible and give a non-trivial $\phi^4$ model, which is currently under investigation \cite{Magnen:2007}.

Then, the Orsay group around V.~Rivasseau presented another renormalizable model pre\-ser\-ving translational invariance \cite{Gurau:2009}, which we will refer to as the \emph{$1/p^2$ model}. The {\uim} is solved by a {\nl} additional term of the form $\phi  \frac 1{\Box}  \phi$.

There are attempts to generalize both of these models to the case of {\nc} gauge theory, which will be discussed in Sections~\ref{sec:canon:induced} and~\ref{sec:canon:p2inv}, respectively. In the former approach (the so-called {\emph induced gauge theory}), the starting point is the renormalizable, scalar Grosse--Wulkenhaar model. In a f\/irst step, the scalar f\/ield is coupled to an external gauge f\/ield. The dynamics of the gauge f\/ield can be extracted form the divergent contributions of the one-loop ef\/fective action \cite{Wallet:2007c,Wohlgenannt:2008zz}. This model contains explicit tadpole terms and therefore gives rise to a non-trivial vacuum. This problem has to be solved before the quantization and the renormalizability properties of the model can be studied. Recently, also a simplif\/ied version of the model has been discussed \cite{Blaschke:2007b,Blaschke:2009aw}. This model includes an oscillator potential for the gauge f\/ield, other terms occurring in the induced action, such as the tadpole terms, are omitted. Hence, the considered action is not gauge invariant, but BRST invariance could be established. Although the tadpoles are not present in the tree-level action, they appear as UV-counter terms at one-loop. Therefore, the induced action appears to be the better choice to study. Yet another approach in this direction exists, see~\cite{Buric:2010}. The scalar Grosse--Wulkenhaar model can be interpreted as the action for the scalar f\/ield on a curved background space~\cite{Wohlgenannt:2009}. In~\cite{Buric:2010}, a model for gauge f\/ields has been constructed on
the same curved space.

In the latter approach, dif\/ferent ways of implementing the $1/p^2$ damping behaviour have been advertised. The quadratic divergence of a {\nc} $U(1)$ gauge theory is known to be of the form
\begin{gather}
\label{eq:intro_main:ir-sing-gf}
\Pi^{\rm IR}_{\mu\nu} \sim \frac {\tilde k_\mu \tilde k_\nu}{(\tilde k^2)^2},
\end{gather}
where $\tilde k_\mu = \cth_{\mu\nu} k^\nu$. There are several possibilities to implement such a term in a gauge invariant way. In \cite{Blaschke:2008a}, the additional term
\begin{gather}
\label{eq:intro_main:inverse-covariant-ds}
F_{\mu\nu}   \frac 1{\tilde D^2 D^2} F_{\mu\nu} ,
\end{gather}
where $F_{\mu\nu}$ denotes the f\/ield strength and $D_\mu \cdot = \partial_\mu \cdot - \ig\starco{A_\mu}{\cdot}$ the {\nc} covariant derivative,
has been introduced in order to accommodate the IR divergences in the vacuum polarization. Since the covariant derivative contains gauge f\/ields, equation~\eqref{eq:intro_main:inverse-covariant-ds} is well def\/ined only as a power-series in the gauge f\/ield. This would produce vertices with an arbitrary number of f\/ields. Therefore, attempts have been made to localize the action by coupling them to unphysical auxiliary f\/ields. There are several ways to implement this, resulting in models with dif\/ferent properties, and even a modif\/ied physical content \cite{Blaschke:2009a,Blaschke:2009b,Blaschke:2009d,Vilar:2009}. In this respect one is led to the conclusion that only minimal couplings and the consequent construction of BRST doublet structures for all auxiliary f\/ields result in a stable theory.
In a recent development \cite{Blaschke:2009e}, the IR damping behaviour was consistently implemented in the so-called ``soft breaking'' term~-- a~method which is well known from the Gribov--Zwanziger approach to QCD~\cite{Gribov:1978,Zwanziger:1989,Zwanziger:1993}. In QCD, the soft-breaking is introduced in order to restrict the gauge f\/ields to the f\/irst Gribov horizon which removes any residual gauge ambiguities, and thereby cures the Gribov problem, see \secref{sec:canon:gribov}. In other words, one introduces an additional gauge f\/ixing in the IR without modifying the UV region, see also~\cite{Baulieu:2009}. In the {\nc} case we have to deal with a~similar problem: the IR region of the model requires a modif\/ication due to {\uim} while the symmetries, which ef\/fectively contribute to the renormalizability in the UV, shall not be altered.

\looseness=1
These two approaches resemble, in our opinion, the most promising candidates for a renormalizable model of {\nc} gauge theory. But since they are restricted to the canonical case, they have to be considered as toy models still. And certainly there are a lot more proposals which are interesting and will also be discussed here. Perhaps the most straightforward approach is given by an expansion in the small non-commutativity parameters. On canonically deformed spaces this will be discussed in \secref{sec:canon:theta-exp}, but also for $\kappa$-deformed spaces in \secref{sec:noncanon:kappa}. Seiberg--Witten maps~\cite{Seiberg:1999} relating the {\nc} f\/ields to the commutative ones (see \secref{sec:canon:theta-exp:SW}) are used. On canonically deformed space-time and to f\/irst order in~$\cth$, non-Abelian gauge models have been formulated e.g.\ in~\cite{Jurco:2001a}, and also in~\cite{Barnich:2003wq} with special emphasis on the ambiguities originating from the Seiberg--Witten map. As a success of this approach, a~{\nc} version of the Standard Model could be constructed in \cite{Wohlgenannt:2001,Melic:2005am,Wohlgenannt:2005b}. By consi\-de\-ring the expansion of the star products and the Seiberg--Witten maps only up to a certain order, the obtained theory is local. The model has the same number of coupling constants and f\/ields as the commutative Standard Model. A perturbative expansion in the {\nc} parameter was considered up to f\/irst order. To zeroth
order, the usual SM is recovered. At higher orders, new interactions occur. The renormalizability has been studied up to one loop. The gauge sector by its own turns out to be renormalizable~\cite{Buric:2006wm}, but the fermions spoil the picture and bring non-renormalizable ef\/fects into the game~\cite{Buric:2002}. This is also true for {\nc} QED (see \cite{Bichl:2001d,Wulkenhaar:2001}). Remarkably, for GUT inspired models~\cite{Martin:2007,Tamarit:2008,Martin:2009} one-loop multiplicative renormalizability of the matter sector could be established, at least on-shell.

$\cth$-expanded theories resemble a systematic approach to physics beyond the Standard Model and to Lorentz symmetry breaking and opens up a vast f\/ield of possible phenomenological applications, see e.g.~\cite{Trampetic:2002eb,Melic:2005am,Melic:2005su,Buric:2007qx,Schupp:2002up,Horvat:2009cm}. Ef\/fects intrinsically {\nc} in nature, such as the~{\uim}, are absent. Only when one considers the Seiberg--Witten maps to all orders in~$\cth$ those ef\/fects reappear \cite{Schupp:2008,Tureanu:2010}.

A dif\/ferent approach has been suggested by A.A.~Slavnov \cite{Slavnov:2003,Slavnov:2004} and will be discussed in some in detail in \secref{sec:canon:slavnov}. Additional constraints are introduced for pure gauge theory. This approach has been explored in detail and developed further in~\cite{Blaschke:2005c,Blaschke:2006a,Blaschke:2007a}.

The construction of models on $x$-dependent deformations is much more involved than the canonical case. Therefore, less results are known. In \secref{sec:noncanon:kappa}, we will discuss gauge models formulated on $\kappa$-deformed spaces. The Seiberg--Witten approach was applied in \cite{Dimitrijevic:2003,Dimitrijevic:2005c}, where f\/irst order corrections to the undeformed models could be computed for an arbitrary compact gauge group. In a recent work~\cite{Bolokhov:2008}, phenomenological implications have been studied by genera\-li\-zing that approach in a rather {\naiv} way to the Standard Model, thereby f\/inding bounds for the non-commutativity scale. Furthermore, the modif\/ication of the classical Maxwell equations have been discussed in~\cite{Harikumar:2010}.

On the fuzzy spaces e.g., f\/ields are represented by f\/inite matrices. Dif\/ferent approaches to gauge theory on the fuzzy sphere have been proposed in \cite{Klimcik:1997, Balachandran:1999, Karabali:2001, Iso:2001, Steinacker:2003, CastroVillarreal:2004}. Also non-perturbative studies are available, see e.g.~\cite{O'Connor:2006}, where Monte Carlo simulations have been performed. These approaches will be discussed in \secref{sec:noncanon:fuzzy}. Related to the fuzzy sphere, also fuzzy $\mathds{C}P^2$~\cite{Alexanian:2001qj,Grosse:2004c} has been considered.

Gauge theory on $q$-deformed spaces have been discussed in \cite{Boulatov:1996, Majid:1996, Mesref:2002, Schraml:2002fi, Meyer:2003wj} and will be reviewed brief\/ly in \secref{sec:noncanon:other}.

In this review, we will not cover supersymmetric theories, since that would be a review of its own. We only mention that in general, supersymmetric {\nc} models are ``less divergent'' than their non-supersymmetric counterparts~-- or even f\/inite (e.g. in the case of the IKKT matrix model which corresponds to $N=4$ {\nc} super Yang--Mills theory~\cite{Ishibashi:1996xs,Jack:2001cr,Matusis:2000jf}).
For some recent work on this topic, see e.g.
\cite{SheikhJabbari:1999iw,Rivelles:2000,Schweda:2000b,Zanon:2000,Ferrara:2000mm,Schweda:2002c,Ooguri:2003qp,Araki:2003se,Ooguri:2003tt,Ferrari:2004ex,Martin:2009mu} and references therein.

{\bf Relation with gravity.}
One of the motivations to introduce {\nc} coordinates was the idea to include gravitational ef\/fects into quantum f\/ield theory formulated on such deformed spaces. Having discussed {\nc} gauge models, let us pose the question, how these models are related to gravity. For a start, we will provide a simple example. Considering the {\moyal} product, $U_\star(1)$ gauge transformations\footnote{As explained in \secref{sec:covar}, $U_\star(1)$ denotes the star-deformed extension of the $U(1)$ gauge group.} contain f\/inite translations, see e.g.~\cite{Szabo:2006}:
\begin{gather*}
g_l(x) \star f(x) \star g_l^\dagger(x) = f(x+l) ,
\end{gather*}
where $g_l(x) = e^{-\ri l^i \theta^{-1}_{ij} x^j}$ and
\[
g_l(x) \star g_l^\dagger(x) = \id .
\]
Gauge transformations contain at least some space-time dif\/feomorphisms. The exact relation is still unknown.

However, the close relation with gravity is also studied in the framework of \emph{emergent gravity}\footnote{Other approaches to emergent gravity from {\nc} Yang--Mills models using Seiberg--Witten maps have been discussed e.g.\ in~\cite{Rivelles:2002,Yang:2004vd,Banerjee:2004rs}.} from matrix models, see e.g.~\cite{Yang:2006, Steinacker:2007, Steinacker:2008, Steinacker:2008a,Steinacker:2008ya}. The {\uim} terms are reinterpreted in terms of gravity. The starting point is a matrix model for {\nc} $U(N)$ gauge theory. The mixing results from the $U(1)$-sector and ef\/fectively describes $SU(N)$ gauge theory coupled to gravity. This approach will be brief\/ly described in \secref{sec:matrixmodels}.

Another relation has been discussed in \cite{Freidel:2005,Freidel:2005a}. L.~Freidel and E.R.~Levine could show that a quantum f\/ield theory symmetric under $\kappa$-deformed Poincar\'e symmetry describes the ef\/fective dynamics of matter f\/ields coupled to quantum gravity, after the integration over the gravitational degrees of freedom.

{\bf Outline.}
This review contains two main parts: In the f\/irst part, \secref{sec:covar}, we will discuss gauge models on canonically deformed spaces. Starting from the early approaches in \secref{sec:canon:early}, we will treat $\cth$-expanded theories (\secref{sec:canon:theta-exp}) employing Seiberg--Witten maps, discuss an approach initiated by A.A.~Slavnov (\secref{sec:canon:slavnov}) and end up with the recent developments generalizing the Grosse--Wulkenhaar model (\secref{sec:canon:induced}) and the $1/p^2$ model (\secref{sec:canon:p2inv}) to the realm of {\nc} gauge theories.

The second part, \secref{sec:noncanon}, deals with more general, $x$-dependent deformations. We start with the twisted approach (\secref{sec:noncanon:twisted}), which also includes the canonically deformed case as its simplest example, then we will focus on gauge models on $\kappa$-deformed (\secref{sec:noncanon:kappa}) and fuzzy spaces (\secref{sec:noncanon:fuzzy}), and conclude this section with reviewing the matrix model formulation in \secref{sec:matrixmodels}.

We will then close with some concluding remarks in \secref{sec:conclusion}.

{\bf Conventions.}
Quantities with ``hats'' either refer to operator valued expressions ($\hat x^i,\hat f(\hat x)$, \dots
$\in(\widehat{\mathcal A},\cdot )$) or, in the context of Seiberg--Witten maps, to {\nc} f\/ields and gauge parameters, respectively ($\widehat\psi, \widehat A, \widehat \alpha,\ldots \in(\mathcal A,\star)$) which can be expanded in terms of the ordinary commutative f\/ields and gauge parameters ($\psi,  A,\alpha \in (\mathcal A, \cdot)$). Quantities with a ``tilde'' are contracted with $\cth_{\mu\nu}$: $\tilde b_\alpha = \cth_{\mu\nu} b^\nu$, or for an object with two indices: $\tilde F = \cth_{\mu\nu} F^{\mu\nu}$; except for coordinates, where we def\/ine: $\tilde x_\mu = {\cth^{-1}}_{\mu\nu}x^\nu$. Furthermore, in \secref{sec:canon:p2inv_gauge_brsw} we use the matrix~$\mth$ rather than $\cth$ for contractions, using the def\/inition
\begin{gather*}
\cth_{\mu\nu} = \sth  \mth_{\mu\nu} ,
\end{gather*}
where $\sth$ has mass dimension $-2$.

\section{Canonical deformation}
\label{sec:covar}

In this section, we concentrate on canonically deformed four dimensional spaces. The commutator of space(-time) generators is given by
\begin{gather*}
[\hat x^i,\hat x^j] = \ri \cth^{ij} ,
\end{gather*}
where $\cth^{ij}$ is a real, constant and antisymmetric matrix. In what follows, we usually assume the following form for the deformation matrix
\begin{gather}
\label{eq:canon:theta_matrix}
(\cth^{ij}) = \sth \left(
\begin{array}{cccc}
0&1&0&0\\-1&0&0&0 \\0&0&0&1 \\0&0&-1&0
\end{array}
\right) ,
\end{gather}
for simplicity.
The corresponding star product of functions is the so-called {\moyal} product
\begin{gather}
\label{eq:canon:Weyl-Moyal}
\left(f\star g\right) (x) = e^{ \frac{\ri}{2} \cth^{ij} \partial^x_i \partial_j^y } f(x) g(y)\big|_{y\to x} .
\end{gather}
In general, the star product \eqref{eq:canon:Weyl-Moyal} represents an inf\/inite series. However, attempts have been made to make the star product local by introducing a bifermionic non-commutativity parameter~\cite{Vassilevich:2007}, so that this series becomes a f\/inite one.

The dif\/ferential calculus is unmodif\/ied, and the derivatives therefore commute:
\[
[\partial_i,\partial_j]=0 .
\]
Also, we can use the ordinary integral for the integration, and we note that it has some remarkable properties: First, one star can always be omitted and it shows the trace-property,
\begin{gather}
\int f \star g   =   \int d^4   x \,  (f\star g)(x) = \int d^4   x\,  f(x) g(x) ,\nonumber\\
\int f_1 \star f_2 \star \cdots \star f_n   =   \int f_2\star\cdots \star f_n \star f_1 = \int (f_2\star \cdots \star f_n)\cdot f_1
 .
\label{eq:canon:star-prod-properties}
\end{gather}
Variation with respect to the function $f_2$, e.g.\ is done in the following way:
\begin{gather*}
\nonumber
 \frac{\delta}{\delta f_2(y)} \int d^4 x \, (f_1 \star f_2 \star \cdots \star f_n)(x) = \frac{\delta}{\delta f_2(y)} \int d^4 x \, (f_2\star\cdots \star f_n \star f_1)(x)\\
\qquad{}  = \frac{\delta}{\delta f_2(y)} \int d^4 x\,  f_2(x)(f_3 \star \cdots \star f_n \star f_1)(x) = (f_3 \star \cdots \star f_n \star f_1)(y) .
\end{gather*}

In classical theory, the gauge parameter and the gauge f\/ield are Lie algebra
valued. Gauge transformations form a closed Lie algebra:
\begin{gather}
\label{eg:canon:commutative-gauge-trafo}
\delta_\alpha \delta_\beta - \delta_\beta \delta_\alpha =
\delta_{-\ri\co{\alpha}{\beta}},
\end{gather}
where $-i\co{\alpha}{\beta}=\alpha^a\beta_b f^{ab}_c T^c$, and $T^a$ denote the generators of the Lie group.
However, there is a~striking dif\/ference to the {\nc} case.
Let $M^\alpha$ be some matrix basis of the enveloping algebra of the internal
symmetry algebra. We can expand the gauge parameters in terms of this basis,
$\alpha=\alpha_a M^a$, $\beta=\beta_b M^b$. Then, two subsequent gauge transformations take the form
\begin{gather}
\label{eg:canon:nc-gauge-trafo}
\widehat \delta_\alpha \widehat\delta_\beta - \widehat\delta_\beta \widehat \delta_\alpha =
	\widehat\delta_{-\ri\starco{\alpha}{\beta}} .
\end{gather}
The $\star$-commutator of the gauge parameters is not Lie algebra valued any more:
\[
\starco{\alpha}{\beta} = \inv{2}\starco{\alpha_a}{\beta_b} \co{M^a}{M^b} + \inv{2}\starco{\alpha_a}{\beta_b} \aco{M^a}{M^b}.
\]
The dif\/ference to equation~\eqref{eg:canon:commutative-gauge-trafo} is the anti-commutator $\aco{M^a}{M^b}$, respectively the $\star$-commu\-ta\-tor of the gauge parameters, $\starco{\alpha_a}{\beta_b}$. This term causes the following problem: Let us assume that $M^\alpha$ are the Lie algebra generators. The anti-commutator of two Hermitian matrices is again Hermitian. But the anti-commutator of traceless matrices is in general not traceless. Therefore, the gauge parameter will in general be enveloping algebra valued. It has been shown~\cite{Terashima:2000,Matsubara:2000gr,Bars:2001,Chaichian:2001mu,Bonora:2000td} that only enveloping algebras, such as $U(N)$ or $O(N)$ and $USp(2N)$, survive the introduction of a deformed product (in the sense that commutators of algebra elements are again algebra elements), while e.g.\ $SU(N)$ does not. Despite this fact, star-commutators in general do not vanish. Hence, any {\moyal} deformed gauge theory is of the {\nA} type. In the general case, gauge f\/ields and parameters now depend on inf\/initely many parameters, since the enveloping algebra on {\moyal} space is inf\/inite dimensional. In order to emphasize this fact, we denote such algebras by $U_\star(N)$, $O_\star(N)$, $USp_\star(2N)$, \ldots, i.e.\ with subscript ``$\star$''. But nevertheless the parameters can be reduced to a f\/inite number, namely the classical parameters, by the so-called Seiberg--Witten maps which we will discuss in \secref{sec:canon:theta-exp:SW}.

Some non-perturbative results are available from lattice calculations \cite{Bietenholz:2006cz,Nishimura:2007ix} on the four-torus (i.e.\ periodic boundary conditions). There, space-time non-commutativity is assumed only in the $\{x_1,x_2\}$-plane, i.e.\ $\cth_{12} = - \cth_{21} = \cth$. A f\/irst order phase transition associated with the spontaneous breakdown of translational invariance in the {\nc} directions is observed. The order parameter is the open Wilson line carrying momentum. In the symmetric phase, the dispersion relation for the photon is modif\/ied:
\[
E^2 = \vec p\,{}^2 - \frac c{ (\cth p)^2} ,
\]
where $c$ is a constant. The IR singular contribution is responsible for the phase transition. In the broken phase, the dispersion relations is equal to the undeformed one. It shows the existence of a Goldstone mode associated to the spontaneous symmetry breaking. Non-perturbative results have also been obtained for the fuzzy sphere, see \secref{sec:noncanon:fuzzy}.

In \secref{sec:canon:early}, we will review some early approaches to {\nc} $U_\star(N)$ gauge theories, where in the commutative action the pointwise product has been replaced the {\moyal} product. Feynman rules have been calculated and above all renormalizability properties have been studied to one-loop. There, no expansion in the {\nc} parameters has been performed. Expanded models will be considered in \secref{sec:canon:theta-exp}. The gauge sectors turn out to be renormalizable, at least up to one-loop. But fermions are still not quite under control and introduce non-renormalizable ef\/fects. Then, we turn to the approach introduced by A.A.~Slavnov in \secref{sec:canon:slavnov}. The latest developments (which go in yet other directions) are discussed in Sections \ref{sec:canon:induced} and \ref{sec:canon:p2inv}, respectively. These approaches generalize the strategies which have been successful in the case of scalar theories.

\subsection{Early approaches}
\label{sec:canon:early}

In this section we brief\/ly review what we would like to call the ``{\naiv}'' attempts of introducing {\nc} actions, i.e.\ by considering those known from the commutative world and simply replacing pointwise by star products. We start with the {\nc} scalar $\phi^4$ model and then continue to gauge theories.

\subsubsection{Scalar f\/ield theories}
In replacing the ordinary pointwise product by the star product, a {\nc} extension to the scalar $\phi^4$ model is given by
\begin{gather}
\label{eq:canon:early:naive_act_complete}
S = \int d^4 x \left( \partial_\mu \phi \star \partial^\mu \phi + m^2 \phi \star \phi + \frac\lambda{4!} \phi\star\phi\star\phi\star\phi \right)
.
\end{gather}
The f\/irst one to consider this action was T.~Filk \cite{Filk:1996} who derived the corresponding Feynman rules, noticing that~-- at least in Euclidean space~-- the propagator is exactly the same as in commutative space, i.e.\ $G^{\phi\phi}(k)=1/k^2$, while the vertex gains phase factors (in this case a~combination of cosines) in the momenta. As a consequence, new types of Feynman graphs appear: In addition to the ones known from commutative space, where no phases depending on internal loop momenta appear and which exhibit the usual UV divergences, so-called {\np} graphs come into the game which are regularized by phases depending on internal momenta.
Other authors \cite{Minwalla:1999,Belov:2000,Matusis:2000jf,Micu:2000,Schweda:2002b} performed explicit one-loop calculations and discovered the infamous {\uim} problem:
Due to the phases in the {\np} graphs, their UV sector is regularized on the one hand, but on the other hand this regularization implies divergences for small external momenta instead.

For example the two point tadpole graph (in 4-dimensional Euclidean space) is approximately given by the integral
\[
\Pi(\L,p) \propto \l \intk \frac{2+\cos(k\p)}{k^2+m^2}
\equiv \Pi^{\rm UV}(\L) + \Pi^{\rm IR}(p) .
\]
The planar contribution is as usual quadratically divergent in the UV cutof\/f $\L$, i.e.\ $\Pi^{\rm UV}\sim \L^2$, and the {\np} part is regularized by the cosine to
\begin{gather}
\label{eq:canon:early:phi4-ir}
\Pi^{\rm IR} \sim \frac 1{\tilde p^2},
\end{gather}
which shows that the original UV divergence is not present any more, but reappears when $\p\to0$ (where the phase is 1) representing a new kind of infrared divergence. Since both divergences are related to one another, one speaks of ``{\uim}''. It is this mixing which renders the action \eqref{eq:canon:early:naive_act_complete} non-renormalizable at higher loop orders.

\subsubsection{Gauge f\/ield theories}

{\sloppy The pure star-deformed Yang--Mills (YM) action is given by
\[
 S_{\text{YM}}=\int d^D x \left(-\inv{4} F_{\mu\nu}\star F^{\mu\nu}\right),
\]
where the f\/ield strength tensor is def\/ined by
\begin{gather*}
F_{\mu\nu}   = \partial_\mu A_\nu - \partial_\nu A_\mu - \ri g \starco{A_\mu}{A_\nu}
 = -\ri \starco{\tilde x_\mu}{A_\nu} + \ri \starco{\tilde x_\nu}{A_\mu} -  \ri g \starco{A_\mu}{A_\nu} .
\end{gather*}
The corresponding Feynman rules for gauge f\/ield theories have been f\/irst worked out by C.P.~Mar\-t\'{\i}n and D.~S\'{a}nchez-Ruiz \cite{Martin:1999aq}.
M.~Hayakawa included fermions \cite{Hayakawa:1999,Hayakawa:1999b}, which leads to the action
\[
S_{\text{QED}} = \int d^D x \left(-\inv{4} F_{\mu\nu}\star F^{\mu\nu}+\bar \psi\star\gamma^\mu \ri D_\mu\psi-m\psi\star\bar\psi\right),
\]
with
\begin{gather*}
D_\mu A_\nu = \partial_\mu A_\nu - \ri g \starco{A_\mu}{A_\nu} .
\end{gather*}
Hayakawa's loop calculations showed that {\uim} is also present in gauge theories. Independently, A.~Matusis et al.~\cite{Matusis:2000jf} derived the same result. Further early papers in this context are \cite{Grosse:2000,Martin:2000,Liao:2001,Nakajima:2002}. Explicitly, F.~Ruiz Ruiz could even show that the quadratic and linear IR divergences in the $U(1)$ sector appear gauge independently\footnote{However, as discussed in the introduction one can improve the divergence behaviour by introduction of supersymmetry.}~\cite{Ruiz:2000}, and are therefore no gauge artefact. Furthermore, it was proven by using an interpolating gauge that quadratic IR divergences not only are independent of covariant gauges, but also of axial gauges \cite{Blaschke:2005b}.
As M.~van Raamsdonk pointed out \cite{VanRaamsdonk:2001jd}, the IR singularities have a natural interpretation in terms of a~matrix model formulation of YM theories.

}

Regarding the group structure of the {\nc} YM theory, A.~Armoni stressed the fact that $SU_\star(N)$ theory by itself is not consistent \cite{Armoni:2000xr,Armoni:2001}, and one should rather consider~$U_\star(N)$. In his computations, he showed that the planar sector leads to a $\b$-function with negative sign, i.e.\ a running coupling $g$, and that the infamous {\uim} arises only in those graphs which have at least one external leg in the $U_\star(1)$ subsector. Furthermore, in the limit $\th\to0$, $U_\star(N)$ does not converge to the usual $SU(N)\times U(1)$ commutative theory, which shows that the limit is non-trivial. One reason for this is that the $\b$-function is independent from $\th$, meaning that the $U(1)$ coupling still runs in that limit.

Nevertheless, up to one loop order, $U_\star(N)$ YM theory is renormalizable in a BRST invariant way. Furthermore, the Slavnov--Taylor identity, the gauge f\/ixing equation, and the ghost equation hold \cite{Martin:2000bk}. As in the {\naiv} scalar model of the previous subsection, {\uim} leads to non-renormalizability at higher loop order.

Finally, the {\nc} two-torus has been studied by several authors  \cite{SheikhJabbari:1999iw,Szabo:1998,Szabo:1999,Wulkenhaar:1999}.

A deformation of the Standard Model is discussed in \cite{Chaichian:2001py}. The authors start with the gauge group $U_\star(3)\times U_\star(2) \times U_\star(1)$.  In order to obtain the gauge group of the Standard Model one has to introduce a breaking and hence additional degrees of freedom. An alternative approach using Seiberg--Witten maps will be discussed in \secref{sec:canon:theta-exp:NCSM}.

\subsection[$\cth$-expanded theory]{$\boldsymbol{\cth}$-expanded theory}
\label{sec:canon:theta-exp}

As one generally assumes the commutator $\cth^{\m\n}$ to be very small (as mentioned in the introduction perhaps even of the order of the Planck length squared), it certainly makes sense to also consider an expansion of a {\nc} theory in terms of that parameter. In the expanded approach, {\nc} gauge theory is based on essentially
three principles,
\begin{itemize}\itemsep=0pt
\item Covariant coordinates,
\item Locality and classical limit,
\item Gauge equivalence conditions.
\end{itemize}

Let $\psi$ be a {\nc} f\/ield with inf\/initesimal gauge transformation
\[
\widehat\delta\psi(x) = \ri\alpha \star \psi(x)
,
\]
where $\alpha$ denotes the gauge parameter. The $\star$-product of a f\/ield and a coordinate does not transform covariantly,
\[
\widehat\delta (x\star \psi(x)) = \ri  x \star \alpha(x) \star \psi(x) \ne  \ri  \alpha(x) \star x \star \psi(x)
.
\]
Therefore, one has to introduce covariant coordinates \cite{Jurco:2000fs}
\[
X^\mu  \equiv  x^\mu + g \cth^{\mu\alpha} A_\alpha
,
\]
such that
\[
\widehat \delta(X^\mu \star \psi)=\ri\alpha \star (X^\mu \star \psi)
.
\]
Hence, covariant coordinates and the gauge potential transform under a {\nc} gauge
transformation in the following way
\[
\widehat \delta X^\mu = \ri\starco{\alpha}{X^\mu} ,\qquad
g  \widehat \delta A_\mu = \ri \cth^{-1}_{\mu\alpha} \starco{\alpha}{x^\alpha} + \ri g \starco{\alpha}{A_\mu}  ,
\]
where we have assumed that $\cth$ is non-degenerate. Other covariant objects can be constructed from covariant coordinates, such as the f\/ield strength,
\[
\ri g \cth^{\m \a} \cth^{\n\b}F_{\a\b} = \starco{X^\mu}{X^\nu} - \ri\cth^{\mu\nu} ,\qquad
\widehat \delta F^{\mu\nu} = \ri \starco{\alpha}{F^{\mu\nu}}  .
\]

\subsubsection{Seiberg--Witten maps}
\label{sec:canon:theta-exp:SW}

For simplicity, we will set the coupling constant $g=1$ in this section. The star product can be written as an expansion in a formal parameter $\sth$,
\[
f \star g = f\cdot g + \sum_{n=1}^\infty \sth^n C_n(f,g).
\]
In the commutative limit $\sth\to 0$, the star product reduces to the pointwise product of functions. One may ask, if there is a similar commutative limit for the f\/ields. The solution to this question was given for Abelian gauge groups by \cite{Seiberg:1999},
\begin{gather*}
\widehat A_\mu[A]   = A_\mu + \frac{\sth}{2}\mth^{\sigma\tau} \left(
	A_\tau\partial_\sigma A_\mu + F_{\sigma\mu} A_\tau\right)
	+\mathcal O\big(\sth^2\big) ,\nonumber\\
\widehat\psi[\psi, A]   = \psi + \frac{\sth}{2}\mth^{\mu\nu}A_\nu\partial_\mu \psi
	+\mathcal O\big(\sth^2\big) ,\nonumber\\
\widehat\alpha   = \alpha + \frac{\sth}{2}\mth^{\mu\nu}A_\nu\partial_\mu\alpha
	+\mathcal O\big(\sth^2\big).
\end{gather*}
The origin of this map lies in string theory. It is there that gauge invariance depends on the regularization scheme applied \cite{Seiberg:1999}. Pauli--Villars regularization provides us with classical gauge invariance
\[
\delta A_i=\partial_i\lambda,
\]
whence point-splitting regularization comes up with {\nc} gauge invariance
\[
\widehat \delta_\lambda \widehat A_i = \partial_i \widehat \Lambda
+\ri \starco{\widehat\Lambda}{\widehat A_i}
 .
\]
N.~Seiberg and E.~Witten argued that consequently there must be a local map from ordinary gauge theory to {\nc} gauge theory
\[
\widehat A[A], \quad \widehat\Lambda[\lambda,A] ,
\]
satisfying
\begin{gather}
\label{eq:canon:theta-exp:SW-4}
\widehat A[A + \delta_\lambda A] = \widehat A[A]+\widehat \delta_\lambda
\widehat A[A] ,
\end{gather}
where $\delta_\alpha$ denotes an ordinary gauge transformation and $\widehat \delta_{\alpha}$ a {\nc} one.
The Seiberg--Witten (SW) maps are solutions of the so-called ``gauge-equivalence relation'' \eqref{eq:canon:theta-exp:SW-4}. The solutions are not unique. Their ambiguities have been discussed in detail e.g. in \cite{Barnich:2003wq} using local BRST cohomology.

By locality we mean that in each order in the non-commutativity parameter $\sth$ there is only a f\/inite number of derivatives.
Let us remember that we consider arbitrary gauge groups. The {\nc} gauge f\/ields $\widehat A$ and gauge parameters $\widehat \Lambda$ are enveloping algebra valued. Let us choose a symmetric basis in the enveloping algebra, $T^a$,  $\inv{2}(T^aT^b+T^bT^a)$, $\dots$, such that
\begin{gather*}
\widehat \Lambda(x)   = \widehat \Lambda_a(x)T^a + \widehat \Lambda^1_{ab}(x)\,:T^aT^b: +
	\cdots  , \nonumber\\
\widehat A_\mu(x)   = \widehat A_{\mu a}(x) T^a + \widehat A_{\mu ab}(x)\, :T^aT^b:
	+ \cdots .
\end{gather*}
Equation~\eqref{eq:canon:theta-exp:SW-4} def\/ines the SW maps for the gauge f\/ield and the gauge parameter. However, it is more practical to f\/ind equations for the gauge parameter and the gauge f\/ield alone~\cite{Jurco:2001a}. First, we will concentrate on the gauge parameters $\widehat\Lambda$. We already encountered the consistency condition
\[
 \widehat \delta_\alpha \widehat\delta_\beta - \widehat\delta_\beta \widehat
 \delta_\alpha = \widehat\delta_{-\ri\starco{\alpha}{\beta}}
 ,
\]
which more explicitly reads
\begin{gather}
\label{eq:canon:theta-exp:SW-5}
\ri\widehat\delta_\alpha \widehat\beta[A] - \ri\widehat\delta_\beta \widehat\alpha[A] + \starco{\widehat\alpha[A]}{\widehat\beta[A]} = (\widehat{\co{\alpha}{\beta}})[A]
 .
\end{gather}
We can expand $\widehat\alpha$ in terms of $\sth$,
\[
\widehat\alpha[A] = \alpha + \alpha^1[A] +\alpha^2[A]+{\mathcal O}\big(\sth^3\big)
 ,
\]
where $\alpha^n$ is $\mathcal O(\sth^n)$. The consistency relation \eqref{eq:canon:theta-exp:SW-5} can be solved order by order in $\sth$:
\begin{gather}
0^{\textrm{th}} \textrm{ order}: \ \alpha^0   = \alpha
 , \nonumber\\
\label{eq:canon:theta-exp:SW-parameter}
1^{\textrm{st}} \textrm{ order}: \ \alpha^1   = \frac{\sth}{4}\mth^{\mu\nu}\aco{\partial_\mu \alpha}{A_\nu}
	  = \frac{\sth}{2}\mth^{\mu\nu}\partial_\mu\alpha_a A_{\mu b} :T^aT^b: .
\end{gather}
For f\/ields $\widehat \psi$ the condition
\begin{gather}
\label{eq:canon:theta-exp:SW-6}
\delta_{\alpha}\widehat\psi[A] = \widehat\delta_{\alpha}\widehat \psi[A] = \ri\widehat\alpha[A]\star \widehat\psi[A]
\end{gather}
has to be satisf\/ied. In other words, the ordinary gauge transformation induces a {\nc} gauge transformation. We expand the f\/ields in terms of the non-commu\-ta\-tivity
\[
\widehat\psi =  \psi^0 + \psi^1[A] + \psi^2[A] + \cdots ,
\]
and solve equation~\eqref{eq:canon:theta-exp:SW-6} order by order in $\sth$. In f\/irst order, we have to f\/ind a solution to
\[
\delta_\alpha\psi^1[A]=\ri\alpha\psi^1 + \ri \alpha^1 \psi - \frac{\sth}{2}\mth^{\mu\nu}\partial_\mu\alpha\partial_\nu\psi .
\]
It is given by
\begin{gather}
0^{\textrm{th}} \textrm{ order}: \  \psi^0   = \psi ,\nonumber\\
1^{\textrm{st}} \textrm{ order}: \  \psi^1   = -\frac{\sth}{2}\mth^{\mu\nu} A_\mu\partial_\nu\psi + \frac{\ri\sth}{4}\mth^{\mu\nu}
	A_\mu A_\nu\psi .\label{eq:canon:theta-exp:SW-fermion}
\end{gather}
The gauge f\/ields $\widehat A_\mu$ have to satisfy
\begin{gather}\label{eq:canon:theta-exp:SW-7}
\delta_{\alpha}\widehat A_\mu[A] =
\partial_\mu\widehat\alpha[A]+\ri \starco{\widehat\alpha[A]}{\widehat A_\mu[A]} .
\end{gather}
Using the expansion
\[
\widehat A_\mu[A] = A^0_\mu + A_\mu^1[A] + A_\mu^2[A] + \cdots  ,
\]
and solving \eqref{eq:canon:theta-exp:SW-7} order by order, we end up with
\begin{gather}
0^{\textrm{th}}\textrm{ order} : \  A^0_\mu    = A_\mu ,\nonumber\\
\label{eq:canon:theta-exp:SW-gauge}
1^{\textrm{st}}\textrm{ order} : \  A^1_\mu    = -\frac{\sth}{4}\mth^{\tau\nu} \aco{A_\tau}{\partial_\nu A_\mu + F_{\nu\mu}} ,
\end{gather}
where $F_{\nu\mu}=\partial_\nu A_\mu - \partial_\mu A_\nu - \ri \co{A_\nu}{A_\mu}$. Similarly, we have for the f\/ield strength $\widehat F_{\mu\nu}$
\begin{gather*}
\delta_\alpha \widehat F_{\mu\nu}   = \ri\co{\widehat \alpha}{\widehat F_{\mu\nu}} \qquad \mbox{and} \nonumber\\
\widehat F_{\mu\nu}   = F_{\mu\nu} + \frac{\sth}{2}\mth^{\sigma\tau}
	\aco{F_{\mu\sigma}}{F_{\nu\tau}} - \frac{\sth}{4} \mth^{\sigma\tau} \aco{A_\sigma}{(\partial_\tau+\mathcal D_\tau)F_{\mu\nu}} ,
\end{gather*}
where $\mathcal D_\mu F_{\tau\nu} = \partial_\mu F_{\tau\nu} - \ri\co{A_\mu}{F_{\tau\nu}}$.

\subsubsection{NC Standard Model}
\label{sec:canon:theta-exp:NCSM}

We start with the commutative Standard Model action and replace the respective f\/ields, e.g.\ fermions $\Psi$ and vector potentials $V_\mu$, by their Seiberg--Witten counterparts $\widehat \Psi[\Psi,V_\mu]$, $\widehat V_\mu[V_\nu]$, see \cite{Wohlgenannt:2001,Wohlgenannt:2003a}. Therefore, the {\nc} action reads
\begin{gather}
S_{\rm NCSM} =   \int d^4x \sum_{i=1}^3 \overline{\widehat \Psi}^{(i)}_L \star i
  \widehat{\fmslash D} \widehat \Psi^{(i)}_L
  +\int d^4x \sum_{i=1}^3 \overline{\widehat \Psi}^{(i)}_R \star i
  \widehat{\fmslash  D} \widehat \Psi^{(i)}_R  \nonumber \\
\phantom{S_{\rm NCSM} =}{} -\int d^4x \frac{1}{2 g'}
  \mbox{{\bf tr}}_{\bf 1} \widehat
  F_{\mu \nu} \star  \widehat F^{\mu \nu}
  -\int d^4x \frac{1}{2 g} \mbox{{\bf tr}}_{\bf 2} \widehat
  F_{\mu \nu} \star  \widehat F^{\mu \nu}
  -\int d^4x \frac{1}{2 g_S} \mbox{{\bf tr}}_{\bf 3} \widehat
  F_{\mu \nu} \star  \widehat F^{\mu \nu} \nonumber \\
\phantom{S_{\rm NCSM} =}{}
+ \int d^4x \bigg( \rho_0(\widehat D_\mu \widehat \Phi)^\dagger
  \star \rho_0(\widehat D^\mu \widehat \Phi)
 -\mu^2 \rho_0(\widehat {\Phi})^\dagger \star  \rho_0(\widehat \Phi)  \nonumber
	\\
\phantom{S_{\rm NCSM} =}{} - \lambda
  \rho_0(\widehat \Phi)^\dagger \star  \rho_0(\widehat \Phi)\star
  \rho_0(\widehat \Phi)^\dagger \star  \rho_0(\widehat \Phi)   \bigg) \nonumber \\
\phantom{S_{\rm NCSM} =}{} + \int d^4x \bigg ( -\sum_{i,j=1}^3  \bigg(
  W^{ij} ( \bar{ \widehat L}^{(i)}_L \star \rho_L(\widehat \Phi))
  \star  \widehat e^{(j)}_R
  + {W^\dagger}^{ij} \bar {\widehat e}^{(i)}_R \star (\rho_L(\widehat \Phi)^\dagger \star \widehat
  L^{(j)}_L) \bigg )  \nonumber \\
\phantom{S_{\rm NCSM} =}{}
 -\sum_{i,j=1}^3  \bigg(
  G_u^{ij} ( \bar{\widehat Q}^{(i)}_L \star \rho_{\bar Q}(\widehat{\bar\Phi}))\star
  \widehat u^{(j)}_R
  + {G_u^\dagger}^{ij} \bar {\widehat u}^{(i)}_R \star
  (\rho_{\bar Q}(\widehat{\bar\Phi})^\dagger
  \star \widehat Q^{(j)}_L) \bigg )   \nonumber \\
\phantom{S_{\rm NCSM} =}{} - \sum_{i,j=1}^3  \bigg(
  G_d^{ij} ( \bar{ \widehat Q}^{(i)}_L \star \rho_Q(\widehat \Phi))\star
  \widehat d^{(j)}_R
  + {G_d^\dagger}^{ij} \bar{ \widehat d}^{(i)}_R \star (\rho_Q(\widehat \Phi)^\dagger
  \star \widehat Q^{(j)}_L) \bigg ) \bigg) .\label{eq:canon:theta-exp:action}
\end{gather}
There is a lot of new notation which we now will gradually introduce. We have to emphasize that there is an ambiguity in the choice of the kinetic terms for the gauge f\/ields. In the commutative case, gauge invariance and renormalizability uniquely determine the dynamics. However, a~principal like renormalizability is not applicable here. Before we come back to this problem, let us brief\/ly def\/ine the particle content and some of the symbols. Left handed fermions are denoted by $\Psi_L$, leptons by $L$ and quarks by $Q$, $\Psi_R$ stands for the right handed fermions:
\[
\Psi^{(i)}_L = \left( \begin{array}{c}  L^{(i)}_L \\ Q^{(i)}_L \end{array} \right) ,\qquad
\Psi^{(i)}_R = \left( \begin{array}{c}  e^{(i)}_R \\ u^{(i)}_R \\   d^{(i)}_R \end{array} \right) ,\qquad
{\Phi} = \left(\begin{array}{c}  \phi^+ \\  \phi^0  \end{array} \right ) .
\]
The index $(i)\in\{1,2,3\}$ denotes the generations, and $\phi^+$ and $\phi^0$ are the complex scalar f\/ields of the scalar Higgs doublet. The gauge group of the Standard Model is $SU(3)_C \times SU(2)_L \times U(1)_Y$. The Seiberg--Witten map of a tensor product of gauge groups is not uniquely def\/ined \cite{Aschieri:2002mc}. We will discuss here only the most symmetric choice. The commutative gauge f\/ield is given by
\begin{gather*}
V_\mu = g' A_\mu (x) Y + \frac{g}{2} \sum_{a=1}^3 B_{\mu a}\sigma^a + \frac{g_S}{2} \sum_{a=1}^8 G_{\mu a} \lambda^a ,
\end{gather*}
where $g' A_\m(x)$ corresponds to the hypercharge symmetry $U(1)_Y$, $B_\mu (x)=\frac{g}{2}B_{\mu a}(x)\sigma^a$
to the weak $SU(2)_L$, and $G_\mu(x) = \frac{g_S}{2}G_{\mu a}(x)\lambda^a$ to the strong interaction $SU(3)_C$. The Pauli matrices are denoted by $\sigma^b$, $b=1,2,3$ and the Gell-Mann matrices by $\lambda^a$, $a=1,\dots,8$. The according gauge parameter has the form{\samepage
\begin{gather*}
{\Lambda} = g' \alpha (x)Y+\frac{g}{2} \sum_{a=1}^{3} \alpha^L_{a}(x) \sigma^a + \frac{g_S}{2} \sum_{b=1}^{8} \alpha^S_{b}(x) \lambda^b .
\end{gather*}
The Seiberg--Witten maps are given by equations~\eqref{eq:canon:theta-exp:SW-parameter}, \eqref{eq:canon:theta-exp:SW-fermion} and \eqref{eq:canon:theta-exp:SW-gauge}, respectively.}

Let us now consider the Yukawa coupling terms in equation~\eqref{eq:canon:theta-exp:action} and their behaviour under gauge transformations. They involve products of three f\/ields, e.g.
\begin{gather}
\label{eq:canon:theta-exp:Yukawa-sample}
-\sum_{i,j=1}^3  \bigg( W^{ij} \big( \bar{ \widehat L}^{(i)}_L \star \rho_L(\widehat \Phi)\big) \star  \widehat e^{(j)}_R + {W^\dagger}^{ij} \bar {\widehat e}^{(i)}_R \star \big(\rho_L(\widehat \Phi)^\dagger \star \widehat L^{(j)}_L\big) \bigg )
 .
\end{gather}
Only in the case of commutative space-time does $\Phi$ commute with the generators of the $U(1)_Y$ and $SU(3)_C$ groups. Therefore, the Higgs f\/ield needs to transform from both sides in order to ``cancel charges'' from the f\/ields on either side (e.g., $\bar{ \widehat L}^{(i)}_L$ and $\widehat e^{(j)}_R$ in \eqref{eq:canon:theta-exp:Yukawa-sample}). The expansion of~$\widehat\Phi$ transforming on the left and on the right under arbitrary gauge groups is called hybrid SW map~\cite{Wohlgenannt:2001},
\begin{gather*}
\widehat\Phi[\Phi,A,A']    =  \phi + \frac{1}{2}\theta^{\mu\nu} A_\nu \left(\partial_\mu\phi -\frac{i}{2} (A_\mu \phi + \phi A'_\mu)\right)\nonumber\\
\phantom{\widehat\Phi[\Phi,A,A']    =}{}  - \frac{1}{2}\theta^{\mu\nu} \left(\partial_\mu\phi - \frac{i}{2} (A_\mu \phi + \phi A'_\mu)\right)A'_\nu
 + \mathcal{O}(\theta^2) ,
\end{gather*}
with gauge transformation $\widehat \delta \widehat\Phi = \ri\widehat\Lambda \star \widehat\Phi-\ri\widehat\Phi \star \widehat\Lambda'$. In the above Yukawa term \eqref{eq:canon:theta-exp:Yukawa-sample}, we have $\rho_L(\widehat \Phi)= \widehat
\Phi[\phi,V,V']$, with
\begin{gather*}
V_\mu  = - \inv{2}g' \mathcal A_\mu + g B_\mu^aT_L^a, \qquad
V'_\mu  = g' \mathcal A_\mu.
\end{gather*}
We need a dif\/ferent representation for $\widehat\Phi$ in each of the Yukawa couplings:
\begin{gather*}
\rho_Q(\widehat\Phi)   = \widehat \Phi \left[\phi, \frac{1}{6} g' {\cal A}_\mu + g B^a_\mu T^a_L + g_S G_\mu^a T^a_S, \frac{1}{3} g' {\cal A}_\nu - g_S G_\nu^a T^a_S  \right] ,\\
\rho_{\bar Q}(\widehat\Phi)   = \widehat\Phi\left[\phi, \frac{1}{6} g' {\cal A}_\mu + g B^a_\mu T^a_L + g_S G_\mu^a T^a_S, -\frac{2}{3} g' {\cal A}_\nu - g_S G_\nu^a T^a_S\right] .
\end{gather*}
The respective sum of the gauge f\/ields on both sides gives the proper quantum
numbers for the Higgs f\/ield.

As we have mentioned earlier, the kinetic terms for the gauge f\/ield in the classical theory are determined uniquely by the requirements of gauge invariance and renormalizability. In the {\nc} case, we do not have a principle like renormalizability at hand. Gauge invariance alone does not f\/ix these terms in the Lagrangian. Therefore, the representations to be used in the trace of the kinetic terms for the gauge bosons are not uniquely determined. For the simplest choice~-- leading to the so-called {\it Minimal Non-Commutative Standard Model}, we have the form displayed in the action~\eqref{eq:canon:theta-exp:action},
\[
-\int d^4x \frac{1}{2 g'} \mbox{{\bf tr}}_{\bf 1} \widehat F_{\mu \nu} \star \widehat F^{\mu \nu} - \int d^4x \frac{1}{2 g} \mbox{{\bf tr}}_{\bf 2} \widehat F_{\mu \nu} \star \widehat F^{\mu \nu} - \int d^4x \frac{1}{2 g_S} \mbox{{\bf tr}}_{\bf 3} \widehat F_{\mu \nu} \star \widehat F^{\mu \nu} ,
\]
where $\mbox{{\bf tr}}_{\bf 1}$ denotes the trace over the $U(1)_Y$ sector with
\[
Y = \inv{2} \left( \begin{array}{cc} 1 & 0 \\ 0 & -1 \end{array} \right) ,
\]
$\mbox{{\bf tr}}_{\bf 2}$ and $\mbox{{\bf tr}}_{\bf 3}$ are the usual $SU(2)_L$ and $SU(3)_C$ matrix traces, respectively.
On the other hand, in considering a Standard Model originating from a $SO(10)$ GUT theory~\cite{Aschieri:2002mc}, these terms are f\/ixed uniquely.

A perhaps more physical (\emph{non-minimal}) version of the Non-Commutative Standard Model is obtained, if we consider a charge matrix $Y$ containing all the f\/ields of the Standard Model with covariant derivatives acting on them. For the simplicity of presentation we will only consider one family of fermions and quarks. Then the charge matrix has the form
\begin{gather*}
Y = \left( \begin{array}{cccccccc} -1 &&&&&&& \\ & -1/2 &&&&&& \\ && -1/2 &&&&& \\ &&& 2/3 &&&& \\ &&&& 2/3 &&& \\ &&&&& 2/3 & \\
&&&&&& -1/3 & \\ &&&&&&& \ddots \end{array} \right) .
\end{gather*}					
The kinetic term for the gauge f\/ield is then given by
\[
S_{\rm gauge} = - \int d^4x\, \mbox{Tr}\, \inv{2G^2} \widehat F_{\mu\nu} \star \widehat F^{\mu\nu},
\]
where $\widehat F_{\mu\nu}=\partial_\mu \widehat V_\nu -\partial_\nu \widehat V_\mu - \ri\starco{\widehat V_\mu}{\widehat V_\nu}$.
The operator $G$ encodes the coupling constants of the theory.

The last missing ingredient to equation~\eqref{eq:canon:theta-exp:action} is the representation $\rho_0$ of the Higgs f\/ield:
\[
\rho_0(\widehat \Phi)=\phi+\rho_0(\phi^1)+\mathcal{O}\big(\sth^2\big) ,
\]
with
\[
\rho_0(\phi^1) = - \frac{1}{2} \cth^{\alpha \beta} (g'{\cal A}_\alpha+g B_\alpha)  \partial_\beta \phi + \frac{\ri}{8} \cth^{\alpha \beta}  \co{ g'{\cal A}_\alpha+g B_\alpha}{g'{\cal A}_\beta+g B_\beta}   \phi .
\]
The full action expanded up to f\/irst order in the {\nc} parameters and the respective Feynman rules can be found in \cite{Wohlgenannt:2001,Melic:2005am,Wohlgenannt:2005b}. The expansion up to second order has been discussed in \cite{Moller:2004qq,Alboteanu:2007bp,Trampetic:2007hx}.

Let us emphasize here, that there is no problem with dif\/ferent charges. Because of its {\nA} nature, the {\nc} photon can only couple to particles with charges $\pm q$ and~$0$ \cite{Hayakawa:1999b,Wohlgenannt:2003a}. Hence, for a particle with charge $q'$ dif\/ferent from~$+q$ or~$-q$ another {\nc} photon has to be introduced. But due to the Seiberg--Witten map, no new degrees of freedom are added, since the expansions of all {\nc} photons only depend on the {\it one} commutative f\/ield.

The special of case of $\cth$-deformed QED has been discussed in \cite{Bichl:2001b} and \cite{Carlson:2002wj}. In the latter reference, $\cth^{\m\n}$ has been promoted to a Lorentz tensor.

Some results on the renormalizability of $\cth$-expanded theories are also available. In general, we can say that the gauge sector alone is much better behaved than the situation where matter is included. Already for QED, evidence is found that the gauge sector is renormalizable. The photon self energy turns out to be renormalizable to all orders both in $\cth$ and $\hbar$ \cite{Bichl:2001d}, see also \cite{Martin:2006gw}. Heavy use is made of the enormous freedom available in the Seiberg--Witten maps. However, if one tries to include matter f\/ields the renormalizability is lost \cite{Wulkenhaar:2001,Buric:2002}.

The same holds true in the case of the {\nc} Standard Model, at least to one-loop and f\/irst order in $\cth$. The renormalizability of the gauge sector of a non-minimal {\nc} Standard Model was studied in \cite{Buric:2006wm}, whereas pure $SU(N)$ gauge theory was discussed in \cite{Buric:2005xe,Latas:2007eu}. In both cases, the model is one-loop renormalizable. The freedom in the Seiberg--Witten maps is f\/ixed~-- to this order~-- by the renormalizability condition. One further encouraging step could be performed in \cite{Buric:2007ix}, where the authors could show that in {\nc} chiral~$U(1)$ and $SU(2)$ gauge theory the 4-fermion vertex is UV f\/inite, again to one-loop and f\/irst order in $\cth$. In previous models with Dirac fermions, this vertex resembled one reason for their non-renormalizability. The same result was obtained for GUT inspired models \cite{Martin:2009sg}. First steps to include the fermionic sector have been performed in~\cite{Tamarit:2008} in the case of the {\nc} Standard Model. GUT inspired theories have been studied in~\cite{Tamarit:2008,Martin:2009}, where the authors computed the UV divergent contributions to the one-loop background f\/ield ef\/fective action. Remarkably, they could show by explicit calculations that even the matter sector is one-loop multiplicatively renormalizable, at least on-shell.

{\Nc} anomalies have been calculated in \cite{Banerjee:2001un,Martin:2005jy}, in the latter reference for {\nc} $SU(N)$; there, the anomaly could be related to the Atiyah--Singer index theorem, whereas in \cite{Brandt:2003fx} it could be showed that Seiberg--Witten expanded gauge theories have the same one-loop anomalies as their commutative counterparts.

As we have mentioned earlier, the Seiberg--Witten maps give rise to new couplings and decay modes, which might be forbidden or highly suppressed in the commutative Standard Model~\cite{Trampetic:2002eb}. As an example let us mention the coupling of photons to neutral particles, and the decay $Z \to \gamma \gamma$. From the study of such processes one can obtain bounds on the non-commutativity scale \cite{Melic:2005am,Melic:2005su,Buric:2007qx}. For some general references on {\nc} particle phenomenology, see e.g.~\cite{Hewett:2000zp,Carlson:2001sw,Kersting:2001zz,Carone:2004wt} and references therein. The Seiberg--Witten map has also been applied to astrophysical scenarios. In~\cite{Schupp:2002up,Minkowski:2003jg}, left and right-handed neutrinos are coupled to photons. Bounds for the {\nc} scale are presented from estimates for the induced energy loss in stars \cite{Schupp:2002up} and from comparison of Dirac/Majorana neutrino dipole moments \cite{Minkowski:2003jg}. Big bang nucleosynthesis is used in \cite{Horvat:2009cm} in order to constrain the scale of {\nc} ef\/fects.

In the following sections, we will discuss some {\nc} gauge models formulated \emph{without} explicit expansions in the non-commutativity parameter $\cth^{\m\n}$, where the main goal is to overcome the {\uim} problem.

\subsection{The Slavnov approach}
\label{sec:canon:slavnov}

In 2003, A.A.~Slavnov~\cite{Slavnov:2003,Slavnov:2004} suggested a way of dealing with arising IR singularities in {\nc} gauge theories by adding a further term in the action. This {\em Slavnov term} has the form
\begin{gather*}
\inv{2}\intx\l\star\cth^{\mu\nu}F_{\mu\nu} ,
\end{gather*}
where $\cth^{\mu\nu}$ is once again the deformation parameter of {\nc} space-time, $F_{\mu\nu}=\pa_\m A_\n-\pa_\n A_\m-\ig\starco{A_\m}{A_\n}$ is the f\/ield strength tensor, and $\l$ is a {\em dynamical} multiplier f\/ield\footnote{We will clarify what is meant by ``dynamical multiplier f\/ield'' in a moment.} leading to a new kind of constraint. This constraint modif\/ies the gauge f\/ield propagator $G^A_{\mu\nu}(k)$ in such a way that it becomes transverse with respect to $\k^\mu=\cth^{\mu\nu}k_\nu$. This is important, since the vacuum polarization $\Pi^{\mu\nu}$ of (4-dimensional) gauge theories is characterized by the quadratically IR singular structure given in equation~\eqref{eq:intro_main:ir-sing-gf}, which is proportional to $\sim \k^\mu\k^\nu / (k^2)^2$ (where~$k_\mu$ represents the external momentum). Higher loop insertions of the IR divergent $\Pi^{\mu\nu}_{\text{IR-div}}$ into internal gauge boson loops therefore vanish. Slavnov's idea was motivated by the results of one loop calculations of {\nc} gauge theories previously done by M.~Hayakawa~\cite{Hayakawa:1999} and others revealing that the leading IR divergent term has the form (\ref{eq:intro_main:ir-sing-gf}), which incidentally is gauge independent~\cite{Ruiz:2000,Blaschke:2005b}~-- and this gauge independence survives after adding the Slavnov term~\cite{Blaschke:2005c}.

Furthermore, it was shown~\cite{Blaschke:2006a,Blaschke:2007a} that the Slavnov term may be identif\/ied with a topological term similar to the BF models~\cite{Schweda:1996,Schweda:1998,Schweda:1999,Gieres:2000}, e.g.:
\[
\Act_{\text{2-dim-BF}} =\int d^2x\, B\vare^{\m\n}F_{\m\n}.
\]
However, the Slavnov term leads to new Feynman rules involving propagators and vertices of the multiplier f\/ield $\l$ (which is why we previously have emphasized that it is a dynamical f\/ield). This means one has to deal with additional (and potentially divergent) Feynman graphs.

\subsubsection{The Slavnov-extended action and its symmetries}

In~\cite{Blaschke:2006a}, the following action in 3+1 dimensional Minkowski space with commuting time, i.e.\ $\cth^{0i}=0$ (and for simplicity also $\cth^{ij}=\cth\e^{ij}$ where $\e^{ij}$ is the 2 dimensional Levi-Civita symbol), was considered:
\begin{gather}\label{eq:slavnov_action-2dim}
\Act = \intx \left( -\inv{4}F_{\mu\nu}\star F^{\mu\nu} +\frac{\cth}{2}\l \star\e^{ij}F_{ij} +b\star n^{i}A_{i}-\bc\star n^{i}D_{i}c \right) .
\end{gather}
The axial gauge f\/ixing was chosen to coincide with the {\nc} plane ($x_1,x_2$), i.e.\ $i\in \{1,2\}$. With these choices the Slavnov term, together with the gauge f\/ixing terms, have the form of a 2-dimensional topological BF model (cf.~\cite{Blaschke:2006a} and references therein). This action is invariant under the BRST transformations
\begin{alignat*}{3}
&sA_\mu = D_\mu c   ,  \qquad &&s \bar c  = b   , &   \\
 &s\l  = - \ri g \, [\l, c ]   ,\qquad  &&sb = 0   , & \\
&sc  =   \frac{\ri g}{2}    [c , c ]   ,\qquad  &&s^2=0 , &
\end{alignat*}
and additionally the gauge f\/ixed action is invariant under a (non-physical) \emph{linear vector supersymmetry $($VSUSY$)$}, whose f\/ield transformations are
\begin{alignat}{3}
&\d_iA_\mu = 0   ,\qquad  &&\d_ic=A_i  ,& \nonumber\\
&\d_i\bc=0   ,\qquad &&\d_ib=\partial_i\bc   , & \nonumber\\
&\d_i\l  = \frac{ \e_{ij} }{\cth} n^j\bc   ,\qquad  &&\d^2=0 . & \label{eq:slavnov_susy-vsusy-orig}
\end{alignat}
Since the operator $\d_i$ lowers the ghost-number by one unit, it represents  an antiderivation (very much like the BRST operator $s$ which raises the  ghost-number by one unit). One has to note, that only the interplay of appropriate choices for $\cth^{\mu\nu}$ and $n^\mu$ lead to the existence of the VSUSY.

In contrast to the pure topological theories, there is an additional vectorial symmetry:
\begin{gather*}
 \hd_iA_J = -F_{iJ}   , \qquad\hd_i\l=-\eth{ij} D_K F^{Kj}   , \qquad
 \hd_i\Phi = 0  \qquad   \mbox{for all other f\/ields}  .
\end{gather*}
This further symmetry (which does not change the ghost number) is in fact a (non-linear) symmetry of the gauge invariant action. Its existence is due to the presence of the Yang--Mills part of the action which, in contrast to the BF-type part, involves also $A_0$ and $A_3$. Notice that the algebra involving $s$, $\d_i$, $\hd_i$ and the ($x_1,x_2$)-plane translation generator $\partial_i$ closes on-shell (cf.~\cite{Blaschke:2006a}). (The reader not interested in the technical details of deriving the total action and related Ward identities, may proceed directly to their consequences on page~\pageref{jump:slavnov}.)

In order to derive a Slavnov--Taylor (ST) identity expressing the invariance of an appropriate total action $\St$ under the symmetries discussed above, one can combine the various symmetry operators into a generalized BRST operator that we denote by $\tri$:
\begin{gather}\label{eq:slavnov_combined-op}
\tri\equiv s +\xid + \vare ^i \d_i+\mu^i\hd_i\qquad \text{with}\quad  \xid \equiv \xi^i\partial_i .
\end{gather}
Here, the constant parameters $\xi^i$ and $\mu^i$ have ghost number $1$, and $\vare^i$
has ghost number $2$. The induced f\/ield variations read
\begin{gather}
 \tri A_i=D_ic+\xid A_i  ,\nonumber\\
 \tri A_J=D_Jc+\xid A_J+\mu^iF_{Ji}  ,\nonumber\\
 \tri\l=- \ri g \, {[ \l, c]} + \xid \l+\vare^i \eth{ij}n^j\bc +\mu^i\eth{ij}D_KF^{jK}  ,\nonumber\\
 \tri c= \frac{\ri g}{2} \, {[c,  c] } +\xid c+\vare^i A_i  ,\nonumber\\
 \tri\bc=b+\xid \bc ,\nonumber\\
 \tri b=\xid b+\eid\bc ,\label{eq:slavnov_trafo-combined}
\end{gather}
and imposing that the parameters $\xi^i$, $\vare^i$ and $\mu^i$   transform as
\begin{gather}\label{eq:slavnov_transpara}
\tri\xi^i=\tri\mu^i=-\vare^i   , \qquad \tri\vare^i=0   ,
\end{gather}
one concludes that the operator (\ref{eq:slavnov_combined-op}) is nilpotent on-shell.
Finally, one has to introduce an external f\/ield $\Phi^*$ (i.e.\ an antif\/ield in the terminology of Batalin and Vilkovisky~\cite{Batalin:1981,Batalin:1983}) for each f\/ield $\Phi \in \{ A_{\mu} ,\l,c \}$ since the latter transform non-linearly under the BRST variations~-- see e.g.~\cite{Piguet:1995}. In view of the transformation laws (\ref{eq:slavnov_trafo-combined}) and (\ref{eq:slavnov_transpara}), the \emph{ST identity} then reads
\begin{gather}
0= \mathcal{S}(\St) \equiv \intx
\Bigg\{ \sum_{\Phi \in \{ A_{\mu} ,\l,c \}} \var{\St}{\Phi^*}\var{\St}{\Phi} + \left( b+ \xid \bc \right) \var{\St}{\bc} \nonumber\\
\phantom{0= \mathcal{S}(\St) \equiv}{}  + \left( \xid b + \eid \bc\right) \var{\St}{b} \Bigg\}
-\vare^i \left( \diff{\St}{\xi^i} + \diff{\St}{\mu^i} \right)  .\label{eq:slavnov_slavnov-id}
\end{gather}
This functional equation is supplemented with the \emph{gauge-fixing condition}
\begin{gather}\label{eq:slavnov_gauge-cond}
\var{\St}{b} = n^i A_i  .
\end{gather}

{\bf Total action.}
By dif\/ferentiating the ST identity with respect to the f\/ield $b$, one f\/inds
\[
0 = \var{}{b} \mathcal{S}(\St) = \mathcal{G} \St - \xid  \var{\St}{b} , \qquad \text{with} \quad \mathcal{G} \equiv \var{\ }{\bc} + n^i \var{\ }{A^{*i}} ,
\]
i.e., by virtue of (\ref{eq:slavnov_gauge-cond}), the so-called \emph{ghost equation}:
\begin{gather}\label{eq:slavnov_ghost-eq}
\mathcal{G} \St = \xid   (n^i A_i)  .
\end{gather}
The associated homogeneous equation $\mathcal{G} \bar S =0$ is solved by functionals which we denote $\bar S [ \hAstar^i , \dots ]\!$ and which depend on the variables $A^{*i}$ and $\bc$ only through the shifted antif\/ield
\begin{gather}\label{eq:slavnov_hat-A-star}
\hAstar^i  \equiv A^{*i}  -n^i\bc .
\end{gather}
Thus, the functional $\St[ A,\l,c,\bc, b   ;A^*,\l^*,c^* ;\xi,\mu,\vare] $ which solves both the ghost equation~(\ref{eq:slavnov_ghost-eq}) and the gauge-f\/ixing condition~(\ref{eq:slavnov_gauge-cond}) has the form
\begin{gather}\label{eq:slavnov_S-bar}
\St = \intx\, (b + \xid \bc ) n^i A_i
  +   \bar S[A,\l,c   ; \hAstar^i, A^{*J}  ,\l^*,c^*;   \xi,\mu,\vare]  ,
\end{gather}
where the $b$-dependent term ensures the validity of condition (\ref{eq:slavnov_gauge-cond}).

By substituting expression (\ref{eq:slavnov_S-bar}) into the ST identity (\ref{eq:slavnov_slavnov-id}), one concludes that the latter equation is satisf\/ied if $\bS$ solves the \emph{reduced ST identity}
\begin{gather}\label{eq:slavnov_reduced-ST}
0 = \BB (\bS) \equiv
\sum_{ \Phi \in \{ A_{\mu} ,\l,c \} } \intx
\var{\bS}{{\hat\Phi}^*}\var{\bS}{\Phi}
  -   \vare^i \, \left( \diff{\bS}{\xi^i}   +   \diff{\bS}{\mu^i} \right)  .
\end{gather}
Here, ${\hat\Phi}^*$ collectively denotes all antif\/ields, but with $A^{*i}$ replaced by the shifted antif\/ield (\ref{eq:slavnov_hat-A-star}). Following standard practise~\cite{Piguet:1995}, we introduce the following notation for the external sources in order to make the formulae clearer:
\[
\rho^{\mu} \equiv A^{*\mu}    , \qquad
\gamma \equiv \l ^*     , \qquad
\sigma  \equiv c^*  ,\qquad {\hat\rho}^i = \hAstar^i  .
\]
It can be verif\/ied in the usual way (e.g.\ see~\cite{Piguet:1995}) that the \emph{solution of the reduced ST iden\-ti\-ty}~(\ref{eq:slavnov_reduced-ST}) is given by\footnote{Simply insert (\ref{eq:slavnov_Sbar}) into (\ref{eq:slavnov_reduced-ST}) to check that it really solves the ST identity.}
\begin{gather}
\bS   = \intx  \Bigg\{ {-}\inv{4}F_{\mu\nu} F^{\mu\nu}
+ \frac{\cth}{2} \l \e^{ij}F_{ij}   +   \hat\rho^i \left( D_ic+\xid A_i \right)
+\rho^{J} \left( D_Jc+\xid A_J+\mu^iF_{Ji} \right) \nonumber \\
\phantom{\bS   =}{}  +   \gamma \left( -\ri g {[\l ,c] } +\xid \l
+\mu^i \eth{ij} D_KF^{jK} \right)
+\s \left( \frac{\ri g}{2}   {[c,c]}+\xid c+\vare^i A_i \right) \nonumber \\
\phantom{\bS   =}{} + \left( \frac{\mu^i\mu^j}{2}  \eth{ij} (D_J \rho^{J} ) + \vare^i \eth{ij} \hat\rho^j-\vare^i\inv{2\cth^2} (D_i\gamma)\right)\gamma \Bigg\} .\label{eq:slavnov_Sbar}
\end{gather}
Note that
\[
\bS = S_{\rm inv} + S_{\rm antif\/ields} + S_{\rm quadratic}  ,
\]
where $S_{\rm inv}$ is the gauge invariant part (i.e.\ the f\/irst two terms) of the action (\ref{eq:slavnov_action-2dim}), $S_{\rm antif\/ields}$ represents the linear coupling of the shifted antif\/ields  ${\hat\Phi}^*$ to the generalized BRST transformations~(\ref{eq:slavnov_trafo-combined}) (the $\bc$-dependent term being omitted) and $S_{\rm quadratic}$, which is quadratic in the shifted antif\/ields, ref\/lects  the contact terms appearing in the closure relations $\tri^2\Phi$.

{\bf Ward identities.}
The Ward identities describing the (non-)invariance of $\St$ under the VSUSY variations $\d_i$, the vectorial symmetry transformations $\hat{\d} _i$ and the translations $\partial _i$ can be derived from the ST identity~(\ref{eq:slavnov_slavnov-id}) by dif\/ferentiating this identity with respect to the corresponding constant ghosts $\vare^i, \, \mu^i$ and $\xi^i$, respectively.

For instance, by dif\/ferentiating (\ref{eq:slavnov_slavnov-id}) with respect to $\xi^i$ and by taking the gauge-f\/ixing condition (\ref{eq:slavnov_gauge-cond}) into account, one obtains the \emph{Ward identity for translation symmetry}:
\[
0=\diff{}{\xi^i}\mathcal{S}(\St) =\sum_{\varphi}
\intx \,  \partial_i \varphi   \var{\St}{\varphi}  ,
\]
where $\varphi \in \{ A_{\mu} , \l, c , \bar c, b; A^*_{\mu} , \l^*, c^* \}$. By dif\/ferentiating (\ref{eq:slavnov_slavnov-id}) with respect to $\vare^i$, we obtain a~\emph{broken Ward identity for the VSUSY:}
\begin{gather*}
\mathcal{W}_i \St= \Delta_i  ,
\end{gather*}
with
\begin{gather}
\mathcal{W}_i \St= \intx  \Bigg\{ \partial_i\bc \, \var{\St}{b}+A_i   \var{\St}{c} +\left(\eth{ij}\left(n^j\bc-\rho^j\right)+\inv{\cth^2} D_i\gamma\right)\var{\St}{\l}\nonumber\\
\phantom{\mathcal{W}_i \St=}{}  + \gamma \eth{ij}   \var{\St}{A_j}
+\left(\s+ \frac{\ri g}{\cth^2}\gamma\gamma\right)\var{\St}{\rho^i}\Bigg\} ,\label{eq:slavnov_ward22}
\end{gather}
and
\begin{gather*}
\Delta_i  = \Delta_i \Big|_{\xi=\mu=0} + b_i [\xi , \mu] ,\nonumber\\
\Delta_i \Big|_{\xi=\mu=0}  =\intx   \left\{\s\partial_ic -\rho^{\mu} \partial_i A_{\mu} -\gamma\partial_i\l-\rho^J  F_{Ji}+\gamma\eth{ij}\left(n^jb
- D_K F^{jK}\right) \right\} , \nonumber \\
b_i [\xi , \mu]  = \intx   \left\{ \xid \bc   \eth{ij} n^j \gamma
+ \eth{ij} \mu^j \left(D_J \rho^J\right) \gamma \right\} .
\end{gather*}
Note that the f\/ield variations given by (\ref{eq:slavnov_ward22}) extend the VSUSY transformations (\ref{eq:slavnov_susy-vsusy-orig}) by source dependent terms. It is the presence of the sources which leads to a breaking $\Delta_i$ of the VSUSY.

In the same spirit, the broken Ward identity for the bosonic vectorial symmetry $\hd_i$ is obtained by dif\/ferentiating the ST identity (\ref{eq:slavnov_slavnov-id}) with respect to $\mu^i$. One f\/inds:
\begin{gather*}
\intx\Bigg\{ -F_{iJ}   \var{\St}{A_J}-\eth{ij}\left(D_KF^{Kj}
+\mu^j   D_K\rho^K\right)\var{\St}{\l}+D_K\rho^K   \var{\St}{\rho^i}\nonumber\\
\qquad{} + \eth{ij}D_KD^K\gamma   \var{\St}{\rho_j}-\left( D_i\rho^I+\eth{ij}D^jD^I \gamma +\ri g\eth{ij} \co{F^{Ij}}{\gamma}\right)\var{\St}{\rho^I}\nonumber\\
\qquad{} + \ri g \eth{ij} \mu^j \co{\rho^I}{\gamma}\var{\St}{\rho^I}\Bigg\}  = -\intx  \eth{ij}   \vare ^j \left(D_K\rho^K\right) \gamma .
\end{gather*}

\begin{figure}[t]
\centering
\includegraphics[scale=0.8]{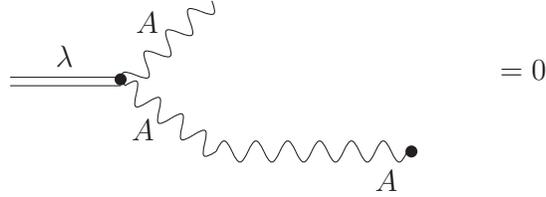}
\caption{The $\l AA$-vertex contracted with a photon propagator vanishes.}
\label{fig:slavnov_vertex-prop}
\end{figure}

{\bf Consequences.}
\label{jump:slavnov}
The linear VSUSY, in particular, has some important consequences which shall now be discussed: The generating functional $Z^c$ of the connected Green functions is given by the Legendre transform of the generating functional $\Gamma$ of the one-particle irreducible Green functions. At the classical level (tree graph approximation) one has $\Gamma\sim\Act$, and hence for vanishing antif\/ields the Ward identity describing the linear vector supersymmetry in terms of~$Z^c$ in the tree graph approximation is given by
\[
\mathcal{W}_i Z^c=\int d^4x \left\{j_b  \partial_i \var{Z^c}{j_{\bc}} - j_c  \var{Z^c}{j_A^i}
+ \eth{ij} n^j  j_{\l}   \var{Z^c}{j_{\bc}}  \right\}=0 ,
\]
where $\{j_A^\mu,j_\l,j_b,j_c,j_{\bc}\}$ are sources of $\{A_\mu,\l,b,c,\bc\}$, respectively. Varying this expression with respect to $j_c$ and $j^\mu_A$ yields for the gauge f\/ield propagator:
\begin{gather}\label{eq:slavnov_prop-relation}
G_{A_iA_\mu}=0 .
\end{gather}
In other words, as soon as one of its indices is either 1 or 2, the gauge f\/ield propagator is zero. As the $\l AA$-vertex is proportional to $\cth_{ij}$, which in this model is non-vanishing only in the ($x_1,x_2$)-plane, relation (\ref{eq:slavnov_prop-relation}) has the following important consequence for the Feynman graphs: \emph{The combination of gauge boson propagator and $\l AA$ vertex is zero} (see Fig.~\ref{fig:slavnov_vertex-prop}).

Furthermore, it is impossible to construct a closed loop including a $\l AA$-vertex without having such a combination somewhere. \emph{Hence, all loop graphs involving the $\l AA$-vertex vanish}.

In particular, dangerous vacuum polarization insertions involving the additional Feynman rules (i.e.\ the $\l$-propagator, the mixed $\l A$-propagator and the $\l AA$-vertex) vanish. This is the reason, why the model is free of the most dangerous, i.e.\ the quadratic, infrared singularities, as pointed out by Slavnov~\cite{Slavnov:2004} for the special case of $n^\mu=(0,1,0,0)$.

\subsubsection{Further generalization of the Slavnov trick}

Now the question arises whether one can show the cancellation of IR singular Feynman graphs for a more general choice of $\cth^{\mu\nu}$ and $n^\mu$. The answer is yes, but one has to impose stronger Slavnov constraints. The initial Slavnov constraint was $\cth^{12}F_{12}+\cth^{13}F_{13}+\cth^{23}F_{23}=0$ and with ``stronger'' we mean that each term in the sum should vanish separately. Upon imposing these stronger conditions one may write for the action (cf.~\cite{Blaschke:2007a}):
\begin{gather*}
\Act_{\text{inv}} =\intx\left[-\inv{4}F_{\mu\nu} F^{\mu\nu}+ \inv{2}\e^{ijk}F_{ij}\l_k\right],
\end{gather*}
with $i,j,k\in\{1,2,3\}$. This action looks like a 3 dimensional BF model coupled to Maxwell theory. As in the pure BF-case, the action has two gauge symmetries
\begin{alignat*}{3}
&\d_{g1}A_\mu=D_\mu\Lambda ,\qquad&& \d_{g2}A_\mu=0, & \nonumber\\
&\d_{g1}\l_k=-\ig\co{\l_k}{\Lambda} ,\qquad&& \d_{g2}\l_k=D_k\Lambda' . &
\end{alignat*}
Similar to the previous model, we have an additional bosonic vector symmetry of the gauge invariant action:
\begin{gather*}
 \hd_iA_0 = -F_{i0}  , \qquad\hd_i\l_j=\e_{ijk} D_0 F^{0k}   ,\qquad
 \hd_iA_j = 0  .
\end{gather*}
There is, however, a dif\/ference to the previous case: The additional vectorial symmetry is broken when f\/ixing the second gauge symmetry $\d_{g2}$.

If one considers a space-like axial gauge f\/ixing of the form\footnote{$d'=d-\ri g\co{\bphi}{c}$ is the redef\/ined multiplier f\/ield f\/ixing the second gauge freedom $\d_{g2}$.}
\[
S_{{\rm gf}} = \intx \left[ b n^iA_i+d'n^i\l_i-\bc n^iD_ic-\bphi n^iD_i\phi\right],
\]
the gauge f\/ixed action is invariant under the linear VSUSY
\begin{gather*}
 \d_ic=A_i   , \qquad\d_i\l_j=-\e_{ijk} n^k \bc  ,\qquad \d_ib=\partial_i\bc  ,\nonumber\\
\d_i\Phi=0\qquad\text{for all other f\/ields},
\end{gather*}
in addition to the usual BRST invariance. The Ward identity describing the linear vector supersymmetry in terms of~$Z^c$ at the classical level is given by
\begin{gather*}
\mathcal{W}_iZ^c=\int d^4x\left[  j_b\partial_i\var{Z^c}{j_{\bc}}-j_{c} \var{Z^c}{j_A^i} +\e_{ijk}n^j j_\l^k\var{Z^c}{j_{\bc}}\right]=0.
\end{gather*}
Hence, the same arguments as before show the absence of IR singular graphs. However, the model exhibits numerous further symmetries which have been discussed in~\cite{Blaschke:2007a}.

One should also note, that a generalization to higher dimensional models is possible. For example if $\l$ had $n$ indices the VSUSY would become
\begin{gather*}
 \d_ic=A_i , \qquad \d_i\l_{j_1\cdots j_n}=\e_{ikj_1\cdots j_n}n^k\bc ,\qquad
 \d_ib=\partial_i\bc ,
\end{gather*}
after appropriate redef\/initions of Lagrange multipliers.

In conclusion, one can state that Slavnov-extended Yang--Mills theory can be shown to be free of the worst infrared singularities, if the Slavnov term is of BF-type. Furthermore, supersymmetry, in the form of VSUSY, seems to play a decisive role in theories which are not Poincar\'e supersymmetric. Another open question is what role the VSUSY plays with respect to {\uim} in topological NCGFT in general.

However, a general proof of renormalizability for this type of models is still missing. Furthermore, the Slavnov-extended models have a major drawback: The Slavnov constraint reduces the degrees of freedom of a gauge model (see~\cite{Slavnov:2004}) and hence it seems that it does not describe {\nc} ``photons''.

\subsection{Models with oscillator term}
\label{sec:canon:induced}

To avoid the {\uim} problem, several models which involve an oscillator like counter term have been put forward. On the one hand such models break translation invariance due to the explicit $x$-dependence of the action,
but on the other hand they in general show a much better divergence behaviour at higher loops or are even (in the case of the scalar Grosse--Wulkenhaar model) proven to be renormalizable. In the following, we will present the Grosse--Wulkenhaar model followed by three gauge models based on similar ideas.

\subsubsection{The Grosse--Wulkenhaar model}
\label{sec:canon:induced_gw}

In 2004, the f\/irst renormalizable {\nc} scalar f\/ield model (in Euclidean $\R_\th^4$) was introduced by H.~Grosse and R.~Wulkenhaar \cite{Grosse:2004b} (for a Minkowskian version see reference \cite{Szabo:2008}). Their trick was to add a harmonic oscillator-like term to the action
\begin{gather}\label{eq:induced_Grosse Wulkenhaar action}
  S[\phi]=\int d^4x\left(\frac{1}{2}\partial_\mu\phi\star\partial_\mu\phi+\frac{\mu_0^2}{2}\phi\star\phi
 +\frac{\Omega^2}{4}(\tilde x\phi)\star(\tilde x\phi)+\frac{\lambda}{4!}\phi\star\phi\star\phi\star\phi\right),
\end{gather}
with $\tilde x_\mu=\left(\cth_{\mu\nu}\right)^{-1}x_\nu$ ($\cth_{\mu\nu}$ constant and antisymmetric). This action cures the infamous {\uim} problem. Indeed, for the bad IR-behaviour found in the {\naiv} model (triggered by the kinetic part of the action), the oscillator term acts as a sort of counter term. By exchanging $\tilde x\leftrightarrow p$ one can see that the action stays form invariant:
\[
 S[\phi;\mu_0,\lambda,\Omega]\mapsto\Omega^2S\left[\phi;\frac{\mu_0}{\Omega},\frac{\lambda}{\Omega^2},\frac{1}{\Omega}\right] .
\]
This symmetry is called Langmann--Szabo duality \cite{Langmann:2002}, and at the self dual point, $\Omega=1$, it is even exact.

The propagator of the model is the inverse of the operator $(-\Delta+\Omega^2\tilde x^2+\mu_0^2)$, and is called the Mehler kernel \cite{Grosse:2003}. It takes the form
\begin{gather*}
 K_M(x,y)=\int\limits_0^\infty d\alpha\,\frac{1}{4\pi^2\omega\sinh^2\alpha}
 e^{-\frac{1}{4\omega}\left(u^2\coth\frac{\alpha}{2}+v^2\tanh\frac{\alpha}{2}\right)-\omega\mu_0^2\alpha} ,
\end{gather*}
with  $\omega=\frac{\theta}{\Omega}$, $u=x-y$ being a \textit{short variable} and $v=x+y$ being a \textit{long variable}. This notation has been introduced by V.~Rivasseau et al.~\cite{Rivasseau:2005a}. They conf\/irmed the renormalizability of the model by making use of a technique called Multiscale Analysis, additionally to the renormalization proof of H.~Grosse and R.~Wulkenhaar which has been given in the matrix base employing the Polchinski approach.

The Mehler kernel features a damping behaviour for high momenta (UV) as well as for low momenta (IR). One can see this by comparison with the heat kernel, which is the inverse of $H_0=-\Delta+\mu_0^2$ and has the form
\[
 H_0^{-1}=
 \int\limits_0^\infty d\alpha\,\frac{1}{16\pi^2\alpha^2}e^{-\frac{(x+y)^2}{2\alpha}-\mu_0^2\alpha}.
\]
For $\mu_0=0$, one f\/inds the well-known form of the undamped propagator after integrating over $\alpha$
\[
 H_0^{-1}=\frac{1}{8\pi^2(x-y)^2}.
\]
When setting $y=0$ and $\mu_0=0$ in the Mehler kernel, one can perform the integration over the auxiliary Schwinger parameter and obtain
\[
 K_M(x)=\frac{e^{-\frac{x^2}{4 \omega
   }}}{\pi ^2 x^2} ,
\]
which shows that the Mehler kernel has a much stronger convergence behaviour for large values of~$x$, corresponding to small values of $p$. However, the price to pay seems to be that translation invariance is broken, which can be seen directly in the action, because of the explicit $x$-dependence of the oscillator term $\tilde x^2\phi^2$. Recently, as discussed in \secref{sec:canon:geometrical}, it has been shown that this term can be interpreted as a coupling to the curvature of a background space, giving it a nice geometrical interpretation.

The renormalizability of the model can also be beautifully illustrated by a quick glance at the beta function for $\lambda$, which can be found for example in \cite{Disertori:2006uy}.
In contrast to the {\naiv} scalar model (without oscillator term) the beta function becomes constant for high energies. Hence it does not diverge, and is therefore free of the Landau ghost problem \cite{Landau:1955,Grosse:2004a,Rivasseau:2006b}.

\subsubsection{Extension to gauge theories}
\label{sec:canon:induced_ext2gauge}

The aim is to obtain propagators for gauge models with a damping behaviour similar to the Mehler kernel in the scalar case. Since an oscillator term $\Omega^2\tilde x^2A^2$ is not gauge invariant, there are more or less two possible ways to construct the model: either one adds further terms in order to make the action gauge invariant (which will be discussed in the following section) or one views the oscillator term as part of the gauge f\/ixing. H.~Grosse, M.~Schweda and one of the present authors (D.~Blaschke) put forward a model which follows the latter approach \cite{Blaschke:2007b}. The action is given by
\begin{gather}
\Gamma^{(0)} =S_{\text{inv}}+S_{\text{m}}+S_{\text{gf}} ,\nonumber\\
S_{\text{inv}} =\inv{4}\int d^4x  F_{\mu\nu}\star F_{\mu\nu} ,\nonumber\\
S_{\text{m}} =\frac{\Omega^2}{4}\int
d^4x\left(\inv{2}\staraco{\tilde x_\mu}{A_\nu}\star\staraco{\tilde x_\mu}{A_\nu}+\staraco{\tilde x_\mu}{\bc}\star\staraco{\tilde x_\mu}{c} \right)
=\frac{\Omega^2}{8}\int d^4x\left(\tilde x_\mu\star\mathcal{C}_\mu\right),\nonumber\\
S_{\text{gf}} =\int d^4x\left[b\star\partial_\mu A_\mu-\inv{2}b\star b-\bc\star\partial_\mu sA_\mu-\frac{\Omega^2}{8} \widetilde c_\mu\star s\,\mathcal{C}_\mu\right],\label{eq:induced_new-action}
\end{gather}
with
\begin{gather*}
F_{\mu\nu} =\partial_\mu A_\nu-\partial_\nu A_\mu-\ig\starco{A_\mu}{A_\nu},\nonumber\\
\mathcal{C}_\mu =\Big(\staraco{\staraco{\tilde x_\mu}{A_\nu}}{A_\nu}+\starco{\staraco{\tilde x_\mu}{\bc}}{c}+\starco{\bc}{\staraco{\tilde x_\mu}{c}}\Big),\nonumber\\
\tilde x_\mu =\left(\cth^{-1}\right)_{\mu\nu}x_\nu.
\end{gather*}
The gauge f\/ield $A_\mu$ transforms under the {\nc} generalization of a $U(1)$ gauge transformation which is inf\/inite by construction of the {\nc} algebra. Once more, we denote the gauge group by $U_\star(1)$ in order to distinguish it from the commutative $U(1)$ gauge group.

The multiplier f\/ield $b$ implements a non-linear gauge f\/ixing\footnote{Notice, that in the limit $\W\to0$ this becomes the Feynman gauge.}:
\begin{gather*}
\var{\Gamma^{(0)}}{b}
 =\partial_\mu A_\mu-b+\frac{\Omega^2}{8}\Big(\starco{\staraco{\tilde x_\mu}{c}}{\widetilde c_\mu}
 -\staraco{\tilde x_\mu}{\starco{\widetilde c_\mu}{c}}\Big)=0.
\end{gather*}
The f\/ield $\widetilde c_\mu$ is an additional multiplier f\/ield which guarantees the BRST-invariance of the action. The BRST-transformations are given by
\begin{alignat}{3}
&sA_\mu=D_\mu c=\partial_\mu c-\ig\starco{A_\mu}{c}, \qquad && s\bc=b, & \nonumber\\
&sc=\ig{c}\star{c}, \qquad && sb=0, & \nonumber\\
&s\widetilde c_\mu=\tilde x_\mu, \qquad && s\tilde x_\mu=0, & \nonumber\\
&s^2\varphi=0\ \forall\ \varphi\in\left\{A_\mu,b,c,\bc,\widetilde c_\mu\right\},\qquad &&& \label{eq:induced_BRST}
\end{alignat}
Since $\widetilde c_\mu$ transforms into $\tilde x_\mu$, the part of the action including the Lagrange-multiplier f\/ield $\widetilde c_\mu$ exactly cancels with $S_m$ under the application of the BRST-operator $s$ onto the whole action.
With these BRST transformations the action \eqref{eq:induced_new-action} can be written in the following beautiful form:
\begin{gather*}
\Gamma^{(0)} = \int d^4x \left(
\frac 14 F_{\mu\nu}\star F_{\mu\nu} + s\left( \frac{\Omega^2}8 \tilde c_\mu \star \mathcal{C}_\mu +
\bar c\star \partial_\mu A_\mu - \frac 12 \bar c \star b \right) \right).
\end{gather*}

{\bf Feynman rules.}
When we assume $\cth_{\mu\nu}$ to be antisymmetric and constant, i.e.\ \[
( \cth_{\mu\nu} )
=\sth\left(\begin{array}{cccc}
0&1&0&0\\
-1&0&0&0\\
0&0&0&1\\
0&0&-1&0
\end{array}
\right)   ,
\]
as def\/ined at the beginning of \secref{sec:covar}, the following property holds:
\[
 \staraco{A_\mu}{\tilde x_\mu}=2\tilde x_\mu A_\mu ,
\]
which can be directly verif\/ied by inserting the def\/inition of the star product~\eqref{eq:canon:Weyl-Moyal}. It is therefore possible to reduce the bilinear parts of the action to one single star. The latter can be removed by the cyclic permutation property of the star product~\eqref{eq:canon:star-prod-properties},
and therefore the non-interacting part of the action is the same as in an undeformed model. Hence the propagators are more or less just the Mehler kernels, like in the scalar case. In momentum space they are given by
\[
 G^{AA}_{\mu\nu}(p,q) =(2\pi)^4 \widetilde{K}_M(p,q) \delta_{\mu\nu},
\qquad
 G^{\bar cc}(p,q) =(2\pi)^4 \widetilde{K}_M(p,q) ,
\]
with the Mehler kernel in momentum representation
\begin{gather}\label{eq:induced_Mehler kernel in momentum space}
 \widetilde{K}_M(p,q)=\frac{\omega^3}{8\pi^2}\int\limits_0^\infty d\alpha\frac{1}{\sinh^2\alpha}
 e^{-\frac{\omega}{4}(p-q)^2\coth\frac{\alpha}{2}-\frac{\omega}{4}(p+q)^2\tanh\frac{\alpha}{2}} .
\end{gather}

The $\widetilde cbc$-vertex involving the multiplier f\/ield $\widetilde c_\mu$ does not contribute to Feynman diagrams since a propagator connecting to that f\/ield does not exist. Similarly, a propagator does exist for~$b$, but the corresponding vertex as stated do not contribute to loop diagrams. Hence, we will omit the related Feynman rules.

The vertices following from the action are just the usual {\nc} ones, as can be found for example in \cite{Hayakawa:1999b}.
Equipped with the complete Feynman rules we can start deriving a power counting formula to estimate the worst degree of divergence of our graphs, which via {\uim} is directly related to the degree of {\nc} IR divergence. We will not give a detailed derivation here but instead quote only the f\/inal result. (For further details we refer the interested reader to~\cite{Blaschke:2009aw}.) Given the number of external legs for the various f\/ields (denoted by $E_\varphi$, $\forall\, \varphi\in\left\{A_\mu,b,c,\bc,\widetilde c_\mu\right\}$) the degree of UV divergence for an arbitrary graph in 4-dimensional space can be up-bounded by
\begin{gather}
 d_\gamma=4-E_A-E_{c/\bar c}-E_{\widetilde c}-2E_b.
\label{eq:induced_powercounting}
\end{gather}
This bound, however, represents merely a crude estimate. The true degree of divergence can (for certain graphs) be improved by gauge invariance. For example, for the one-loop boson self-energy graphs the power counting formula would predict at most a quadratic divergence, but gauge invariance usually reduces the sum of those graphs to be only logarithmically divergent. In our case we will show, however, that this does not happen due to a violation of translation invariance. The corresponding Ward identity will be worked out more explicitly in the next subsection.

\begin{figure}[t]
\centering
\includegraphics[scale=0.8]{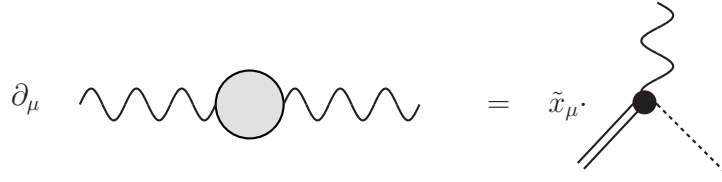}
\caption{Ward identity replacing transversality.}
\label{fig:induced_almost_transversality_relation}
\end{figure}

{\bf Symmetries.}
In this subsection, we will take a closer look at the Ward identities (describing translation invariance) and the Slavnov--Taylor identities (describing BRST invariance). Every symmetry in general implies a conservation operator that gives zero when applied to the action. In the case of the BRST symmetry this is $s$. Regarding $s$ as a total derivation of $\Gamma^{(0)}$ we can write
\begin{gather*}\nonumber
 s\Gamma^{(0)}[A_\mu,b,c,\bar c,\widetilde c_\mu]
 \\
\qquad{} =\int d^4x\,\left( sA_\mu\star\dif{\Gamma^{(0)}}{A_\mu}+sb\star\dif{\Gamma^{(0)}}{b}
 +sc\star\dif{\Gamma^{(0)}}{c}+s\bar c\star\dif{\Gamma^{(0)}}{\bar c}
 +s\widetilde c_\mu\star\dif{\Gamma^{(0)}}{\widetilde c_\mu}\right).
\end{gather*}
By introducing external sources $\rho_\mu$ and $\sigma$ for $sA_\mu$ and $sc$, respectively
\begin{gather*}
 \Gamma =\Gamma^{(0)}+\Gamma_{\text{ext}} ,\qquad
\Gamma_{\text{ext}} =\intx\left(\rho_\mu \star sA_\mu+ \sigma \star sc\right)\,,
\end{gather*}
and making use of \eqref{eq:induced_BRST} we can write the Slavnov--Taylor identity in a more convenient form:
\begin{gather*}
 \text{S}(\Gamma)=\int d^4x
 \left(\dif{\Gamma}{\rho_\mu}\star\dif{\Gamma}{A_\mu}+\dif{\Gamma}{\sigma}\star\dif{\Gamma}{c}+b\star\dif{\Gamma}{\bar{c}}
 +\tilde x_\mu\star\dif{\Gamma}{\tilde c_\mu}\right)=0 .
\end{gather*}
To arrive now at the Ward identity describing translation invariance, one has to take as usual the functional derivative of the Slavnov--Taylor identity with respect to $A_\rho$ and $c$. One immediately recognizes that the $\tilde x_\mu$-term which originates from the oscillator term in the action gives an additional contribution. The usual translation invariance is explicitly broken:
\begin{gather}\label{eq:induced_transversality_relation}
 \partial^z_\mu\vvar{\Gamma}{A_\rho(y)}{A_\mu(z)}
 =\int d^4x\left(\tilde x_\mu\frac{\delta^3\Gamma}{\delta c(z)\delta A_\rho(y)\delta\tilde{c}_\mu(x)}\right)
 =\frac{1}{2\omega^2}\staraco{y_\rho}{\delta^4(y-z)}\neq 0 .
\end{gather}
Graphically this can be depicted as shown in Fig.~\ref{fig:induced_almost_transversality_relation}.

\begin{figure}[t]
\centering
\includegraphics[scale=0.65]{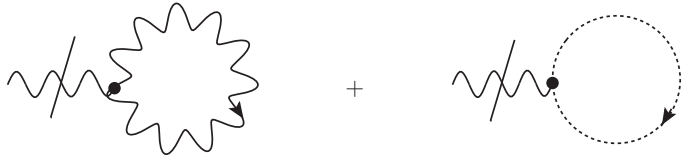}
\caption{Tadpole graphs.}
\label{fig:induced_tadpoles}
\end{figure}

{\bf Loop calculations.}
The simplest graphs one may construct are the (one-point) tadpoles, consisting of just one vertex and one internal propagator.
They consist of two graphs which are depicted in Fig.~\ref{fig:induced_tadpoles}.
According to the Feynman rules, their sum is straightforwardly given by
\[
\Pi_\mu(p) =2\ig\intk \intk'\d^4\left(p+k'-k\right)\sin\left(\frac{k\p}{2}\right)K_M(k,k')\left[2k_\mu+3k'_\mu\right] .
\]
We may now transform to ``long and short'' variables
\[
  u=k-k',\quad v=k+k'\quad \Rightarrow\quad k=\frac{v+u}{2},\quad k'=\frac{v-u}{2} ,
\]
with functional determinant $\frac{1}{16}$. Moreover,
we make use of
\[
\sin\left(\frac{k\tilde p}{2}\right)=\sum_{\eta=\pm1}\frac{\eta}{2\ri}\exp\left(\frac{\ri\eta}{2}k\tilde p\right),
\]
and plug in the explicit expression for the Mehler kernel~\eqref{eq:induced_Mehler kernel in momentum space}. Altogether this leads to
\begin{gather*}
\Pi_\mu^\vare(p) =\frac{g\omega^3}{2^{8}\pi^2}\sum\limits_{\eta=\pm1}\int d^4v\left[5v_\mu-p_\mu\right]\int\limits_{\vare}^{\infty}d\a\frac{\eta \re^{\frac{\ri\eta}{4}v\p}}{\sinh^2\a}\exp\left(-\tfrac{\omega}{4}\left[\coth\left(\tfrac{\a}{2}\right)p^2+\tanh\left(\tfrac{\a}{2}\right)v^2\right]\right)\nonumber\\
\phantom{\Pi_\mu^\vare(p)}{}
=\frac{5\ig\p_\mu}{64}\int\limits_{\vare}^{\infty}d\a\frac{\cosh\left(\frac{\alpha}{2}\right)}{\sinh^5\left(\frac{\a}{2}\right)}
 \exp\left[-\inv{4}\coth\left(\frac{\a}{2}\right)\left(\omega+\frac{\th^2}{4\omega}\right)p^2\right] ,
\end{gather*}
where in the last step the Gaussian integral has been solved and trigonometric identities have been used. Furthermore we have introduced a cutof\/f $\vare=1/\Lambda^2$ which regularizes the integral.

Na{\"i}vely, one could simply integrate out $\alpha$ and discover a divergence structure of higher degree than expected, since it still contains a ``smeared out'' delta function. To make this clear, consider the usual commutative propagator, which depends on a second momentum only through a delta function, i.e.\ $G(k,k')\propto G(k)\d^4(k-k')$. In the present case, due to the breaking of translational invariance, the delta function is replaced by something which might be described by a smeared out delta function, which is contained in the Mehler kernel, and hence one cannot simply split that part of\/f.
However, by integrating over one external momentum one can extract the divergence one is actually interested in. In some sense one can interpret this procedure as an expansion around the usual momentum conservation. This is the general procedure we will use to calculate the Feynman graphs. The 1-point tadpoles however are an exception: since they have only one external momentum, integrating the latter out would equally mean to set $p=0$. (One can see this by noticing that the integrand is antisymmetric in $p$, and the integration over the symmetric interval from $-\infty$ to $\infty$ would thus give zero.) With this procedure we would just hide the divergences. In conclusion, one can state that the integration over an external momentum is applicable for graphs with more than one external leg.

For the 1-point graphs, we use the trick of coupling an external f\/ield to the graph and expanding it around $p=0$:
\begin{gather*}
\int \frac{d^4p}{(2\pi)^4} \Pi^\vare_\mu(p) \bigg[A_\mu(0)+p_\nu\left(\partial^p_\nu A_\mu(p)\big|_{p=0}\right)+\frac{p_\nu p_\rho}{2}\left(\partial^p_\nu\partial^p_\rho A_\mu(p)\big|_{p=0}\right)  \nonumber\\
\qquad{}
 +\frac{p_\nu p_\rho p_\s}{6}\left(\partial^p_\nu\partial^p_\rho\partial^p_\s A_\mu(p)\big|_{p=0}\right)+\cdots\bigg] .
\end{gather*}
After smearing out the graph by coupling it to an external f\/ield, an integration over $p$ is allowed. All terms of even order are zero for symmetry reasons.  Of the other terms, we now show that only the f\/irst two, namely orders 1 and 3, diverge in the limit $\vare\to 0$:
\begin{itemize}\itemsep=0pt
 \item \emph{order 1:}
With the external f\/ield, we obtain a counter term of the form
\begin{gather}
\left(\partial^p_\nu A_\mu(p)\big|_{p=0}\right)  \int \frac{d^4p}{(2\pi)^4}\,p_\nu\Pi^\vare_\mu(p)  \nonumber\\
\qquad{} =
\frac{5 g \W^2}{32\pi^2 \omega\left(1+\frac{\W^2}{4}\right)^3}
\left[ \inv{\vare}-1+\mathcal{O}(\vare) \right] \int d^4x  \, \tilde x_\mu A_\mu(x) .\label{eq:induced_tadpole-order1}
\end{gather}

\item \emph{order 3:}
We get the counter term
\begin{gather}
\nonumber
\left(\partial^p_\a \partial^p_\b \partial^p_\gamma A_\mu(p)\big|_{p=0}\right)
\int \frac{d^4p}{(2\pi)^4}\, \frac{p_\a p_\b p_\gamma}{6} \Pi^\vare_\mu(p)  \\
\qquad{}
= \frac{ 5g } { 8\pi^2 } \frac { \W^4 } { \left(1+\frac{\W^2}{4}\right)^4}
\left[ \ln\vare + \mathcal{O}(0) \right] \int d^4x\, \tilde x_\mu {\tilde x}^2 A_\mu (x)
.\label{eq:induced_tadpole-order3}
\end{gather}

\item \emph{order 5 and higher:} These orders are \emph{finite}. The contribution to order $5 + 2n$, $n\ge 0$ is proportional to
\[
 \int\limits_{0}^{\infty} d\a \frac{ \sinh^n \frac \a 2 } { \cosh^{n+4} \frac \a 2 } = \frac 4 { (n+1) (n+3) } .
\]
\end{itemize}

Notice, that all tadpole contributions vanish in the limit $\W\to0$ as expected. However when~$\Omega\neq0$ the unphysical tadpole contributions are non-zero. Since this can certainly not describe nature, we must have started with a wrong vacuum. Furthermore, since we get additional counter terms of mathematical structure which were not initially present in the original action, we certainly need a new theory. Apparently this is the case here because equations~\eqref{eq:induced_tadpole-order1} and~\eqref{eq:induced_tadpole-order3} reveal counter terms linear in $A_\mu$.
Ultimately this means that we will have to consider a whole new model, which will be the induced gauge theory, but more on that in \secref{sec:canon:induced}.

\begin{figure}[t]
\centering
\includegraphics[scale=0.65]{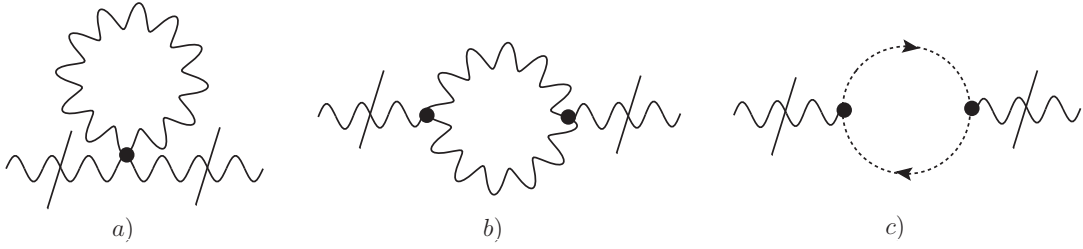}
\caption{Gauge boson self-energy~-- amputated graphs.}
\label{fig:induced_photonself}
\end{figure}

{\bf Two-point functions at one loop level.}
Here we analyze the divergence structure of the gauge boson self-energy at one-loop level. The relevant graphs are depicted in Fig.~\ref{fig:induced_photonself}.

As explained in the previous paragraph, we do not need to couple an external f\/ield and expand around it in this case. The notion of long and short variables has proven to be very useful and we will use it again here. In order to be able to calculate the three graphs, we use the following simplif\/ications:
\begin{itemize}\itemsep=0pt
 \item For the cosine we use
 \[
  \cos\left(\frac{k\tilde p}{2}\right)=\sum_{\eta=\pm1}\frac{1}{2}\exp\left(\frac{\ri\eta}{2}k\tilde p\right) .
 \]
 \item We approximate the hyperbolic functions of the Mehler kernel:
 \[
  \coth\left(\frac{\alpha}{2}\right)\simeq\frac{2}{\alpha},
  \qquad
  \tanh\left(\frac{\alpha}{2}\right)\simeq\frac{\alpha}{2},
  \qquad
  \sinh(\alpha)\simeq\alpha
 \]
 for the dangerous region $\alpha=0$, where the kernel has a quadratic pole, in order to extract the divergent parts of our Feynman graphs.
 \item We will, in addition to the inner momenta, integrate over the external momentum $p'$ in order to reveal the divergence structure of the general result without the ``smeared out delta function'' of the Mehler kernel.
 \item For the parameter integrals $\alpha$ (one per Mehler kernel) we perform a useful change of variables, which can be found e.g.\ in~\cite[page~15]{Blaschke:2008b}.
\end{itemize}

In this form, we can easily sum up all three graphs Fig.~\ref{fig:induced_photonself}a), b) and c). The sum yields the f\/inal result
\begin{gather}
\Pi_{\mu\nu}^{\text{div}}(p) =
\frac{g^2 \d _{\mu\nu} \left(1-\tfrac{3}{4}\W^2\right)}{4\pi^2 \omega\,\vare \left(1+\tfrac{\W^2}{4}\right)^3}+
\frac{3 g^2 \delta_{\mu\nu} \W^2}{8\pi^2 \p^2 \left(1+\tfrac{\W^2}{4}\right)^2}+
\frac{2 g^2 \p_\mu \p_\nu }{\pi ^2 (\p^2)^2 \left(1+\tfrac{\W^2}{4}\right)^2}\nonumber\\
\phantom{\Pi_{\mu\nu}^{\text{div}}(p) =}{} +\text{logarithmic UV divergence} .\label{eq:induced_one-loop_divergences}
\end{gather}
In the limit $\W\to0$ (i.e.\ $\omega\to\infty$), this expression reduces to the usual transversal result
\[
\lim\limits_{\W\to0}\Pi_{\mu\nu}^{\text{div}}(p) =\frac{2g^2}{\pi^2} \frac{\p_\mu\p_\nu}{(\p^2)^2}+\text{logarithmic UV divergence} ,
\]
which is quadratically IR divergent\footnote{In fact, this term is consistent with previous results~\cite{Blaschke:2005b,Hayakawa:1999,Ruiz:2000} calculated in the {\naiv} model, i.e.\ without any additional $x$-dependent terms in the action.} in the external momentum $p$ and logarithmically UV divergent. The single graphs a), b) and c), however, do not show this behaviour, only the sum of all 3 graphs is transversal in the limit $\Omega\to0$. When not taking this limit we can see from the general result~\eqref{eq:induced_one-loop_divergences} that not only transversality is broken due to the f\/irst two terms, but also that it has an ultraviolet divergence parameterized by $\vare$, whose degree of divergence is higher compared to the (commutative) gauge model without oscillator term. Both properties are due to the term $\Act_{\text{m}}$ in the action which breaks gauge invariance (cf.~\eqref{eq:induced_transversality_relation}).

\begin{figure}[t]
 \centering
 \includegraphics[scale=0.8]{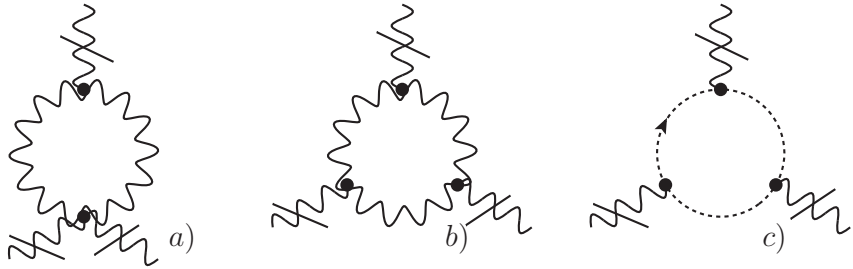}
 \caption{One loop corrections to the 3A-vertex.}
 \label{fig:induced_1loop_3A_all}
\end{figure}

{\bf Vertex corrections at one-loop level.}
Due to the vast amount of terms that arise when calculating these graphs it is practicable to use a computer.
This, in fact, was done in~\cite{Blaschke:2009aw} in order
to calculate the graphs depicted in Fig.~\ref{fig:induced_1loop_3A_all}.
The sum of these graphs yields
\begin{gather}
 \Gamma^{\text{3A,IR}}_{\mu\nu\rho}(p_1,p_2,p_3)
 =\frac{-8\ig^3}{\pi^2\left(4{+}\W^2\right)^3}\!\sum\limits_{i=1}^{3}\!\bigg[
 \frac{16\p_{i,\mu}\p_{i,\nu}\p_{i,\rho}}{\p_i^4}\!+\frac{3\W^2}{\p_i^2}
 \left(\delta_{\mu\nu}\p_{i,\rho}\!+\d_{\mu\rho}\p_{i,\nu}\!+\d_{\nu\rho}\p_{i,\mu}\right)\bigg],\!\!\!\!\label{eq:induced_3Avert}
\end{gather}
which is linearly divergent.
Once more, this expression is not transversal due to the non-vanishing oscillator term parametrized by $\W$. However, in the limit $\W\to0$ transversality is recovered, and~\eqref{eq:induced_3Avert} reduces to the well-known expression~\cite{Matusis:2000jf, Armoni:2000xr,Ruiz:2000}{\samepage
\[
 \lim\limits_{\W\to0}V^{\text{1loop}}_{\mu\nu\rho}(p_1,p_2,p_3)=\frac{-2\ig^3}{\pi^2}\sum\limits_{i=1}^{3}\bigg[
 \frac{\p_{i,\mu}\p_{i,\nu}\p_{i,\rho}}{\p_i^4}\bigg].
\]
In the ultraviolet, the graphs of Fig.~\ref{fig:induced_1loop_3A_all} diverge only logarithmically.}

Of course corrections to the 4A-vertex exist too, but those show only a logarithmic divergence according to the power counting~\eqref{eq:induced_powercounting}.

\subsubsection{Induced gauge theory}
\label{sec:canon:induced_induced}

Since in the previous section it has been shown that additional counter terms arise which were not present in the original action it is natural to start with an action that has those terms already built in, instead. Such an action is the ``induced gauge theory'' of~\cite{Wallet:2007c,Wohlgenannt:2008zz}. Its major advantage is that it is, by construction, completely gauge invariant. Let us review how this action is derived.

One starts with the Grosse--Wulkenhaar model \eqref{eq:induced_Grosse Wulkenhaar action}:
\begin{gather*}\nonumber
 \Gamma^{(0)}[\phi]=\int d^4x \bigg( \frac{1}{2}\phi\star\starco{\tilde x_\nu}{\starco{\tilde x_\nu}{\phi}}
 +\frac{\Omega^2}{2}\phi\star\staraco{\tilde x_\nu}{\staraco{\tilde x_\nu}{\phi}}
 \\
\phantom{\Gamma^{(0)}[\phi]=}{}
 -\frac{\mu^2}{2}\phi\star\phi+\frac{\lambda}{4!}\phi\star\phi\star\phi\star\phi\bigg)(x),
\end{gather*}
where, in order to write the action in the previous form, the following important property has been used:
\[
 \starco{\tilde x_\mu}{\phi}=\ri\partial_\mu \phi .
\]
Now, one introduces external gauge f\/ields by generalizing the ordinary coordinates $x_\mu$ to co\-va\-riant ones\footnote{Notice the slight dif\/ference to the $\cth$-expanded case where one usually introduces covariant coordinates without tilde, see \secref{sec:canon:theta-exp}.} $\wtild X_\mu$, with
\[
 \wtild X_\mu=\tilde x_\mu + gA_\mu .
\]
These coordinates have the nice property that they gauge transform covariantly, which is why they are named likewise. Therefore, the Grosse--Wulkenhaar action is gauge invariant by construction:
\begin{gather*}
 \int d^4x \bigg( \frac{1}{2}\phi\star\starco{\wtild X_\nu}{\starco{\wtild X_\nu}{\phi}}
 +\frac{\Omega^2}{2}\phi\star\staraco{\wtild X_\nu}{\staraco{\wtild X_\nu}{\phi}}
-\frac{\mu^2}{2}\phi\star\phi+\frac{\lambda}{4!}\phi\star\phi\star\phi\star\phi\bigg)(x) .
\end{gather*}
It can be shown either by performing a heat kernel expansion \cite{Wohlgenannt:2008zz}, or by explicit loop calcula\-tions~\cite{Wallet:2007c} that to one loop order the action becomes
\begin{gather}
\Gamma^{(1l)}[A_\mu]= \int d^4x  \bigg\{
\frac{3}{\th} \left(1-\rho^2\right) \left(\tilde \mu^2-\rho^2\right)\left(\wtild X_\nu \star \wtild X_\nu -\tilde x^2\right)
\nonumber\\
\phantom{\Gamma^{(1l)}[A_\mu]=}{}
+ \frac{3}{2}\left(1-\rho^2\right)^2 \left( \left(\wtild X_\mu\star \wtild X_\mu\right)^{\star 2}-\left(\tilde x^2\right)^2 \right)
+ \frac{\rho^4}{4}  F_{\mu\nu}\star F_{\mu\nu}
\bigg\},\label{eq:induced_induced1loop}
\end{gather}
where
\[
\rho = \frac{1-\W^2}{1+\W^2},\qquad
\tilde \mu^2 = \frac{\mu^2\th}{1+\W^2}.
\]
Notice also, that the f\/ield strength tensor $F_{\m\n}=\pa_\m A_\n-\pa_\n A_\m-\ig\starco{A_\m}{A_\n}$ can be written in terms of the covariant coordinates as
\[
\ri\starco{\wtild X_\m}{\wtild X_\n}=\th^{-1}_{\m\n}-gF_{\m\n}.
\]
The f\/ield $\phi$ has been integrated out in order to arrive at the ef\/fective action~\eqref{eq:induced_induced1loop}, and $A_\mu$ has been considered as a background f\/ield. However, through its coupling to~$A_\mu$, the scalar f\/ield ``induces'' the ef\/fective one-loop action~\eqref{eq:induced_induced1loop} above and $A_\mu$ becomes dynamical.

As already mentioned in \secref{sec:canon:induced_ext2gauge}, all (UV-divergent) terms that arise in the loop calculations of the previous model are present in the induced action. Hence, the chance that any unexpected new contributions arise during loop calculations is improbable, especially in the light that the whole action is gauge invariant. This gives good hope concerning the renormalizability of the model. However, the problem that the tadpole graphs do not vanish, and that we therefore have a non-trivial vacuum, is still present. Furthermore, calculating the propagator of the induced gauge theory is a non-trivial enterprise since the operator which has to be inverted is non-minimal (i.e.\ no Lorentz scalar). Additionally, calculating the propagator from the pure bilinear part seems not to be suf\/f\/icient because, as already mentioned, linear (tadpole) terms in~$A_\mu$ are also present in the action. All those severe problems need to be taken into account in the future work on this action.

\subsubsection{Geometrical approach}
\label{sec:canon:geometrical}

Another way to generalize the Grosse--Wulkenhaar model to gauge theories is via geometry. In a~recent paper \cite{Wohlgenannt:2009}, it has been shown that the renormalizable Grosse--Wulkenhaar action \cite{Grosse:2003,Grosse:2004b}
\[
S = \int d^2x \left(
\frac 12 \partial_\mu \phi   \partial_\mu \phi + \frac{m^2}2 \phi^2 + \frac{\Omega^2}2 \tilde x_\mu \phi  \tilde x^\mu \phi + \frac{\lambda}{4!} \phi^4
\right)
\]
can be interpreted as the action for a scalar f\/ield on a curved background space, namely
\[
S' = \int d^2x  \sqrt{g} \left(
\frac 12 \partial_\mu \phi   \partial_\mu \phi + \frac{m^2}2 \phi^2 - \frac \xi 2  R \phi^2 + \frac{\lambda}{4!} \phi^4
\right) ,
\]
where $m$ denotes the mass of the scalar f\/ield, $R$ the scalar curvature of the background space and $\xi$ an arbitrary constant. This constitutes another remarkable connection between gravity and {\nc} geometry. Let us stick to two dimensions. The four dimensional case is straight forward~\cite{Wohlgenannt:2009}. The starting point is the so-called {\it truncated Heisenberg algebra} of $n\times n$ matrices satisfying the relation
\[
[x,y]= \ri \alpha \m^{-2} (1-\m nP_n),
\]
where $P_n$ denotes a projector. Def\/ining $z\equiv nP_n$, we obtain a three dimensional algebra:
\begin{gather*}
 [x,y] = \ri \alpha \m^{-2} ( 1 - \m z ), \qquad [x,z] = \ri \alpha ( yz + zy ) , \qquad
  [y,z] = - \ri \alpha ( xz + zx ) ,
\end{gather*}
where the parameter $\alpha$ is dimensionless and def\/ined such that $\alpha\to 0$ gives the commutative limit. There are two relevant length scales in the problem. One of them is $\sqrt{\sth}$, the {\nc} scale. The other scale is the gravitational one denoted by $\m^{-1}$ (the Schwarzschild radius or the cosmological constant for example). It is assumed that $\alpha = \m^2 \sth$.

In the limit $n\to \infty$, the usual Heisenberg algebra is recovered; this corresponds to $z\to 0$. Using the frame formalism \cite{Madore:2000aq}, the geometry of this space can be computed. In the limit $z\to 0$, the space is still curved, and remarkably the cotangent space is still three dimensional\footnote{This is also true for e.g.\ the fuzzy sphere, where the algebra is also two dimensional whereas the cotangent space is three dimensional, see \secref{sec:noncanon:fuzzy}.}. The scalar curvature is given by
\[
R = \frac{15\mu^2}{2} - 8 \mu^4 \big(x^2 + y^2\big) .
\]
The f\/irst term just renormalizes the mass. On this curved geometry, also the algebra of $n$-forms is deformed. In \cite{Buric:2010}, the resulting gauge action for NC $U_\star(1)$ has been found. It reads
\begin{gather*}
S_{\rm YM}   = \frac 12  \int d^2x \Big( (1-\alpha^2) (F_{12})^{\star 2} -2 ( 1-\alpha^2 ) \mu F_{12}\star \phi + ( 5-\a^2 ) \mu^2 \phi^{\star 2} + 4\ri\a F_{12}\star \phi^{\star 2}
 \nonumber\\
\phantom{S_{\rm YM}   =}{}
 + (D_1 \phi)^{\star 2} +(D_2 \phi)^{\star 2} -\a^2\{ p_1 +A_1, \phi\}_\star^{\star 2} -\a^2\{ p_2 +A_2, \phi\}_\star^{\star 2} \Big)   ,
\end{gather*}
where $p_1=\ri\frac{\mu^2}\a y$, $p_2=-\ri\frac{\mu^2}\a x$, and $F_{12}$ denotes the $12$-component of the f\/ield strength. The star product is given by the {\moyal} product. Similar to the approach before, we can express the action in terms of covariant coordinates, $p_i + A_i$. In a next step, the renormalizability properties of this action have to be studied.

\subsection[Benef\/iting from damping~-- the $1/{p^2}$ approach]{Benef\/iting from damping~-- the $\boldsymbol{1/{p^2}}$ approach}
\label{sec:canon:p2inv}

The success of the Grosse--Wulkenhaar model with its oscillator term drew a lot of attention from the community but problems, such as the explicit breaking of translation invariance, could not be solved in an entirely satisfactory way.  An alternative approach to tackle the problem of {\uim} was proposed by Gurau et al.~\cite{Gurau:2009}. The main idea is to add a non-local term
\begin{gather}
 \Act_{\text{nloc}}[\phi]=-\intx \, \phi(x)\star\frac{a^2}{\theta^2\square_x}\star\phi(x),
 \label{eq:canon:p2inv_gurau_act_nloc}
\end{gather}
to the action \eqref{eq:canon:early:naive_act_complete}, where $a$ is a dimensionless constant. The practical motivation for this is clearly to provide a counter term for the expected quadratic IR divergence in the external momentum, a mechanism which has explicitly been demonstrated in~\cite{Blaschke:2008b}. \emph{A priori} the physical interpretation of the operator $\inv{\square}$ is dif\/f\/icult -- especially in $x$-space one faces the inverse of a~derivative. In momentum space the situation becomes more intuitive since the inverse of the scalar function $k^2$ is well known. A~sensible interpretation of the new operator $\square^{-1}$ is to regard it as the `Green operator' of $\square \equiv \partial_\m\partial_\m$.

The action including the non-local insertion reads\footnote{Note, that the interaction term is written as a generic Fourier transformed quantity $\mathcal{F}\left(\phi^{\star4}\right)$, without stating the explicit form of the phase factors.}
\begin{gather}
\Act[\phi]=\intk \left[ \inv{2}\left(k_\m\phi(-k) k_\m \phi(k)+m^2\phi^{2}+a^2\phi(-k)\inv{\k^2}\phi(k)\right) +\frac{\l}{4!}\mathcal{F}\big(\phi^{\star4}\big)\right],
\label{eq:canon:p2inv_gurau_act_pspace}
\end{gather}
with $m$ and $\l$ being parameters of mass dimension 1 and 0, respectively.
Variation of the bilinear part of the action~\eqref{eq:canon:p2inv_gurau_act_pspace} with respect to $\phi$ immediately leads to the propagator
\begin{gather}
\raisebox{-1.5pt}[1pt][0pt]{\includegraphics[scale=0.8, trim=0 0 0 10,clip=true]{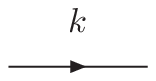}}= \; G(k)=
\inv{k^2+m^2+\frac{a^2}{\k^2}} .
\label{eq:canon:p2inv_gurau_propagator}
\end{gather}
This Green function is the core achievement of the approach by Gurau et al.\ since it features a damping behaviour in the IR while not af\/fecting the UV region, i.e.
\[
\lim\limits_{k \to 0}G(k)=\lim\limits_{k \to \infty}G(k)=0,\qquad
\forall\, a\neq0 .
\]
In Multiscale Analysis~\cite{Rivasseau:1991ub,Gurau:2009}, this also allows the propagator to be bounded from above by a~constant which is a basic ingredient leading to the renormalizability of the model.
In contrast to the propagator, the vertex functional
is not altered in comparison to the {\naiv} implementation of $\phi^{\star4}_4$ theory\footnote{We will refer to the $\phi^4$ theory on {\nc} Euclidean space which is simply generated by insertion of star products \eqref{eq:intro_main:star_prod_def} into the interaction term as `{\naiv}' implementation.}.
The \emph{damping} ef\/fect of the propagator \eqref{eq:canon:p2inv_gurau_propagator} becomes obvious when considering higher loop orders. An $n$-fold insertion of the divergent one-loop result\footnote{Note that, for the sake of simplicity, we neglect any ef\/fects due to recursive renormalization, and approximate the insertions of irregular single loops by the most divergent (quadratic) IR divergence. See also~\cite{Minwalla:1999}.} \eqref{eq:canon:early:phi4-ir} into a single large loop can be written as
\begin{gather}
\label{eq:canon:p2inv_gurau_n-loop-int}
\Pi^{n \text{ np-ins.}}(p) \approx\l^2\intk\,
  \frac{\re^{\ri\wprod{k}{p}}}{\left(\k^2\right)^n\left[k^2+m^2+\frac{a'^2}{k^2}\right]^{n+1}} .
\end{gather}
For the {\naiv} model (where  $a=0$), the integral of equation~\eqref{eq:canon:p2inv_gurau_n-loop-int} involves an IR divergence for $n\geq2$, because the integrand scales as $(k^2)^{-n}$ for $k^2\to 0$. In contrast, for the $1/p^2$ model (where  $a\neq 0$), the integrand behaves like
\begin{gather*}
\inv{\big(\k^2\big)^n\left[\frac{a'^2}{k^2}\right]^{n+1}}=\frac{\k^2}{\left(a'^2\right)^{n+1}},
\end{gather*}
which is independent of the order.

\subsubsection{Gribov's problem and Zwanziger's solution}
\label{sec:canon:gribov}

\begin{figure}[t]
 \centering
 \includegraphics[width=6.5cm]{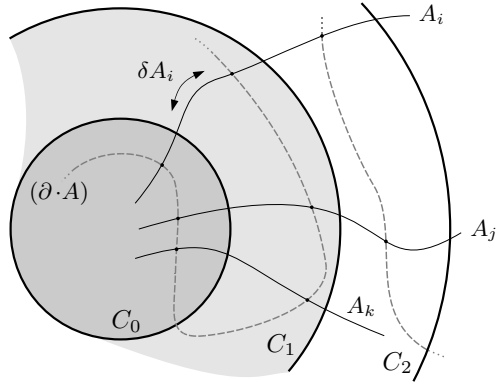}
 \caption{Visualization of the gauge conf\/iguration space with Gribov horizons $C_n$.}
 \label{fig:gribov_horizons}
\end{figure}

As has been f\/irst indicated by Gribov \cite{Gribov:1978} in 1978 and was reviewed for example in~\cite{Sorella:2005} and \cite[pp.~145--174]{Kummer-memorial}, in {\nA} theories the gauge is not f\/ixed uniquely by a condition of the form $\partial A = f$, with $f$ being some function or constant. This can be understood when considering two f\/ields $A_\m,\,A'_\m$, being elements of some general gauge group\footnote{For the sake of simplicity, we suppress any group indices or additional notation here and in the following.}, which are connected by the transformation
\[
 A'_\m=U^\dagger\left(\partial_\m+A_\m\right)U=A_\m+U^\dagger\left(\partial_\m U+\co{A_\m}{U}\right)=A_\m+\delta A_\m ,
\]
with $U=\re^{\a}$, and $\a$ being the algebra valued gauge parameter.
For some f\/ixed $A_\m$, we may f\/ind some $A'_\m$ fulf\/illing the same gauge condition, and therefore being \emph{equivalent} to the original one. Such \emph{Gribov copies} are solutions of the equation
\begin{gather}
 \partial_\m A'_\m = \partial_\m A_\m = f \quad
 \Rightarrow\quad \partial_\m\big[U^\dagger\left(\partial_\m U+\co{A_\m}{U}\right)\big]=f, \label{eq:canon:canon:gribov_gribov_copy_cond}
\end{gather}
and give rise to divergences in the corresponding path integral.
Obviously, the operator on the left hand side of equation~\eqref{eq:canon:canon:gribov_gribov_copy_cond} is the Faddeev--Popov operator $\mathcal{M}(A)=-\partial_\m D_\m$ (acting on $\a$), whose determinant appears in the functional integral upon integrating out the ghost f\/ields. We therefore recognize the latter relation as an eigenvalue equation
\[
 \mathcal{M}(A)\psi = \epsilon(A)\psi.
\]
Intuitively, the form of $\mathcal{M}$ admits comparison with the Schr\"odinger operator. Proceeding in this parallel picture, $A_\m$ takes the role of a potential. For small $|A_\m|$ all $\epsilon(A)$ will be positive, while with rising $|A_\m|$ more and more eigenvalues will vanish, and then become negative. The idea is to divide the gauge conf\/iguration space into \emph{Gribov spaces} $C_n$, $n\in\N^0$ having $n$ negative eigenvalues. These domains are separated by the \emph{Gribov horizons} $l_n$ which correspond to the solution $\epsilon_n(A)=0$. The situation is depicted in Fig.~\ref{fig:gribov_horizons} for three exemplary conf\/igurations $A_a$, $a\in\{i,j,k\}$ being represented by lines which are generated by variations of the parameter $\a$. The gauge f\/ixing $\partial A = f$ (symbolized by the dashed line) crosses each $A_a$ exactly once\footnote{We should note, that it is generally accepted that $C_0$ is \emph{not} free of Gribov copies due to the appearance of multiple eigenvalues $\epsilon(A)>0$. It is possible to restrict the domain of integration further in order to remedy this problem. However, this is beyond the scope of this review and we shall refer to the literature~\cite{vanBaal:2000zc} for further discussion.} in the domain $C_0$. The same is true for any further $C_n$.

The important point to note is~\cite{Gribov:1978,Sorella:2005} that for each $A_\m$ in $C_0$ with $\partial A = f$ we f\/ind an equivalent $A'_\m$ in $C_n$ (for at least some $n$). This is the motivation to restrict the domain of integration in the path integral to $C_0 = \{A:{\rm Tr}[\mathcal{M}(A)]>0\}$. According to Gribov, this restriction shall be implemented by inserting a Heaviside weighting function $\nu(k,A)\equiv\theta(1-\sigma(k,A))$, yielding (for YM theory on $D=4$ with $\mathcal{N}$ being the vacuum normalization factor)
\begin{gather}
\label{eq:canon:gribov_part_fun_limited}
 Z=\mathcal{N}\int {\mathcal{D}A}\mathcal{D}c\mathcal{D}\bc \re^{-S_{\rm YM}-\intx\partial_\m D_\m c}\nu(0,A)=\mathcal{N}\int \mathcal{D}A \re^{-S_{\rm YM}}\nu(A) .
\end{gather}
The function $\sigma(f,A)$ appears in the perturbative expansion of the ghost propagator~\cite{Sorella:2005}, and takes the form \[
\sigma(k,A)=\lim\limits_{V\to\infty}\inv{3}\frac{N}{N^2-1}\frac{k_\m k_\n}{k^2}\inv{V}\sum\limits_q\frac{A_\l(q)A_\l(-q)}{(k-q)^2}\left(\delta_{\m\n}-\frac{q_\m q_\n}{q^2}\right) ,
\]
where $N$ is the dimensionality of the underlying gauge group.
Basically, the pole of the propagator $G^{\bc c}(k)\approx\big[k^2(1-\s(k,A))\big]^{-1}$ appears on the horizons $l_i$, corresponding to $\s(k,A)\to 1$. When restricting the range of functional integration to $C_0$, we can state a condition to ensure it is free of poles, namely $\s(k,A)<1$, which immediately leads back to the expression for $\nu(k,A)$ stated above.

Starting from equation~\eqref{eq:canon:gribov_part_fun_limited}, we can insert the explicit form for $\nu(0,A)$ (and the step function), and pick out the quadratic part.
\begin{gather*}
 Z_{\text{quadr.}} =\mathcal{N}\int\frac{{\rm d}\b}{2\ri \pi\b}\mathcal{D}A\,\re^{-\inv{2g^2}\sum\limits_q A_\m(q) \mathcal{Q}_{\m\n}A_\n(-q)}
 =\mathcal{N}\int\frac{{\rm d}\b}{2\ri \pi\b}\left(\det \mathcal{Q}\right)^{-\inv{2}}
 \equiv\mathcal{N}\int\frac{{\rm d}\b}{2\pi \ri}\re^{f(\b)} ,
\end{gather*}
where the operator of the quadratic part $Q_{\m\n}=\big[q^2+\frac{\b N g^2}{N^2-1}\inv{2V} \inv{q^2}\big]\d_{\m\n}-\big(1-\tinv{\a}\big)q_\m q_\n$ has been introduced. In the last step we have pulled the logarithm of the determinant into the exponential, and introduced a short hand $f(\b)$ for the resulting expression.
Here, we can already see the most important ef\/fect of the restriction to $C_0$, which is that the bilinear part is modif\/ied with respect to the unconstrained case. In fact, the expression for $Q_{\m\n}$ now contains not only positive powers of the momentum $q$ but also a term with negative ones. This gives rise to a dramatic change of the behaviour of the theory in the IR. Before discussing this aspect in more detail, we def\/ine the coef\/f\/icient of the negative powers as
\[
 \g^4\equiv\left.\frac{\b_0 N g^2}{N^2-1}\inv{2V}\right|_{V\to\infty}\quad\Rightarrow\quad\frac{3N g^2}{4}\int\frac{{\rm d}^Dq}{(2\pi)^D}\inv{q^4+\g^4}=1 ,
\]
where $\b_0$ is the solution of the equation $\partial_\b f(\b)|_{\b\to\b_0}=0$, and the right hand side is generally recognized as \emph{gap equation} which def\/ines the value of the so-called \emph{Gribov parameter} $\g$.
Finally, we can derive the gluon ($A_\m$) two-point function to be (in Landau gauge $\a\to 0$),
\begin{gather}
\label{eq:canon:gribov_prop_AA}
 G^{AA}_{\m\n}(q)=g^2\frac{q^2}{q^4+\g^4}\left(\d_{\m\n}-\frac{q_\m q_\n}{q^2}\right) ,
\end{gather}
which obviously exhibits the nice property
\[
 \lim\limits_{|q|\to 0}G^{AA}_{\m\n}(q) = \lim\limits_{|q|\to \infty}G^{AA}_{\m\n}(q) = 0
 ,
\]
i.e.\ an IR damping behaviour\footnote{Notice, the similarity with propagators~\eqref{eq:canon:p2inv_gurau_propagator} and~\eqref{eq:canon:p2inv_gauge_brsw_prop_AA}.}.

It was realized by D. Zwanziger~\cite{Zwanziger:1989} that the restriction of the path integral to the f\/irst Gribov horizon can be implemented by a special type of term in the action. In order to see this, we have to reconsider the weighting function $\nu(0,A)$ we have introduced in equation~\eqref{eq:canon:gribov_part_fun_limited}. For large $N$, corresponding to the thermodynamic limit $V\to\infty$, the volume described by the Heaviside function is concentrated at the surface, as $V_{C_0}/V_{\partial C_0}\sim R/N$. Therefore, one may replace $\theta(1-\s(k,A))\to \delta(1-\s(k,A))$. Due to the equivalence of canonical and microcanonical ensembles in the same limit ($V\to\infty$), we can furthermore replace $\d(1-\s(k,A))\to\re^{-\g^4H}$, where $H$ is the Hamiltonian of the theory~\cite{Dudal:2009xh}
\[
 H=\intx\, h(x)=\langle A|\mathcal{M}^{-1}A\rangle = -\Act_{\text{nloc.}} ,
\]
and $h(x)$ is called \emph{horizon function}. Obviously, $H$ is {\nl} since $\mathcal{M}$ contains derivatives. For this reason, Zwanziger applied the method of \emph{localization} by introducing auxiliary f\/ields and ghosts (see \secref{sec:canon:p2inv_loc} for an intuitive discussion). His approach can be summarized as the replacement
\begin{gather}
 \Act_{\text{nloc.}} \to\Act_{\text{loc.}}, \nonumber\\
 \re^{-\g^4\langle A|\mathcal{M}^{-1}A\rangle} \to\int\mathcal{D}\omega\mathcal{D}\bar{\omega}\mathcal{D}\phi\mathcal{D}\bar{\phi} \re^{\langle \bar{\omega}|\mathcal{M}\omega\rangle-\langle \bar{\phi}|\mathcal{M}\phi\rangle-\gamma^2\langle A|\phi-\bar{\phi}\rangle},\label{eq:canon:gribov_zwanziger_localization}
\end{gather}
applied to the partition function \eqref{eq:canon:gribov_part_fun_limited} after performing the approximation $\nu(0,A)\to\Act_{\text{nloc.}}$. Equivalence can be seen by performing the integration with respect to the complex conjugated ghost f\/ields $\{\bar{\omega}, \omega\}$, then completing the square over the auxiliary f\/ields $\{\bar{\phi}, \phi\}$, and f\/inally integrating out the latter. The action $\Act_{\text{loc.}}$ is def\/initively local, and leads to the same propaga\-tor~\eqref{eq:canon:gribov_prop_AA} (up to a factor~$g^2$) as the original approach by Gribov discussed above.

It can be shown~\cite{Zwanziger:1993,Dudal:2008} that almost all parts of $\Act_{\text{loc.}}$ containing auxiliary ghosts and f\/ields can be written in a BRST-exact way, and are thereby physically irrelevant. The only exception is the term $-\gamma^2\langle A|\phi-\bar{\phi}\rangle$ being parametrized by the Gribov parameter $\g$. Bearing in mind that the behaviour $G^{AA}(q)|_{|q|\to 0}$ vitally depends on $\g$, it now becomes obvious that the exponential factor being introduced in equation~\eqref{eq:canon:gribov_zwanziger_localization} in order to rewrite the restriction to the f\/irst Gribov horizon, which now takes the form of a BRST breaking term, ef\/fectively changes physics. Remarkably, the exclusion of Gribov copies, giving rise to divergences in the path integral, does not alter the UV region of the theory but only the low momentum limit. Details of the implementation in the form of a \emph{soft breaking}, and a discussion of the impact on symmetries and the renormalization are given in \secref{sec:canon:p2inv_loc}.

\subsubsection{The long way to consistent gauge models}
\label{sec:canon:p2inv_gauge}

Motivated by the rather simple mathematical structure of the model~\eqref{eq:canon:p2inv_gurau_act_nloc} ef\/forts have been started to implement the damping behaviour of the $1/p^2$-model in respective {\nc} gauge theories. However, it turned out soon that there are some peculiarities which frustrate a~straightforward proceeding. Let us brief\/ly review these in a little more detail. As the following discussion is of rather technical nature (which could contribute to a better understanding of the details), the more experienced reader should directly proceed to page~\pageref{page:canon:p2inv_gauge:after_details}.

Let us start from the simplest possible model -- a free photon f\/ield, described by a $U_\star(1)$ symmetry. As for the scalar case, there exists a {\naiv} approach which is def\/ined by the action
\begin{gather}
 \Act_{\rm YM}^{\text{\naiv}}=\intx F_{\m\n}\star F_{\m\n} ,
 \label{eq:canon:p2inv_gauge_naive_act}
\end{gather}
with the def\/initions
\begin{gather}
 F_{\m\n}  = \partial_\m A_\n -\partial_\n A_\m -\ig \starco{A_\m}{A_\n},\qquad
 D_{\m}\phi  = \partial_\m \phi - \ig\starco{A_\m}{\phi} \qquad \forall \; \phi,
\label{eq:canon:p2inv_gauge_fstrength_d_def}
\end{gather}
Aiming to construct a physical theory, the following BRST transformations\footnote{In {\nc} theory the well-known principle applies that a gauge boson propagator only exists if the gauge is explicitly broken by a f\/ixing term. As can be found in many text books on the subject~\cite{Ryder:1996,Piguet:1995,Schweda-book:1998} the f\/ixing requires the additional introduction of Grassmann-valued (Faddeev--Popov) ghost f\/ields in order to leave invariant the functional integral. As has been recognized by Becchi, Rouet, Stora and Tyutin the resulting action remains invariant with respect to a nilpotent supersymmetric non-linear transformation, represented by the BRST operator $s$ with $s^2=0$.} are imposed
\begin{alignat}{3}
& sA_\m = D_\m c , \qquad  && sc = \ri c\star c , & \nonumber\\
& s\bar{c}=b ,  \qquad && sb =0 ,&\nonumber\\
&  s^2\phi =0  \qquad \forall\, \phi\mathrel{\in}\{A,b,c,\bc\} . \qquad &&&
\label{eq:canon:p2inv_gauge_naive_brst}
\end{alignat}
From these the properties
\[
 s F = \ig \starco{c}{F}, \qquad s D^2 F =\ig \starco{c}{D^2F},\qquad s \inv{D^2} F =\ig \starco{c}{\inv{D^2}F},
\]
follow, which are proven elsewhere~\cite{Sedmik:2009}. It is well known that the model~\eqref{eq:canon:p2inv_gauge_naive_act} gives rise to IR divergences similar to equation~\eqref{eq:intro_main:ir-sing-gf}, namely
\begin{gather*}
 \Pi_{\m\n}\propto\frac{\p_\m\p_\n}{(\p^2)^2} .
\end{gather*}
From the form of this divergence one is led intuitively to the insertion
\[
 \Act_{\text{nloc}}^{\text{1st try}}[A]=\intx \, A_\m(x)\star\frac{\tilde{\partial}_\m\tilde{\partial}_\n}{\wsq^2}\star A_\n(x) .
\]
However, $\Act_{\text{nloc}}^{\text{1st try}}$ is not invariant under the BRST transformations \eqref{eq:canon:p2inv_gauge_naive_brst}.
Noting that
\begin{gather*}
\intx \, A_\m(x)\star\frac{\tilde{\partial}_\m\tilde{\partial}_\n}{\wsq^2}\star A_\n(x)=-\intx \, \tilde{\partial}_\m A_\m(x)\star\frac{\tilde{\partial}_\n}{\wsq^2}\star A_\n(x)\qquad \text{and}\\
\tilde{\partial}_\m A_\m=\theta_{\m\r}\partial_\r A_\m=\inv{2}\theta_{\m\r}\left(\partial_\m A_\r-\partial_\r A_\m\right)\overset{\text{bilin.}}{\approx}\inv{2}\F,
\end{gather*}
the next proposal is the insertion~\cite{Putz:2003}
\[
\Act_{\text{nloc}}^{\text{2nd try}}[A]=\intx\, \F(x)\inv{\wsq^2}\F(x) .
\]
Again, gauge invariance is fulf\/illed but the $\inv{\wsq^2}$ operator is not compatible with the BRST transformations \eqref{eq:canon:p2inv_gauge_naive_brst}. The only way to remedy this problem seems to be the replacement $\wsq\to \Dt^2=\Dt_\m\Dt_\m=\theta^2D^2$.
Hence, in momentum space one has
\begin{gather}
\Act_{\text{nloc}}^{\text{3rd try}}[A]=\intk\, \F(k)\inv{\big(\Dt^2\big)^2}\F(-k) .
\label{eq:canon:p2inv_gauge_insertion3}
\end{gather}
This insertion is completely invariant under all demanded symmetries, and features the right dimension. However, as has been argued in~\cite{VanRaamsdonk:2001jd, Blaschke:2008a}, the resulting gauge propagator shows a~quadratically IR divergent overall factor, i.e.\ $G^{AA}_{\m\n}(k)\propto\inv{k^2}P_{\m\n}(k)$, where $P_{\m\n}(k)$ denotes the tensor structure which is not specif\/ied here. Hence, the term~\eqref{eq:canon:p2inv_gauge_insertion3} cannot be utilized to implement the desired damping behaviour (which would require an overall factor $\left(\smash[b]{k^2+\tfrac{\text{const}}{\k^2}}\right)^{-1}$).

Finally, the solution seems to be
\begin{gather*}
\Act_{\text{nloc}}^{\text{f\/inal}}[A]=\intx\,  F_{\m\n}(x)\inv{D^2\Dt^2}F_{\m\n}(x).
\end{gather*}
The full tree-level action in position space then takes the form,
\begin{gather}
\Act^{(0)} =S_{\text{inv}}+S_{\text{gf}} ,\nonumber\\
S_{\text{inv}} =\intx\left[\inv{4}F_{\m\n}\star F_{\m\n}+\inv{4}F_{\m\n}\inv{D^2\Dt^2}F_{\m\n}\right]\,,\nonumber\\
S_{\text{gf}} =s\intx\, \bar{c}\star\left[\left(1+\frac{1}{\square\wsq}\right)\partial_\m A_\m-\frac{\alpha}{2}b\right]\nonumber\\
\phantom{S_{\text{gf}}}{}
=\intx\left[b\star\left(1+\frac{1}{\square\wsq}\right)\partial_\m A_\m-\frac{\a}{2}b\star b -\bar{c}\star\left(1+\frac{1}{\square\wsq}\right)\partial_\m D_\m c\right] ,
\label{eq:canon:p2inv_gauge_model1_act}
\end{gather}
where the parameter $\a$ and the unphysical Lagrange multiplier f\/ield $b$ have been introduced in order to f\/ix the gauge. The insertion of the operators $\left(\smash[b]{1+\tfrac{1}{\square\wsq}}\right)^{-1}$ (which are of the same type as in $S^{\text{inv}}$) in the gauge sector $S_{\text{gf}}$ is motivated by the expectation of a~damping for the ghost propagator $G^{\bc c}$.

However, the question arises how to interpret the new operator $\inv{D^2}$. In contrast to the scalar version $\inv{\square}=\inv{\partial_\m\partial_\m}$, the covariant derivative \eqref{eq:canon:p2inv_gauge_fstrength_d_def} includes the gauge f\/ield. Since the inverse of a~f\/ield cannot be def\/ined in a reasonable way, an alternative representation for the new operator has to be found. Such is given by the redef\/inition
\[
 \F =D^2\inv{D^2}\F\equiv D^2\mathcal{Y} ,
\]
leading to the relation
\begin{gather}
 \label{eq:canon:p2inv_gauge_infinite_recursion}
 \mathcal{Y} =\inv{\square}\F+\ri g\frac{\partial_\m}{\square}\starco{A_\m}{\mathcal{Y}}+\ri g\inv{\square}\starco{A_\m}{\partial_\m \mathcal{Y}}+g^2\inv{\square}\staraco{A_\m}{\staraco{A_\m}{\mathcal{Y}}} ,
\end{gather}
which can be rewritten in the form of a recursion~\cite{Blaschke:2008a}, and indicates that no closed solution to the problem is possible. In fact, equation~\eqref{eq:canon:p2inv_gauge_infinite_recursion} is mathematically well def\/ined (since $\square^{-1}$ is) but it represents an inf\/inite number of gauge boson vertices, which in turn correspond to an inf\/inite number of parameters,and thereby renders the theory power counting non-renormalizable~\cite{Piguet:1995}. In addition, only the complete recursion which cannot be reached in practise, is gauge invariant. Therefore, any computations being based on a truncated form of \eqref{eq:canon:p2inv_gauge_infinite_recursion} will contain an unintended breaking of the symmetry, and can, strictly speaking, not be considered to be a gauge theory.
\label{page:canon:p2inv_gauge:after_details}

\subsubsection{Localization}
\label{sec:canon:p2inv_loc}

As was discussed in \secref{sec:canon:p2inv_gauge}, in the $1/p^2$ model one is forced to introduce the inverse of covariant derivatives which can only be interpreted in terms of an inf\/inite series in the gauge f\/ield~$A_\m$, thereby inevitably leading to a 
non-renormalizable theory. However, it turns out that there are alternative representations which ``quasi localize'' the problematic terms by coupling them to unphysical auxiliary f\/ields\footnote{Notice that even the ``quasi localized'' action remains non-local due to the star products.}. There are several ways to implement this, resulting in models with dif\/ferent properties, and even a modif\/ied physical content. In this respect we will be led to the insight that only minimal couplings and the consequent construction of BRST doublet structures for all auxiliary f\/ields result in a stable theory (even at tree level). Moreover, the consistent implementation of the damping behaviour of the $1/p^2$ model requires the insertion of a~so-called ``soft breaking'' term into the action~-- a method which is well known from the Gribov--Zwanziger approach to QCD (see~\cite{Gribov:1978,Zwanziger:1989,Zwanziger:1993}). In the following, the developments leading to a~consistent gauge model are sketched step by step.

In the f\/irst ansatz \cite{Blaschke:2008a} to the construction of a renormalizable $U_\star(1)$ gauge version of the $1/p^2$ scalar model the operator $\big(D^2\Dt^2\big)^{-1}$ (in the action~ \eqref{eq:canon:p2inv_gauge_model1_act}, being denoted here by $\Act_{\text{inv}}^{\text{nloc}}$) was coupled to an auxiliary real-valued antisymmetric f\/ield $\oldB_{\m\n}$ of mass dimension two. This was achieved by replacing
\begin{gather}
\label{eq:canon:p2inv_loc_real_localization}
 \Act_{\text{inv}}^{\text{nloc}} \to \Act_{\text{inv}}^{\text{loc}},\\
 \intx \left[\inv{4}F_{\mu\nu} F_{\mu\nu}
+\inv{4}F_{\mu\nu}\frac{a^2}{D^2\Dt^2} F_{\mu\nu}\right] \to\intx \left[\inv{4}F_{\mu\nu} F_{\mu\nu} +a\oldB_{\mu\nu} F_{\mu\nu} -\oldB_{\mu\nu}D^2\Dt^2 \oldB_{\mu\nu}\right],\nonumber
\end{gather}
in the action, where (here, and in what follows) all products between f\/ields are understood to be of the {\moyal} form, and $a$ is a dimensionless constant, motivated by the fact that a similar parameter was renormalized in the scalar $1/p^2$ model~\cite{Blaschke:2008b}. Obviously, the action~$\Act_{\text{inv}}^{\text{loc}}$ contains only local terms which is the reason why the process is called ``localization''. Equivalence of localized and non-localized actions can immediately be seen by reinserting $\oldB_{\m\n}$ (which can be expressed from the equation of motion being obtained from the right hand side of equation~\eqref{eq:canon:p2inv_loc_real_localization}) into $\Act_{\text{inv}}^{\text{loc}}$. The resulting model features the desired damping behaviour in the gauge and ghost f\/ield propagators $G^{AA}_{\m\n}$ and $G^{\bc c}$, respectively. However, there are also mixed and pure propagators $G^{A\oldB}_{\m,\r\s}$, $G^{\oldB A}_{\r\s,\m}$, and $G^{\oldB\oldB}_{\m\n,\r\s}$ of the new auxiliary f\/ield which diverge in the limit of vanishing momentum. For the UV power counting it was obtained that the superf\/icial degree of divergence behaves like
\[
 d_\g=4-E_A-E_{c/\bc}-2E_a,
\]
for $E_A$ and $E_{c/\bc}$ counting external f\/ields of gauge and ghost/antighost f\/ields, respectively, and~$E_a$ counting the overall powers of the parameter $a$ which parametrizes the term $a\oldB_{\mu\nu} F_{\mu\nu}$ in the action \eqref{eq:canon:p2inv_loc_real_localization}. In fact, $a$ appears in all propagators containing at least one auxiliary f\/ield. Therefore, the ef\/fect of the damping mechanism is obvious since $d_\g$ is lowered by any appearance of~$\oldB$.

However, it turned out that the auxiliary f\/ield has not been introduced in a physically inva\-riant way. A f\/irst indication is that for the limit $a\to 0$, $\oldB_{\m\n}$ is not eliminated from the equations of motion. In fact, part of the propagator $G^{\oldB\oldB}_{\m\n,\r\s}$ as well as vertices with at least two $\oldB$ f\/ields and 1--4 gauge f\/ields remain unaltered in this case, and give rise to respective interactions. The actual cause, however, lies in the mathematical scheme used for the localization and can be revealed by integrating out the auxiliary f\/ield in the path integral formalism
\begin{gather}
\int\mathcal{D} A \mathcal{D}\oldB \exp\bigg\{{-}\intx\left[\inv{4} F_{\m\n}F_{\m\n}+a\oldB_{\m\n}F_{\m\n}-\oldB_{\m\n}D^2\Dt^2\oldB_{\m\n}\right]\bigg\}\nonumber\\
\qquad = \int\mathcal{D} A \mathcal{D}\oldB \exp\bigg\{{-}\intx\bigg[\inv{4} F_{\m\n}F_{\m\n}\nonumber\\
\qquad\quad{} -\bigg(\oldB_{\m\n} -\frac{a}{2}\inv{D^2\Dt^2}F_{\m\n}\bigg)D^2\Dt^2\Big(\oldB_{\m\n} -\frac{a}{2}\inv{D^2\Dt^2}F_{\m\n}\Big)
+\frac{a^2}{4}F_{\m\n}\inv{D^2\Dt^2}F_{\m\n}\bigg] \bigg\}\nonumber\\
\qquad = \int\mathcal{D} A \big(\det D^2\Dt^2\big)^{-2}\exp\bigg\{{-}\intx\inv{4} F_{\m\n}\left(1+\frac{a^2}{D^2\Dt^2}\right)F_{\m\n}\bigg\}.\label{eq:canon:p2inv_loc_real_loc-nloc_equiv}
\end{gather}
Similar to the case of QED, where the ghost f\/ields $c$ and $\bc$ are required in order to counterbalance the non-vanishing functional determinant after integrating out the Lagrange multiplier f\/ield which implements the gauge constraint, the non-vanishing factor $\big(\det D^2\Dt^2\big)^{-2}$ in equation~\eqref{eq:canon:p2inv_loc_real_loc-nloc_equiv} indicates that the integrated and non-integrated versions ($\Act_{\text{inv}}^{\text{loc}}$ and $\Act_{\text{inv}}^{\text{nloc}}$, respectively) of the action are \emph{not} equivalent, and additional ghosts would be required to restore the original physical content.

A solution to this problem was proposed by Vilar et al.~\cite{Vilar:2009} who replaced the real-valued f\/ield~$\oldB$ by two pairs of complex conjugated f\/ields ($B$, $\bB$ and $\chi$, $\bar{\chi}$) which are assigned to appropriate ghosts. The respective localization reads
\begin{gather}
\label{eq:canon:p2inv_loc_vilar_localization}
\Act_{\text{nloc}}\to\Act_{\text{loc}} =\Act_{\text{loc,0}}+\Act_{\text{break}}\\
\phantom{\Act_{\text{nloc}}\to\Act_{\text{loc}}}{}
  = \intx \left(\bar{\chi}_{\m\n} D^2B_{\m\n}+\bar{B}_{\m\n} D^2\chi_{\m\n} + \gamma^2\bar{\chi}_{\m\n}\chi_{\m\n}\right)
 +\ri \frac{\g}{2}\intx \left(\bB_{\m\n}-B_{\m\n}\right) F_{\m\n} ,\nonumber
\end{gather}
with $\gamma$ being a parameter of mass dimension one. The term $\Act_{\text{nloc}}$ is now split into a BRST invariant part $\Act_{\text{loc,0}}$, and a breaking term $\Act_{\text{break}}$ as can be seen by explicit calculation with the def\/initions in~\cite{Vilar:2009}. The additional degrees of freedom are eliminated by following the ideas of Zwanziger~\cite{Zwanziger:1989} (see~\cite{Dudal:2008} for a more comprehensive review of the topic) to add a ghost for each auxiliary f\/ield in such a way that BRST doublet structures are formed. This results in a trivial BRST cohomology for $\Act_{\text{loc,0}}$ from which follows~\cite{Baulieu:2009} that
\[
s\Act_{\text{loc,0}}=0\quad \Rightarrow\quad \Act_{\text{loc,0}} = s \hat{\Act}_{\text{loc,0}} ,
\]
i.e.\ the part of the action depending on the auxiliary f\/ields and their associated ghosts can be written as an exact expression with respect to the nilpotent BRST operator $s$.

In contrast to that, the breaking term $\Act_{\text{break}}$ does not join this nice property due to a non-trivial cohomology (i.e.\ $s\Act_{\text{break}}\neq0)$. However, it is constructed in such a way, that the mass dimension of its f\/ield dependent part is smaller than $D=4$, the dimension of the underlying Euclidean space $\R_{\th}^4$. Such a breaking is referred to as `soft' (c.f.~\cite{Baulieu:2009}), and does not spoil renormalizability~\cite{Zwanziger:1993}. This latter fact becomes intuitively clear if we consider that a theory with vertices $v$ having a canonical dimension $d_v<D$ is known to be super-renormalizable. Since the breaking term also features this dimensional property, it seems reasonable that it does not inf\/luence higher order quantum corrections corresponding to the high energy limit. Additionally, $\Act_{\text{break}}$ is the actual origin of the suppression of UV/IR mixing featured by this theory, as it alters the IR sector while not af\/fecting the UV part. The mechanism of soft breaking in combination with UV renormalization will be discussed subsequently below for the BRSW model.

Another important aspect of the model by Vilar et al.\ is the splitting of the operator $D^2\Dt^2$ into two separate parts, and an overall constant factor carrying the mass dimension of the parameter $\th$, i.e.\ $D^2\Dt^2\to\th^2(D^2)^2$. Such a splitting, however, is only possible in Euclidean space\footnote{In Minkowski space, non-commutativity with time leads to dif\/f\/iculties in the interpretation of time ordering and unitarity, and hence to rather new types of Feynman rules (see~\cite{Denk:2003,Bahns:2003} and references therein). Generally, the trend is therefore to restrict non-vanishing components of $\th$ to the spacial part of the metric.} if $\th_{\m\n}$ has full rank, as has the special form of $\th_{\m\n}$ being def\/ined in equation~\eqref{eq:canon:theta_matrix}, and allows for $\Dt^2\equiv\th^2D^2$. Therefore, the proposed solution~\eqref{eq:canon:p2inv_loc_vilar_localization} will only exist in special cases, and cannot be considered as a general solution to the localization problem.

Yet another aspect which comes into play with the approach of Vilar et al.\ is the question of a general proof of renormalizability. It has been argued~\cite{Vilar:2009} that the symmetry content of the model would satisfy the Quantum Action Principle known from commutative theory. Therefore, the method of Algebraic Renormalization should be applicable. As has been discussed extensively in~\cite{Blaschke:2009c,Sedmik:2009,Rofner:2009} there are serious concerns if the mathematical basics and presumptions of Algebraic Renormalization are valid on {\nc} spaces. However, no f\/inal conclusion has yet been achieved in this respect.

An alternative version of the model by Vilar et al.\ was proposed in~\cite{Blaschke:2009b}. The main idea was to keep the operator $D^2\Dt^2$ in its original form and to not split it in two. In this way, the number of f\/ields, sources and ghosts which are necessary for the localization could be reduced (from~30 to~22) without signif\/icantly lowering the symmetry content of the theory. However, as was shown by explicit computations in~\cite{Blaschke:2009d}, the total number of Feynman graphs which need to be considered (even at one loop order) in the perturbative renormalization procedure is still rather high. Similar to the model by Vilar  et al.\ the damping is implemented in a breaking term. Since the scheme is quite general it may be interesting to discuss it in a little more detail. First of all, the localization is now given by
\begin{gather}
\label{eq:canon:p2inv_loc_our_sb_localization}
\Act_{\text{nloc}} \longrightarrow \Act_{\text{loc}},\\
\int\! d^4x F_{\m\n}\inv{D^2\Dt^2}F_{\m\n}   \longrightarrow \int\! d^4x \!\left[\frac{\l}{2} \left(B_{\m\n}\! + \bB_{\m\n}\right) F^{\m\n} \! -\mu^2\bB_{\m\n}D^2 \Dt^2B^{\m\n}\! +\mu^2\bpsi_{\m\n}D^2 \Dt^2\psi^{\m\n}\right],\!
\nonumber
\end{gather}
where again the auxiliary f\/ields $B$ and $\bB$ are coupled to respective ghost and antighost f\/ields~$\psi$ and~$\bpsi$. As can easily be checked as sketched above in equation~\eqref{eq:canon:p2inv_loc_real_loc-nloc_equiv} the localized version is mathematically and physically identical to the initial version of $\Act_{\rm inv}$ in~\eqref{eq:canon:p2inv_gauge_model1_act}. The new f\/ields obey the following BRST transformation rules
\begin{alignat}{3}
& s\bpsi_{\mu\nu}=\bB_{\m\n}+\ri g\aco{c}{\bpsi_{\m\n}} ,  \qquad               && s\bB_{\m\n}=\ig\co{c}{\bB_{\m\n}}, & \nonumber\\
& sB_{\m\n}=\psi_{\m\n}+\ri g\co{c}{B_{\m\n}},  \qquad                  && s\psi_{\m\n}=\ig\aco{c}{\psi_{\m\n}}, &
\label{eq:canon:p2inv_loc_our_sb_BRST_auxiliary_fields}
\end{alignat}
which have to be considered in addition to the existing relations of equation~\eqref{eq:canon:p2inv_gauge_naive_brst}. Now we can rewrite equation~\eqref{eq:canon:p2inv_loc_our_sb_localization} in the form
\begin{gather*}
\Act_{\text{loc}}
=\intx\left[s\left(\frac{\l}{2}\bpsi_{\m\n}F^{\m\n}
-\mu^2\bpsi_{\m\n}D^2\Dt^2B^{\m\n}\right)+\frac{\l}{2}B_{\m\n}F^{\m\n}\right],
\end{gather*}
where the last term gives rise to a breaking of BRST invariance, as
\begin{gather}
\label{eq:canon:p2inv_loc_our_sb_act_break_wo_source}
s\Act_{\text{break}} =\intx\, \frac{\l}{2}\psi_{\m\n}F^{\m\n}, \qquad \text{with}\quad \Act_{\text{break}}=\intx\, \frac{\l}{2}B_{\m\n}F^{\m\n} .
\end{gather}
As in the model by Vilar et al.\ the mass dimension $d_m$ of the f\/ield dependent part of $\Act_{\text{break}}$ fulf\/ills the condition $d_m\left(\psi_{\m\n}F^{\m\n}\right)=3<D=4$, and thus can be considered as an implementation of a~{\emph soft breaking}. However, in order to restore BRST invariance in the UV region (as is a~prerequisite for an eventual future application of AR) an additional set of sources
\begin{alignat*}{3}
&s \bar{Q}_{\m\n\a\b}=\bar{J}_{\m\n\a\b}+\ig \aco{c}{\bar{Q}_{\m\n\a\b}}, \qquad && s \bar{J}_{\m\n\a\b}=\ig \co{c}{\bar{J}_{\m\n\a\b}}, &
\nonumber\\
&s Q_{\m\n\a\b}=J_{\m\n\a\b}+\ig \aco{c}{Q_{\m\n\a\b}}, \qquad && s J_{\m\n\a\b}=\ig \co{c}{J_{\m\n\a\b}}, & 
\end{alignat*}
has to be coupled to the breaking term which then takes the form
\begin{gather*}
\Act_{\text{break}} =\intx\,  s\left(\bar{Q}_{\m\n\a\b}B^{\m\n}F^{\a\b}\right)
 =\intx \left(\bar{J}_{\m\n\a\b}B^{\m\n}F^{\a\b}-\bar{Q}_{\m\n\a\b}\psi^{\m\n}F^{\a\b}\right).
\end{gather*}
The original term equation~\eqref{eq:canon:p2inv_loc_our_sb_act_break_wo_source} is reobtained if the sources $\bar{Q}$ and $\bar{J}$ are assigned to their `physical values'
\begin{alignat*}{3}
& \bar{Q}_{\m\n\a\b}\big|_{\text{phys}}=0,\qquad && \bar{J}_{\m\n\a\b}\big|_{\text{phys}}=\frac{\l}{4}\left(\d_{\m\a}\d_{\n\b}-\d_{\m\b}\d_{\n\a}\right), & \nonumber\\
& Q_{\m\n\a\b}\big|_{\text{phys}}=0,\qquad &&  J_{\m\n\a\b}\big|_{\text{phys}}=\frac{\l}{4}\left(\d_{\m\a}\d_{\n\b}-\d_{\m\b}\d_{\n\a}\right). &
\end{alignat*}

The soft breaking term implements the damping mechanism in the limit of low energies (IR) while not af\/fecting the symmetries or divergence structure in the UV. This interplay between the scales should presumably lead to a renormalizable model. However, as has been analyzed in~\cite{Blaschke:2009c,Blaschke:2009d}, there are hidden obstacles. Without going into detail at this point, the problem is that the damping behaviour is not featured by all propagators. Although at one loop level only the expected quadratic divergences appear, a respective renormalization is inhibited. This is due to the fact that the contributions which enter the perturbative series represent dressed graphs which have external propagators attached to them. The IR-divergences of mixed propagators $G^{AB}=G^{A\bB}$ add to the ones of the respective results in the vacuum polarization. In a more formal way,
\begin{gather}
G^{\rm AA,1l-ren}_{\m\n}(p)=G^{\rm AA}_{\m\n}(p) +G^{\rm AA}_{\m\r}(p)\Pi_{\r\s}(p)G^{\rm AA}_{\s\n}(p)
  + G^{\rm AA}_{\m\r}(p)2\Sigma^{\rm AB}_{\r,\s1\s2}(p)G^{\rm BA}_{\s1\s2,\n}(p)\nonumber\\
\phantom{G^{\rm AA,1l-ren}_{\m\n}(p)=}{}
 + G^{\rm AA}_{\m\r}(p)2\Sigma^{\rm A\bB}_{\r,\s1\s2}(p)G^{\rm \bB A}_{\s1\s2,\n}(p)
 + G^{\rm AB}_{\m,\r1\r2}(p)\Sigma^{\rm BB}_{\r1\r2,\s1\s2}(p)G^{\rm BA}_{\s1\s2,\n}(p)\nonumber\\
\phantom{G^{\rm AA,1l-ren}_{\m\n}(p)=}{}
 + G^{\rm AB}_{\m,\r1\r2}(p)2\Sigma^{\rm B\bB}_{\r1\r2,\s1\s2}(p)G^{\rm \bB A}_{\s1\s2,\n}(p) \nonumber\\
 \phantom{G^{\rm AA,1l-ren}_{\m\n}(p)=}{}
 + G^{\rm A\bB}_{\m,\r1\r2}(p)\Sigma^{\rm \bB\bB}_{\r1\r2,\s1\s2}(p)G^{\rm \bB A}_{\s1\s2,\n}(p)+\mathcal{O}\left(g^4\right),\label{eq:canon:p2inv_loc_our_sb_dressed_prop_complete}
\end{gather}
where $\Pi\equiv\Sigma^{AA}$ and $\Sigma^{XY}$ symbolizes the sum of all divergent one-loop contributions with external f\/ields $X$ and $Y$. In the end, the corrections on the right hand side of equation~\eqref{eq:canon:p2inv_loc_our_sb_dressed_prop_complete} behave like $(\p^2)^{-3}$ in the infrared. This, in turn gives rise to respective counter terms to the action. As long as these were the only ones to appear in higher loop orders this would state no problem but the intuitive apprehension has proven to be true~\cite{Blaschke:2009d} that for specif\/ic types of graphs (those with only external $B$ and $\bB$ lines) the divergences rise with the order. Although there is no rigorous proof up to now the model may be considered as problematic with respect to renormalization. It has to be noted that the same ef\/fects are obtained in the model by Vilar et al.\ so the same conclusion applies to it.

After all, the series of attempts~\cite{Blaschke:2008a,Blaschke:2009a,Vilar:2009,Blaschke:2009b,Blaschke:2009c,Blaschke:2009d} for the construction of a renormalizable gauge theory based on the damping mechanism of the $1/p^2$ model has led to several insights which can be considered as prerequisites for the success of any further approach in this direction.
\begin{itemize}\itemsep=0pt
 \item [--] A consistent BRST-invariant and physically sound implementation of the damping can only be achieved by localization with auxiliary f\/ields.
 \item [--] Localization has to be performed such that a soft breaking of the BRST invariance results. Only in this way a damping of the IR singularities can be implemented without af\/fecting the UV region, which is the relevant domain for the symmetry content of the theory.
 \item [--] It is of vital importance that any f\/ield being connected to a physically relevant (gauge) f\/ield by a two-point function (mixed propagator) features the same damping behaviour.
\end{itemize}
Rigorous implementation of these demands has f\/inally led to a (presumably) renormalizable model which is described in \secref{sec:canon:p2inv_gauge_brsw} below.

\subsubsection{BRSW model}
\label{sec:canon:p2inv_gauge_brsw}

A promising attempt for the construction of a renormalizable gauge model on {\nc} space has been published recently~\cite{Blaschke:2009e}. The intention is to start from the localized action~\eqref{eq:canon:p2inv_loc_vilar_localization}, and modify it in order to achieve renormalizability and avoid the problems discussed in Sections~\ref{sec:canon:p2inv_gauge} and \ref{sec:canon:p2inv_loc}. In a f\/irst step, the interplay between terms of the action, and the form and type of propagators is analyzed thoroughly. There are three main ideas leading to success.

First, in order to avoid (or at least restrict) the appearance of dimensionless derivative operators (as is discussed in~\cite{Blaschke:2009c}) it is desirable to remove any explicit appearance of parameters with negative mass dimension from the action. However, this is impossible, since the ef\/fect of {\uim} inevitably leads to divergences being contracted with $\th_{\m\n}$ (as discussed in \secref{sec:canon:p2inv_loc}), which enter the action in the form of counter terms. A viable solution to this problem is to split the parameter of non-commutativity into a dimensionless tensor structure $\mth_{\m\n}=-\mth_{\n\m}$,  and a~dimensionful scalar parameter $\sth$, i.e.
\begin{gather}
 \theta_{\m\n}\to\sth \mth_{\m\n}\qquad  \text{with} \quad d_m(\mth_{\m\n})=0,\qquad \text{and} \qquad d_m(\sth)=-2.
\label{eq:canon:p2inv_gauge_brsw_constr_def_mth}
\end{gather}
In consequence, the appearance of $\sth$ in the tree level action is reduced by modifying our def\/inition of contractions, $\wsq\equiv\mth_{\m\r}\mth_{\n\s}\partial_\r\partial_\s$, $\p_\m\equiv p_\n\mth_{\m\n}$, for any vector $p_\m$, and $\widetilde{O}_{\m_1\,\m_2\ldots\m_n}\equiv{O}_{\n\,\m_2 \ldots\m_n}\mth_{\m_1\n}$ for a tensor with $n$ indices. Hence, the only occurrence of the dimensionful $\sth$ is in the phase associated with the star product, which does not inf\/luence the bi-linear part according to the cyclic invariance of the star product under the integral. In this respect we note that operators such as $\wsq$ or $\Dt$ now come with their usually expected mass dimensions $d_m(\wsq)=2$ and \mbox{$d_m(\Dt)=1$}, respectively.
Starting from the localized part of the action~\eqref{eq:canon:p2inv_loc_our_sb_localization}, the remaining two steps can be written as
\begin{subequations}
\label{eq:brss_constr_evol}
\begin{gather}
\intx \, \frac{a}{2}\left(B_{\m\n}+\bB_{\m\n}\right)F_{\m\n}-\bB_{\m\n}\sth^2\Dt^2D^2B_{\m\n},\label{eq:canon:p2inv_gauge_brsw_constr_step0}\\
\phantom{\text{step 1}}\downarrow\text{step 1}\notag\\
\intx\,  \frac{\gamma^3}{2}\left(B_{\m\n}+\bB_{\m\n}\right)\inv{\wsq}F_{\m\n}+\bB_{\m\n}(\s-D^2) B_{\m\n},\label{eq:canon:p2inv_gauge_brsw_constr_step1}\\
\phantom{\text{step 2}}\downarrow\text{step 2}\notag\\
\intx \, \frac{\gamma^2}{2}\left(B_{\m\n}+\bB_{\m\n}\right)\inv{\wsq}\left(f_{\m\n}+\s\frac{\mth_{\m\n}}{2}\tilde{f}\right)-\bB_{\m\n}B_{\m\n},\label{eq:canon:p2inv_gauge_brsw_constr_step2}\,\,
\end{gather}
\end{subequations}
with several new def\/initions being explained subsequently. To understand the f\/irst step we note that the divergences in the $G^{\{AB,A\bB\}}$, $G^{\{\bB B,BB\}}$, and $G^{\bpsi\psi}$ propagators (see \secref{sec:canon:p2inv_gauge}, equation~\eqref{eq:canon:p2inv_loc_our_sb_dressed_prop_complete} above) are mainly caused by the appearance of the operator $D^2\Dt^2$ sandwiched between $\bB_{\m\n}$ and $B_{\m\n}$. On the other hand this term is crucial to the construction of the correct damping factor for the gauge boson propagator $G^{AA}$. A detailed analysis~\cite{Sedmik:2009} leads to the insight that it is possible to move the problematic operator into the soft breaking term, thereby maintaining the desired damping while eliminating the divergences. Note also that, due to the redef\/inition of $\theta_{\m\n}$ in equation~\eqref{eq:canon:p2inv_gauge_brsw_constr_def_mth} the dimensionful $\sth$ does not appear explicitly after the f\/irst step in equation~\eqref{eq:canon:p2inv_gauge_brsw_constr_step1}. In the resulting action, the correct mass dimensions are restored by the new parameters~$\g$ and $\s$ featuring $d_m(\g)=1$ and $d_m(\s)=2$, respectively.

In step~2, we note that the regularizing ef\/fects are solely implemented in the bi-linear part of the action, therefore opening the option to reduce the f\/ield strength tensor $F_{\m\n}$ in the soft breaking term to its bi-linear part $f_{\m\n}\equiv\partial_\m A_\n-\partial_\n A_\m$\label{eq:canon:p2inv_gauge_brsw_bilin_f_def} (and $\tilde{f}\equiv \mth f_{\m\n}= 2 \tilde{\partial}\cdot A$). Noting furthermore, that the $D^2$ operator in the $\bB/B$ sector is not required any more for the implementation of the damping mechanism we may entirely omit this derivative. Due to this reduction, any interaction (represented by $n$-point functions with $n\geq3$) of $A_\m$ with auxiliary f\/ields and ghosts is eliminated. However, in order to restore the correct mass dimension for the altered terms we have to change $d_m$ of the f\/ields $B_{\m\n}$ and $\bB_{\m\n}$ from~1 to~2.
Furthermore, in order to implement a suitable term to absorb the $\theta$-contracted one loop divergence of the two point function (see~\cite{Blaschke:2009e}) we further modify the breaking by the insertion of the term $\frac{\gamma^2}{4}\s\left(B_{\m\n}+\bB_{\m\n}\right)\inv{\wsq}\mth_{\m\n}\tilde{f}$, resulting in \eqref{eq:canon:p2inv_gauge_brsw_constr_step2}. Additionally, we add a soft-breaking term to absorb the $\theta$-contracted one loop divergence of the three point function~\cite{Blaschke:2009e}.

The complete action hence takes the form,
\begin{gather}
\Act=\Act_{\text{inv}}+\Act_{\text{gf}}+\Act_{\text{aux}}+\Act_{\text{break}}+\Act_{\text{ext}},\nonumber\\
\Act_{\text{inv}}=\intx\, \inv{4}F_{\m\n}F_{\m\n},\nonumber\\
\Act_{\text{gf}}=\intx\,  s\left(\bc \pa_\m A_\m\right)=\intx\left(b \pa_\m A_\m-\bc \pa_\m D_\m c\right),\nonumber\\
\Act_{\text{aux}}=-\intx \, s\left(\bpsi_{\m\n}B_{\m\n}\right)=\intx\left(-\bB_{\m\n}B_{\m\n}+\bpsi_{\m\n}\psi_{\m\n}\right),\nonumber\\
\Act_{\text{break}}=\intx \, s\left[\left(\bar{Q}_{\m\n\a\b}B_{\m\n}+Q_{\m\n\a\b}\bar{B}_{\m\n}\right)\inv{\wsq}\left(f_{\a\b}
+\s\frac{\mth_{\a\b}}{2}\tilde{f}\right)\right]\nonumber\\
\phantom{\Act_{\text{break}}}{}
=\intx\Bigg[\left(\bar{J}_{\m\n\a\b}B_{\m\n}+J_{\m\n\a\b}\bar{B}_{\m\n}\right) \inv{\wsq} \left(f_{\a\b}+\frac{\s\mth_{\a\b}}{2}\tilde{f}\right) \nonumber\\
\phantom{\Act_{\text{break}}=}{}
- \bar{Q}_{\m\n\a\b}\psi_{\m\n}\inv{\wsq} \left(f_{\a\b}+\frac{\s\mth_{\a\b}}{2}\tilde{f}\right)-\left(\bar{Q}_{\m\n\a\b}B_{\m\n}+Q_{\m\n\a\b}\bar{B}_{\m\n}\right)\inv{\wsq}\mathop{s}\left(f_{\a\b}
+\s\frac{\mth_{\a\b}}{2}\tilde{f}\right)\nonumber\\
\phantom{\Act_{\text{break}}=}{}
+ J'\aco{A_\m}{A_\n} \frac{\tilde{\pa}_\m\tilde{\pa}_\n\tilde{\pa}_\r}{\wsq^2}A_\r -
Q's\left(\aco{A_\m}{A_\n} \frac{\tilde{\pa}_\m\tilde{\pa}_\n\tilde{\pa}_\r}{\wsq^2}A_\r\right)\Bigg], \nonumber\\
\Act_{\text{ext}}=\intx\left(\W^A_\m sA_\m+\W^c sc\right),\label{eq:canon:p2inv_gauge_brsw_act_complete}
\end{gather}
where all products are implicitly assumed to be deformed Groenewold--Moyal products, and we have introduced the external sources $\W^A_\m$ and $\W^c$.
Due to the decoupling of the gauge sector the form of the BRST transformations is simpler than the respective counterparts in equation~\eqref{eq:canon:p2inv_loc_our_sb_BRST_auxiliary_fields} being representative for the models\footnote{Since the (anti-)commutator relations can be omitted, thus.} sketched in \secref{sec:canon:p2inv_loc}
\begin{alignat*}{3}
& s\bpsi_{\mu\nu} =\bB_{\m\n},\qquad                 && s \bB_{\m\n} =0,& \nonumber\\
& s B_{\m\n} =\psi_{\m\n},  \qquad                  && s \psi_{\m\n} =0 , & \nonumber\\
& s \bar{Q}_{\m\n\a\b} =\bar{J}_{\m\n\a\b} , \qquad && s  \bar{J}_{\m\n\a\b} =0 , & \nonumber\\
& s Q_{\m\n\a\b} =J_{\m\n\a\b}, \qquad && s J_{\m\n\a\b} =0, & \nonumber\\
& s Q'=J', \qquad && s J'=0. & 
\end{alignat*}
As before, the additional pairs of sources $\{\bar{Q}_{\m\n\a\b},Q_{\m\n\a\b},Q'\}$ and $\{\bar{J}_{\m\n\a\b},J_{\m\n\a\b},J'\}$ have been introduced in order to restore BRST invariance of the action \eqref{eq:canon:p2inv_gauge_brsw_act_complete} in the UV limit, i.e.\ $s\Act= 0$. In the IR limit the physical values
\begin{gather*}
 \bar{Q}_{\m\n\a\b}\big|_{\text{phys}}=Q_{\m\n\a\b}\big|_{\text{phys}}=Q'\big|_{\text{phys}}=0 ,
\qquad J'\big|_{\text{phys}}=\ig\g'^2, \nonumber\\ \bar{J}_{\m\n\a\b}\big|_{\text{phys}}=J_{\m\n\a\b}\big|_{\text{phys}}=\frac{\g^2}{4}\left(\d_{\m\a}\d_{\n\b}-\d_{\m\b}\d_{\n\a}\right),
\end{gather*}
lead back to a breaking term.
The BRSW model yields the following relevant propagators:
\begin{subequations}
\begin{gather}
G^{AA}_{\m\n}(k)=\inv{k^2\left(1+\frac{\g^4}{(\k^2)^2}\right)}\left(\d_{\m\n}-\frac{k_\m k_\n}{k^2}- \frac{\left(\s+\frac{\mth^2}{4}\s^2\right)\g^4}{\left[\left(\s+\frac{\mth^2}{4}\s^2\right)\g^4
+k^2\left(\k^2+\frac{\g^4}{\k^2}\right)\right]}\frac{\k_\m\k_\n}{\k^2}\right),\!\!\label{eq:canon:p2inv_gauge_brsw_prop_AA}\\
G^{\bc c}(k)=\frac{-1}{k^2},\label{eq:canon:p2inv_gauge_brsw_prop_cc}
\end{gather}
\end{subequations}
where the Landau gauge $\a\to0$ has led to the omission of the term $-\a\tfrac{k_\m k_\n}{k^4}$.

Although there also exist two-point functions $G^{\{AB, A\bB\}}$, $G^{\{BB, \bB B\}}$ and $G^{\bpsi\psi}$, they will not contribute to any quantum correction since none of the vertex expressions $V^{3A}_{\rho\s\tau}$, $V^{4A}_{\rho\s\tau\e}$, and $V^{\bc A c}_{\m}$ connects either of these to the gauge f\/ield. At this point we note a remarkable similarity of the Feynman rules of the BRSW model, and the respective expressions of the {\naiv} implementation of NCQED in~\cite{Hayakawa:1999b}. The quadratic divergence for $k\to0$ in the ghost propagator~\eqref{eq:canon:p2inv_gauge_brsw_prop_cc} is typical for the Landau gauge $\a\to0$. Alternatively, as has been done in~\cite{Blaschke:2009a} for the real valued auxiliary f\/ield $\oldB_{\m\n}$ (see page~\pageref{eq:canon:p2inv_loc_real_localization} above) we could add a damping factor to the gauge f\/ixing term~$b(\partial A)$ and the ghost sector $\bc\partial_\m D^\m c$. However, these damping insertions would inevitably appear in vertex expressions with an inverse power relative to the respective propagators and, thus, cancel each other. Moreover, these factors contribute to UV divergences at higher loop orders, and are omitted, hence.

The gauge boson two point function~\eqref{eq:canon:p2inv_gauge_brsw_prop_AA} fulf\/ills all requirements which have been stated at the beginning of this section. It is f\/inite in both, the IR limit $k^2\to0$, and the UV limit $k^2\to\infty$. A simple analysis reveals that
\begin{gather*}
G^{AA}_{\m\n}(k)\approx\begin{cases}
\displaystyle \frac{\k^2}{\g^4}\left[\d_{\m\n}-\frac{k_\m k_\n}{k^2}-\frac{\st^4}{\left(\st^4+\g^4\right)}\frac{\k_\m\k_\n}{\k^2}\right], & \text{for} \ \ \k^2\to0 ,\vspace{1mm}\\
\displaystyle  \inv{k^2}\left(\d_{\m\n}-\frac{k_\m k_\n}{k^2}\right), & \text{for} \ \ k^2\to\infty,
\end{cases}
\end{gather*}
where the abbreviation
\[
 \st^4\equiv2\left(\s+\frac{\mth^2}{4}\s^2\right)\g^4,
\]
has been introduced for convenience\footnote{Note that this requires the property $\k^2=\mth^2k^2$ which follows from the special block-diagonal form of $\mth^{\m\n}$, as has been introduced in equation~\eqref{eq:canon:theta_matrix}. Moreover, since $\mth^2=\mth_{\m\r}\mth_{\r\n}=\d_{\m\n}$, we have indeed $\k^2\equiv k^2$.}. As has been shown explicitly in \cite{Blaschke:2009e,Sedmik:2009} the form of~$G^{AA}$ is stable under quantum corrections since it provides a suitable term $\propto\frac{\k_\m\k_\n}{\k^2}$ to absorb expected divergences.

From the Feynman rules (see~\cite{Blaschke:2009e}), it is straightforward to derive an expression for the UV power counting of the BRSW model. We obtain
\begin{gather}
\label{eq:canon:p2inv_gauge_brsw_powercounting}
d_\gamma=4-E_A-E_{c\bc},
\end{gather}
which, again, shows remarkable agreement with the respective relations for the {\naiv} implementation of {\nc} $U_\star(1)$. Indeed, none of the auxiliary f\/ields or respective parameters inf\/luences the power counting\footnote{In comparison, the results of respective relations in \secref{sec:canon:p2inv_loc} for previous models are ef\/fectively reduced by the number of external legs of auxiliary f\/ields and/or the parameter of the breaking (respectively damping) term.}.

Explicit one-loop calculations for the model \eqref{eq:canon:p2inv_gauge_brsw_act_complete} have been conducted in~\cite{Blaschke:2009e}. As expected, the vacuum polarization\footnote{It shall be remarked that in the BRSW model the one-loop corrections to the photon propagator are contributed by only three graphs, which are similar to those being known from QCD, and can also be found in {\naiv} implementations of NCQED~\cite{Hayakawa:1999b}.} contains a quadratic IR divergence in the external momentum~$p_\m$, and a logarithmic UV divergence in the cutof\/f~$\L$
\begin{gather}
\label{eq:canon:p2inv_gauge_brsw_vac_pol_divergence}
\Pi_{\mu\nu}(p)= \frac{2 g^2}{\pi^2\sth^2}\frac{\p_\m \p_\n}{\left(\p^2\right)^2}+\frac{13 g^2}{3 (4\pi)^2}\left(p^2\delta_{\m\n}-p_\m p_\n\right)\ln\left(\L\right)+\text{f\/inite terms}.
\end{gather}
Note, that the physical requirement of transversality $p_\m \Pi_{\m\n}=0$ is fulf\/illed due to the property $p_\m\p_\m=0$ arising from the antisymmetry of $\mth_{\m\n}$. Further analysis reveals that the f\/irst term of equation~\eqref{eq:canon:p2inv_gauge_brsw_vac_pol_divergence} gives rise to a~renormalized constant $\s$ while the remaining divergences yield a wave function renormalization of the gauge f\/ield, and a redef\/inition of~$\g$. The form of the propagator~\eqref{eq:canon:p2inv_gauge_brsw_prop_AA} remains invariant under these operations.

As expected from the power counting~\eqref{eq:canon:p2inv_gauge_brsw_powercounting} corrections to the $V^{3A}$ vertex ($E_A=3$) are at most linearly divergent. The rather lengthy result can be summarized in the form
\begin{subequations}
\begin{gather}
\label{eq:canon:p2inv_gauge_brsw_3A_IR_divergence}
 \Gamma^{3A,\text{IR}}_{\m\n\r}(p_1,p_2,p_3) =-\frac{2\ri g^3}{\pi^2}\cos\left(\sth \frac{p_1\p_2}{2}\right)\sum\limits_{i=1,2,3}\frac{\p_{i,\m}\p_{i,\n}\p_{i,\r}}{\sth(\p_i^2)^2},\\
\Gamma^{3A,\text{UV}}_{\m\n\r}\!(p_1,p_2,p_3) =\frac{17}{3}\ig^3\pi^2\!\ln(\L)\sin\!\left(\!\sth \frac{p_1\p_2}{2}\!\right)\!\big[(p_1\!-p_2)_\r\d_{\m\n}\!+(p_2\!-p_3)_\m\d_{\n\r} \!
 +(p_3\!-p_1)_\n\d_{\m\r}\big]\!\nonumber\\
 \phantom{\Gamma^{3A,\text{UV}}_{\m\n\r}\!(p_1,p_2,p_3)}{}
 =-\frac{17\,g^2}{6(4\pi)^2}\ln (\Lambda) \widetilde{V}^{3A,\text{tree}}_{\m\n\r}(p_1,p_2,p_3).\label{eq:canon:p2inv_gauge_brsw_3A_UV_divergence}
\end{gather}
\end{subequations}
Similarly, the corrections to the four-point function $V^{4A}$ yield a sole logarithmic singularity\footnote{The correction for $V^{4A}$ can either be obtained by comprehensive explicit computations or from gauge invariance which can intuitively be understood from the fact that the relative factors between the terms $\ig \co{A_\m}{A_\n} \partial_\m A_\n$ and $-g^2{\co{A_\m}{A_\n}}^2$ in the $F^2$ term of the action has to remain the same before and after the renormalization. The latter method has been described explicitly in~\cite{Sedmik:2009}.} in~$\L$,
\begin{gather}
\label{eq:canon:p2inv_gauge_brsw_4A_UV_divergence}
\Gamma^{4A,\text{UV}}_{\m\n\r\s}(p_1,p_2,p_3,p_4)  =
- \frac{5}{8\pi^2} \ln(\L) g^2  \widetilde{V}^{4A}_{\m\n\r\s}(p_1,p_2,p_3,p_4) .
\end{gather}
While equation~\eqref{eq:canon:p2inv_gauge_brsw_3A_UV_divergence} (and \eqref{eq:canon:p2inv_gauge_brsw_4A_UV_divergence}) obviously  represent a renormalization of the coupling constant $g$, the contribution~\eqref{eq:canon:p2inv_gauge_brsw_3A_IR_divergence}
leads to a redef\/inition of $\g'$, thus leaving the action form-invariant.
Finally, the $\b$-function of the model is negative
which indicates asymptotic freedom. This can be understood from the fact that on {\nc} space the gauge group (intentionally $U_\star(1)$) is deformed such that the commutator $\starco{A_\m}{A_\n}\neq 0$. Therefore, any $U_\star(N)$ is ef\/fectively {\nA}.

The BRSW model has proven to be renormalizable at the one-loop level.
A~proof of the renormalizability to all loop orders
is currently being constructed using the method\footnote{Note that the method of Algebraic Renormalization, which is usually conducted in the framework of soft-breaking, requires a \emph{local} f\/ield theory. Since this requirement is never fulf\/illed in {\nc} f\/ield theories due to the star product, Multiscale Analysis seems more fruitful.} of Multiscale Analysis~\cite{workinprogress}.

\subsection{Time-ordered perturbation theory}
\label{sec:canon:other}

Throughout the previous (sub)sections we have either considered Euclidean spaces, or kept time commutative, i.e.\ $\cth^{0\mu}=0$. The dif\/f\/iculty with handling $\cth^{0\mu}\neq0$ lies in the fact that, due to the star products, the interaction part of the Lagrangian depends on inf\/initely many time derivatives acting on the f\/ields. A workaround has been proposed by S.~Doplicher et al.~\cite{Doplicher:1994b} and further developed for {\nc} scalar $\phi^4$ theory by several authors~\cite{Denk:2003,Denk:2004,Liao:2002,Fischer:2002}. It is termed ``interaction point time ordered perturbation theory'' (IPTOPT) and is based on the following idea: Consider the Gell-Mann--Low formula applied to the f\/ield operators $\phi$ of a~scalar $\phi^4$ theory
\begin{gather*}
\left\langle0|T\{\phi_H(x_1)\ldots\phi_H(x_n)\}|0\right\rangle =\sum_{m=0}^{\infty}\frac{(-\ri)^m}{m!}
\int\limits_{-\infty}^{\infty}dt_1\int\limits_{-\infty}^{\infty}dt_2\cdots\int\limits_{-\infty}^{\infty}dt_m\times\nonumber\\
\phantom{\left\langle0|T\{\phi_H(x_1)\ldots\phi_H(x_n)\}|0\right\rangle =}{}
\times\left\langle0|T\{\phi_I(x_1)\cdots\phi_I(x_n)V(t_1)\cdots V(t_m)\}|0\right\rangle.
\end{gather*}
The subscripts $H$ and $I$ denote the Heisenberg picture and the interaction picture, respectively. $V$ is the interaction part of the Hamiltonian
\begin{gather}\label{sec:canon:other_phi4-int}
V(z^0) =\int d^3z\, \frac{\kappa}{4!}\phi(z)\star\phi(z)\star\phi(z)\star\phi(z) .
\end{gather}
The idea is that the time-ordering operator $T$ acts on the time components of the $x_i$ and on the so-called \emph{time stamps} $t_1,\ldots,t_m$. For example, considering the interaction~(\ref{sec:canon:other_phi4-int}) with an alternative representation for the star products
\begin{gather*}
V(z^0) =\frac{\kappa}{4!}\prod_{i=1}^3\int\frac{d^4s_id^4l_i}{(2\pi)^4}e^{\ri s_il_i} \nonumber\\
\phantom{V(z^0) =}{}
\times\phi\left(z-\inv{2}\tilde{l}_1\right)\phi\left(z+s_1-\inv{2}\tilde{l}_2\right)\phi\left(z+s_1+s_2-\inv{2}\tilde{l}_3\right)\phi(z+s_1+s_2+s_3),
\end{gather*}
the time ordering only af\/fects $z^0$ and no other time components (like e.g.~$l_i^0$ etc.).
This leads to modif\/ied Feynman rules. For example, the propagator of $\phi^4$ theory
\begin{gather}\label{sec:canon:other_sl-other-prop1}
G (x,x')=\int\frac{d^4k}{(2\pi)^4}\frac{e^{\ri k(x-x')}}{k^2+m^2-\ri\e} ,
\end{gather}
is generalized to the so-called \emph{contractor}
\begin{gather*}
G_C(x,t;x',t')=\int\frac{d^4k}{(2\pi)^4}\frac{\exp\left[\ri k(x-x')+\ri k^0(x^0-t-(x'^0-t'))\right]}{k^2+m^2-\ri\e} \nonumber\\
\phantom{G_C(x,t;x',t')=}{}
\times\left[\cos\left(\omega_k(x^0-t-(x'^0-t'))\right)-\frac{\ri k^0}{\omega_k}\sin\left(\omega_k(x^0-t-(x'^0-t'))\right)\right] ,
\end{gather*}
which for $x^0=t$ and $x'^0=t'$ (being the case when $\cth^{0\mu}=0$) reduces to~(\ref{sec:canon:other_sl-other-prop1}). This approach seems promising in some respects, meaning that one may extend the formalism to {\nc} gauge f\/ields, although (among many others) the question of unitarity is still unclear~\cite{Ohl:2003}.

Finally, one should also remark that similar work, i.e.\ considerations concerning proper time ordering when dealing with {\nc} time, has been done by D.~Bahns et al.~\cite{Bahns:2002,Bahns:2003}. There even have been claims that in Minkowski space-time with proper time ordering, no inconsistencies related to {\uim} are present~\cite{Bahns:2004}. However, these conjectures still lack a rigorous proof.

\section{Non-canonical deformations}
\label{sec:noncanon}

In the previous sections, we have thoroughly discussed gauge theories formulated on {\moyal} space. The following shall therefore give a brief overview over other approaches, such as $x$-dependent $\cth^{\m\n}$. The topics we will cover are twisted gauge theories, then we will proceed to the case of linear dependence on $x$, i.e.~$\kappa$-deformed spaces and fuzzy spaces, and f\/inally review approaches with the most general $x$ dependence of the commutator, such as quantum groups and matrix models.

\subsection{Twisted gauge theories}
\label{sec:noncanon:twisted}

The approach to the so-called ``twisted gauge theories'' which we present in this subsection goes back to J.~Wess and his group\footnote{In fact, there have been even earlier proposals of twisting physical symmetries, see~\cite{Kulish:1999,Oeckl:2000eg}.}. For a recent review, see~\cite{Wess:2009zza,Aschieri:2006ye,Vassilevich:2006tc} and references therein. The main idea is, in addition to the pointwise product, to also deform the Leibniz rule by using Hopf algebra techniques. Following~\cite{Wess:2009zza}, consider f\/irst the undeformed (i.e.\ commutative) case: We def\/ine a pointwise product as
\begin{gather}
\m\{f\otimes g\} =f\cdot g ,
\label{eq:noncanon:twisted_pointwise}
\end{gather}
and the inf\/initesimal gauge transformation of a f\/ield scalar $\phi$ as
\[
\d_\a\phi(x) =\ri\a(x)\phi(x) ,
\]
where $\a(x)$ is Lie algebra valued (see \secref{sec:covar}).
The co-multiplication $\Delta(\a)$, an essential ingredient for a Hopf algebra (for more details see \secref{sec:noncanon:other:q}), is def\/ined by
\begin{gather*}
\Delta(\a) =\a\otimes\id+\id\otimes\a ,
\end{gather*}
and allows us to write the Leibniz rule for the gauge transformation of a product of f\/ields in the language of Hopf algebras as
\begin{gather}
\d_\a(\phi_1\cdot\phi_2) =(\d_\a\phi_1)\phi_2+\phi_1(\d_\a\phi_2)
 =\m\{\Delta(\a)\phi_1\otimes\phi_2\} .
\label{eq:noncanon:twisted_leibniz}
\end{gather}
In the \emph{deformed} case, on the other hand, one has to replace the pointwise product~\eqref{eq:noncanon:twisted_pointwise} with a deformed version, which in the simplest case could be the {\moyal} product of the previous section, i.e.\ in the Hopf algebra language
\begin{gather*}
\m_\star\{f\otimes g\} =\m\big\{\re^{\frac{\ri}{2}\cth^{\m\n}\pa_\m\otimes\pa_\n}f\otimes g\big\} .
\end{gather*}
The {\nc} gauge transformation $\delta^\star_\alpha$ on a single f\/ield is def\/ined as
\[
\delta^\star_\alpha \phi = \ri \alpha\cdot \phi ,
\]
as in the commutative case. This can be rewritten in terms of the star product \cite{Wess:2009zza},
\[
\delta^\star_\alpha \phi = \ri X^\star_{\alpha^a} \star T^a \phi .
\]
Furthermore, one considers a deformed~-- or ``twisted''~-- co-product
\begin{gather}
\Delta_{\mathcal{F}}(\a) =\mathcal{F}(\a\otimes\id+\id\otimes\a)\mathcal{F}^{-1}
 ,\qquad
\mathcal{F} =\re^{-\frac{\ri}{2}\th^{\m\n}\pa_\m\otimes\pa_\n} ,
\label{eq:noncanon:twisted_coprod-def}
\end{gather}
where $\mathcal{F}$ denotes a ``twist operator'' that has all the properties to def\/ine a Hopf algebra with~\eqref{eq:noncanon:twisted_coprod-def} as a co-multiplication.
Hence, we may write a {\moyal} deformed version of the Leibniz rule \eqref{eq:noncanon:twisted_leibniz} as
\begin{gather*}
\d^\star_\a(\phi_1\star\phi_2) =\ri\mu_\star\{\Delta_{\mathcal{F}}(\d^\star_\a)\phi_1\otimes\phi_2\}
=\ri(\a\phi_1)\star\phi_2+\ri\phi_1\star(\a\phi_2) \\
\phantom{\d^\star_\a(\phi_1\star\phi_2) =}{}
+\ri\sum\limits_{k=1}^{\infty}\inv{k!}\left(\frac{-\ri}{2}\right)^k\th^{\m_1\n_1}\cdots\th^{\m_k\n_k}\Big[(\pa_{\m_1}
 \cdots\pa_{\m_k}\a)\phi_1\star(\pa_{\n_1}\cdots\pa_{\n_k}\phi_2)\nn\\
\phantom{\d^\star_\a(\phi_1\star\phi_2) =}{}
  +(\pa_{\m_1}\cdots\pa_{\m_k}\phi_1)\star(\pa_{\n_1}\cdots\pa_{\n_k}\a)\phi_2\Big]
 .
\end{gather*}
Of course, this formalism can be readily used to include gauge f\/ields as well. As usual, the f\/ield strength (assuming $g=1$) is given by
\[
F_{\mu\nu} = \partial_\mu A_\nu - \partial_\nu A_\mu - \ri \starco{A_\mu}{A_\nu} ,
\]
which transforms covariantly:
\[
\delta^\star_\a F_{\mu\nu} = \ri X^\star_{\alpha^a} \star \co{T^a}{F_{\mu\nu}} = \ri\co{\alpha}{F_{\mu\nu}}.
\]
For the {\moyal} case, the action reads
\begin{gather}
\label{eq:noncanon:twisted:action}
S=-\frac 14 \int d^4x\, F_{\mu\nu}\star F^{\mu\nu} .
\end{gather}
Gauge invariance of this action has been shown explicitly also in~\cite{Banerjee:2006jy}. There is a remarkable dif\/ference to the non-twisted approach: Starting with a Lie algebra valued connection, twisted gauge transformations close in the Lie algebra. However, the consistency of the equations of motion of~\eqref{eq:noncanon:twisted:action} require the introduction of additional new vector potentials. The number of the new degrees of freedom is representation dependent but remains f\/inite for f\/inite dimensional representations.

To summarize, the main idea of this approach is to extend symmetry transformations, \mbox{(co-)products}, etc. by twists $\mathcal{F}$ in a consistent way. This approach can be generalized to $x$-dependent star products, if these products can be expressed in terms of a twist $\mathcal F$ as
\[
(f\star g) (x) = \mu \big( \mathcal F^{-1} f\otimes g \big) .
\]

The group around A.P.~Balachandran has proposed a dif\/ferent approach~-- for a review see~\cite{Balachandran:2007kv,Akofor:2008ae} and references therein: They consider canonically deformed Euclidean space. Non-commutative matter f\/ields are decorated with an additional dressing factor,
\[
\widehat \phi = \phi \re^{\frac 12 \stackrel{\leftarrow}{\partial_\mu} \cth^{\mu\nu} \hat{P}_\nu} ,
\]
where $\hat{P}_\nu$ denotes the total momentum operator, whereas the gauge f\/ields are the undeformed ones. So the {\nc} ef\/fects appear in the coupling of the gauge sector to matter. The dressing factors above lead to a twisted quantum statistics. In formulation of gauge models, consistency of the twisted statistics and the gauge invariance is required. The implications of this interesting approach and renormalizability of the resulting models are not yet fully explored.

\subsection[$\kappa$-deformation]{$\boldsymbol{\kappa}$-deformation}
\label{sec:noncanon:kappa}

Let us consider a $n$ dimensional Euclidean space with coordinates $x^1, \dots, x^n$. In the following, Latin indices range from $1$ to $n-1$, Greek indices from $1$ to $n$. The most general linear quantum space structure compatible with a deformed version of Poincar\'{e} symmetry is given by \cite{Lukierski:2001}
\begin{gather*}
 [ \hat x^\mu, \hat x^\nu ] = \ri \left( a^\mu \delta^\nu_\sigma -
	a^\nu \delta^\mu_\sigma \right) \hat x^\sigma,
\end{gather*}
where $a^\mu$ is a constant $4$-vector ``pointing into the direction of non-commutativity''. Its components also play the role of Lie algebra structure constants. In Euclidean spaces all directions are equivalent. For convenience, the non-commutativity will point into the $n$-direction\footnote{Commonly, the parameter $\kappa$ which gives its name to this approach, is introduced as $\kappa=1/a$.}, i.e.
\begin{gather*}
a^\mu = a   \delta^{n\mu}.
\end{gather*}
The coordinates $\hat x^1, \dots,\hat x^n$ generate the $n$-dimensional $\kappa$-Euclidean space algebra~$\mathcal E_\kappa$, and satisfy the relations
\begin{gather}
\label{eq:noncanon:kappa:commutator}
[\hat x^n,\hat x ^i] = \ri a\hat x^i,\qquad
[\hat x^i, \hat x^j] = 0 .
\end{gather}
The symmetry group of the $\kappa$-Euclidean space is a deformed version of the $n$-dimensional rotation group. It is generated by the rotations $M^{\mu\nu}$. Since the $n$-direction is special, we will denote the generators $M^{n l}$ by $N^l$ and call them boosts, in analogy to the Lorentz algebra. The relations between symmetry generators and coordinates have to be compatible with the algebra structure on the $\kappa$-deformed Euclidean space $\mathcal E_\kappa$ and are supposed to be linear. As a result, one obtains
\begin{alignat}{3}
& M^{rs} \hat x^k   = \delta^{rk} \hat x^s - \delta^{sk} \hat x^r + \hat x^k M^{rs}
 ,\qquad &&
M^{rs} \hat x^n   = \hat x^n M^{rs}
,& \nonumber\\
& N^l \hat x^i   = -\delta^{li} \hat x^n + \hat x^i N^l -\ri a M^{li}
 ,\qquad &&
N^l \hat x^n   = \hat x^l + \left( \hat x^n + \ri a \right) N^l
 .& \label{eq:noncanon:kappa:wirkung}
\end{alignat}
In the commutative limit, $a\to 0$, the usual relations for a $4$-dimensional Euclidean space are recovered. The consistent choice of algebra relations is given by
\begin{gather*}
 [N^l,N^k]   = M^{lk} ,\qquad
 [M^{rs},N^l]   = \delta^{rl}N^s - \delta^{sl}N^l , \\
 [M^{rs},M^{kl}]   = \delta^{rl}M^{ks} - \delta^{sl} M^{kr} - \delta^{rk}
	M^{ls} + \delta^{sk} M^{lr} .
\end{gather*}
These are just the undeformed algebra relations. The dif\/ference arises in the co-algebra structure. The commutation relations \eqref{eq:noncanon:kappa:wirkung} can be generalized to {\nc} functions:
\begin{gather*}
N^l\hat f(\hat x)   = \big(N^l \hat f(\hat x)\big)
	+ \hat f(\hat x^i, \hat x^n + \ri a)N^l - \ri a \big( \hat \partial_b \hat f(\hat x)\big)  M^{lb},\\
M^{rs} \hat f(\hat x)  = \big( M^{rs} \hat f(\hat x)\big) + \hat f(\hat x)  M^{rs}.
\end{gather*}
We can read of\/f the co-product structure of the rotation generators from the above formulae, using the crossed product
\begin{gather*}
T \hat x^\nu = ( T_{(1)}  \hat x^\nu )   T_{(2)} ,
\end{gather*}
and obtain
\begin{gather*}
\Delta N^l   = N^l\otimes\id + e^{ \ri a\hat \partial_n } \otimes N^l - \ri a\hat \partial_b\otimes M^{lb},\qquad
\Delta M^{rs}   = M^{rs}\otimes\id +  \id\otimes M^{rs} .
\end{gather*}
Now, let us def\/ine derivatives on this $\kappa$-Euclidean space. We introduce them by f\/inding a~deformed Leibniz rule compatible with the algebra relations~\eqref{eq:noncanon:kappa:commutator}. Since the coordinate algebra is the freely generated algebra divided by the ideal generated by relations~\eqref{eq:noncanon:kappa:commutator}, the derivatives have to map co-sets onto co-sets. Consistent Leibniz rules are given by
\begin{alignat*}{3}
& \hat \partial_n \hat x^i   = \hat x^i \hat \partial_n ,\qquad && \hat \partial_n \hat x^n  = 1 + \hat x^n \hat \partial_n ,&
\nonumber\\
& \hat \partial_i \hat x^j   = \delta^j_i + \hat x^j \hat \partial_i, \qquad && \hat \partial_i \hat x^n = (\hat x^n + \ri a) \hat \partial_i. &
\end{alignat*}
However, these relations are not unique, cf.~\cite{Dimitrijevic:2004}. Additionally, the derivatives have to form a~module algebra of the deformed rotation algebra, i.e.\ they have to transform like a vector. For the action of the symmetry generator on the derivatives one obtains
\begin{gather*}
 [M^{rs}, \hat \partial_i]   = \delta^{ri} \hat \partial_s -  \delta^{si} \hat \partial_r
 ,\qquad [M^{rs}, \hat \partial_n]   = 0,\nonumber\\
 [N^l, \hat \partial_i]   = \delta^{li}\inv{2\ri a} \big( 1-e^{2ia\hat \partial_n} \big) -
	\frac{\ri a}{2}\delta^{li} \widehat {\Delta}_\kappa + \ri a  \hat \partial_l\hat \partial_i
 ,\qquad
 [N^l, \hat \partial_n]   = \hat \partial_l,
\end{gather*}
where we have def\/ined the $\kappa$-deformed Laplacian $\widehat{\Delta}_\kappa =\sum_i \hat \partial_i\hat \partial_i$. The commutator of derivatives compatible with~(\ref{eq:noncanon:kappa:commutator}) is given by
\begin{gather}
\label{eq:noncanon:kappa:1.2}
 [\hat \partial_\mu,\hat \partial_\nu]=0.
\end{gather}
The Leibniz rule for {\nc} functions reads
\[
\hat \partial_i\hat f(\hat x) =  (\hat \partial_i \hat f (\hat x)) + \hat f (\hat x^i, \hat x^n + \ri a)  \hat \partial_i.
\]
The derivatives $\hat\partial_n$ satisf\/ies the ordinary Leibniz rule. The $\kappa$-deformed Poincar\'{e} algebra $\mathcal P_\kappa$ is generated by rotations $M^{rs}$, boosts $N^l$ and translations~$\hat \partial_\mu$. The co-product of the translation generators reads
\begin{gather*}
\Delta\hat\partial_n   = \hat \partial_n \otimes {\id} + {\id}\otimes\hat
	\partial_n,\qquad
\Delta\hat\partial_i   = \hat \partial_i \otimes {\id} + e^{\ri a \hat \partial_n}
	\otimes \hat\partial_i.
\end{gather*}
The Dirac operator $\hat{D}$ is given by
\begin{gather*}
\hat{D}_n = \left(\frac{1}{a}\sin(a\hat{\partial}_n)+\frac{\ri a}{2}\hat{\partial}_k\hat{\partial}_k
e^{-\ri a\hat{\partial}_n}\right),\qquad
\hat{D}_j =\hat{\partial}_j e^{-\ri a\hat{\partial}_n}.
\end{gather*}
It can be viewed as a derivative as well satisfying the following Leibniz rule:
\begin{gather}
\hat D_n (\hat f \cdot \hat g )   = (\hat D_n \hat f) \cdot
	\big(e^{-\ri a\hat \partial_n} \hat g\big) + \big(e^{\ri a\hat \partial_n} \hat f\big) \cdot (\hat D_n \hat g)
 + \ri a \sum_{i=1}^{n-1}
	\big( \hat D_i e^{\ri a\hat \partial_n} \hat f\big) (\hat D_i \hat g) ,\nonumber\\
\hat D_i (\hat f \cdot \hat g)   = (\hat D_i \hat f) \cdot \big(e^{-\ri a\hat\partial_n} \hat g\big) + \hat f \cdot (\hat D_i \hat g) .
\label{eq:noncanon:kappa:dirac-leibniz}
\end{gather}
Acting on the coordinates, it yields
\begin{gather*}
[\hat{D}_n, \hat{x}^i ] = \ri a\hat{D}_i , \nonumber \\
 [\hat{D}_n, \hat{x}^n ]   =
\sqrt{1-a^2\hat{D}_\mu\hat{D}_\mu}= 1-\frac{a^2}{2}\hat{\square} , \nonumber\\
 [\hat{D}_j, \hat{x}^i ] = \delta^i_j\left(-\ri a\hat{D}_n+
\sqrt{1-a^2\hat{D}_\mu\hat{D}_\mu}\right)=\delta^i_j\left(1-\ri a\hat{D}_n-\frac{a^2}{2}\hat{\square}
\right),\nonumber \\
 [\hat{D}_j, \hat{x}^n ] = 0 .
\end{gather*}
This completes the algebraic setting of $\kappa$-deformed spaces. Let us now introduce the star product using a symmetrical ordering. It is given by \cite{Dimitrijevic:2004}
\begin{gather*}
(f\star g) (x) = \int d^4k \, d^4 p \, \tilde f(k) \tilde g(p)
   e^{\ri(\omega_k + \omega_p)x^1} e^{\ri\vec x( \vec k e^{a\omega_p} A(\omega_k,\omega_p)+ \vec p A(\omega_p,\omega_k))} ,
\end{gather*}
where $k=(\omega_k,\vec k)$, $\vec x=(x^2,x^3,x^4)$, and
\begin{gather*}
A(\omega_k,\omega_p) \equiv  \frac{a(\omega_k+\omega_p)}{e^{a(\omega_k+\omega_p)}-1}
	\frac{e^{a\omega_k}-1}{a\omega_k} .
\end{gather*}
For the star product in arbitrary ordering see \cite{Meljanac:2007xb}. In symmetrical ordering, the action of the deformed derivatives on commutative functions (denoted by $\partial^\star$) can be expressed in terms of the usual derivatives
\begin{gather*}
\partial^\star_i f(x)   = \partial_i   e^{\ri a\partial_n}   f(x)
 ,\qquad
\partial^\star_n f(x)   = \partial_n   f(x) .
\end{gather*}
In the same way, we obtain for the Dirac operator
\begin{gather*}
D^\star_n   = \frac 1a \sin (a\partial_n) + \Delta \frac{\cos (a\partial_n) - 1}{\ri a \partial_n^2} , \qquad
D^\star_i   = \frac{e^{-\ri a\partial_n}-1}{-\ri a\partial_n} \partial_i  ,
\end{gather*}
where $\Delta$ denotes the undeformed Laplacian.

\subsubsection{Deformed Maxwell equations}
\label{sec:noncanon:kappa:Maxwell}

The modif\/ications of the classical Maxwell equations under $\kappa$-deformation are important in order to obtain the correct dispersion relations. Starting from the def\/inition of the deformed $U(1)$ f\/ield strength
\[
[ \dot{\hat x}^\mu, \dot{\hat x}^\nu ] = - [\hat x^\mu, \ddot{\hat x}^\nu] = \frac{\ri q\hbar}{m^2} F^{\mu\nu} ,
\]
where $q$ denotes the charge, $m$ the mass of the charged particle, and a time derivative is denoted by a ``dot'', the deformed Maxwell equations\footnote{We assume natural units with $\hbar=1$ in this review, and hence (in contrast to~\cite{Harikumar:2010}) have omitted $\hbar$ in these expressions.} take the form~\cite{Harikumar:2010}
\begin{gather*}
 \vec{\nabla}\vec{B} + {ma} \vec{v} \partial_0 \vec{B} = 0,\\
 \partial_0 \vec{B} + \vec{\nabla} \times \vec{E} + {ma} \big(
v^i \partial_i \vec{B} + \vec{v}\times \partial_0 \vec{E} \big) = 0
,\\
 \vec{\nabla}\vec{E} + {ma} \vec{v} \partial_0 \vec{E} = \rho_e,\\
 \partial_0 \vec{E} - \vec{\nabla}\times \vec{B} + {ma} \big(
a v^i \partial_i \vec{E} - a \vec{v}\times \partial_0 \vec{B} \big) = -\vec{j}_e.
\end{gather*}
Remarkably, the modif\/ication and therefore the coupling of the particle to the electro-magnetic f\/ield depends on the mass of the particle.

\subsubsection{Seiberg--Witten map}
\label{sec:noncanon:kappa:SW}

In \cite{Dimitrijevic:2003,Dimitrijevic:2005c}, the Seiberg--Witten maps have been calculated in the case of $\kappa$-deformed Minkowski space for arbitrary compact gauge group, up to f\/irst order in $a$ (cf.\ the discussion in \secref{sec:canon:theta-exp:SW}). For convenience, we choose to gauge the Dirac operators $\hat D_\alpha$, since they also form the basis of the derivatives and have transformation properties,
\[
[M^{\mu\nu}, D^\star_\rho] = \delta^\nu_\rho D^{\star \mu} - \delta^\mu_\rho D^{\star \nu}.
\]
The most remarkable new feature of the SW map is that the gauge f\/ield attains a derivative valued contribution. This is due to the modif\/ied Leibniz rule for the covariant derivatives \eqref{eq:noncanon:kappa:dirac-leibniz}. In the quantum group case~\cite{Schraml:2002fi}, the co-product of derivatives $ \hat \partial_\mu$ reads
\[
\Delta \hat \partial_\mu = \hat \partial_\mu \otimes \id + {L_\mu}^\nu \otimes \hat \partial_\nu,
\]
where ${L_\mu}^\nu$ is the so-called L-matrix, which is a linear transformation. In this case covariant derivatives are def\/ined by introducing a vielbein ${E_\mu}^\nu$ with non-trivial transformation properties,
\[
\mathcal D_\mu \Psi = {E_\mu}^\nu \big( \hat \partial_\nu - \ri A_\nu \big) \Psi.
\]
Additionally, the gauge potential attains a derivative valued part. In the present case, this factor is non-linear in the derivatives and cannot be compensated by a vielbein.

We def\/ine the covariant Dirac operator by
\begin{gather*}
\mathcal D^\star_\alpha   = {E_\alpha}^\mu \partial^\star_\mu - \ri \widehat V_\alpha =  D^\star_\alpha - \ri \widehat V_\alpha,
\end{gather*}
where $V_\alpha$ is the enveloping algebra valued gauge potential. Starting from the {\nc} gauge transformation
\begin{gather}
\label{eq:noncanon:kappa:nc-transf}
\widehat \delta_{\alpha} \widehat \psi = \ri \widehat \Lambda_\alpha \star \widehat \psi,\qquad \textrm{with}\quad
(\widehat \delta_\alpha \widehat \delta_\beta - \widehat \delta_\beta\widehat \delta_\alpha)\widehat \psi = \widehat \delta_{-\ri[\alpha,\beta]} \widehat \psi
,
\end{gather}
let us f\/irst consider the Seiberg--Witten map of the gauge parameter $\widehat \Lambda$ to f\/irst order in $a$. The gauge equivalence relation \eqref{eq:canon:theta-exp:SW-5} reads
\begin{gather}
\label{eq:noncanon:kappa:SW-parameter}
 \ri\delta_\alpha \Lambda_\beta^1 - \ri\delta_\beta \Lambda_\alpha^1 +
[\alpha, \, \Lambda_\beta^1] + [\Lambda_\alpha^1,\, \beta]
-\Lambda_{[\alpha,\beta]}^1
  = - \frac{\ri a}{2} \left( x^\mu \{\partial_n \alpha,\, \partial_\mu \beta\} - x^\mu
\{\partial_n\beta,\, \partial_\mu \alpha\} \right),\!\!
\end{gather}
where $\widehat \Lambda_\alpha[A] = \alpha + \Lambda_\alpha^1[A] + \mathcal O(a^2)$. The right hand side of \eqref{eq:noncanon:kappa:SW-parameter} can be written more concisely as
\[
- \frac{\ri a}{2} \left( x^\mu \{\partial_n \alpha,  \partial_\mu \beta\} - x^\mu
\{\partial_n\beta,  \partial_\mu \alpha\} \right) = -\frac{\ri}{2}x^\lambda
C^{\mu,\nu}_\lambda \{\partial_\mu\alpha,  \partial_\nu \beta\} ,
\]
where $C^{\mu\nu}_\lambda$ are the structure constants of the space-time algebra, with $C^{\mu\nu}_\lambda = a  (\delta^\mu_n \delta^\nu_\lambda -\delta^\mu_\lambda \delta^\nu_n)$. The solution of \eqref{eq:noncanon:kappa:SW-parameter} is given by
\begin{gather*}
\Lambda^1_\alpha = - \inv{4}x^\lambda C_\lambda^{\mu\nu} \{ V_\mu,\partial_\nu \alpha \} .
\end{gather*}
Also in higher orders in the expansion, there will occur terms that look similar to those in the canonical case, replacing $\theta^{\mu\nu}$ by $x^\lambda C^{\mu\nu}_\lambda$~\cite{Dimitrijevic:2003}. Both expressions
constitute the respective Poisson structure. Expanding the f\/ield $\widehat \psi$ in terms of $a$,
\[
\widehat \psi = \psi + \psi^1 + \mathcal O\big(a^2\big),
\]
equation~\eqref{eq:noncanon:kappa:nc-transf} becomes
\begin{gather*}
\delta_\alpha \psi^1 = \ri\Lambda_\alpha^1   \psi +\ri  \alpha  \psi^1 - \inv{2} x^\lambda C^{\mu\nu}_\lambda \partial_\mu \alpha   \partial_\nu \psi.
\end{gather*}
A solution is given by
\begin{gather*}
\psi^1[V] = -\inv{2} x^\lambda C^{\mu\nu}_\lambda V_\mu   \partial_\nu \psi + \inv{\ri}{4} x^\lambda C_\lambda^{\mu\nu} V_\mu V_\nu \psi .
\end{gather*}
For the gauge f\/ield $\widehat V_\alpha[V]$, the SW map is much more involved, because of the complicated co-product structure of the derivatives $D^\star_\mu$. Starting from
\[
\widehat \delta_\alpha (D^\star_\mu \widehat \psi ) = \ri\widehat \Lambda \star \mathcal D^\star_\mu \widehat \psi\,,
\]
one obtains
\begin{gather*}
\widehat \delta_\alpha \widehat V_\gamma \star \widehat \psi   = \widehat D_\gamma (\widehat \Lambda_\alpha \star \widehat \psi ) - \widehat \Lambda_\alpha \star \widehat D_\gamma \widehat \psi  - \ri \widehat V_\gamma \star \widehat \Lambda_\alpha \star \widehat \psi + \ri \widehat \Lambda_\alpha \star \widehat V_\gamma \star \widehat \psi.
\end{gather*}
Using the co-product of the derivatives $\widehat D_\mu$, we can eliminate the f\/ield $\widehat \psi$,
\begin{gather*}
\widehat \delta_\alpha \widehat V_c   = ( D^\star_c \hat \Lambda_\alpha ) \star e^{-\ri a\partial^\star_n} - \ri \widehat V_c \star \widehat \Lambda_\alpha + \ri \widehat \Lambda_\alpha \star \widehat V_c
 ,\nonumber\\
\widehat \delta_\alpha \widehat V_n   = (D^\star_n \widehat \Lambda_\alpha) \star e^{-\ri a \partial^\star_n} + ia(D\star_i e^{\ri a \partial^\star_n} \widehat \Lambda_\alpha)\star D^\star_i
  \nonumber\\
  \phantom{\widehat \delta_\alpha \widehat V_n   =}{}
  + \big( (e^{\ri a\partial^\star_n}-1) \widehat \Lambda_\alpha \big) \star D^\star_n - \ri \widehat V_n \star \widehat \Lambda_\alpha + \ri \widehat \Lambda_\alpha \star \widehat V_n  .
\end{gather*}
This leads to derivative valued gauge f\/ields, and a solution is given by
\begin{gather*}
\widehat V_i   = V_i - \ri a V_i \partial^\star_n - \frac {\ri a}2 \partial_n V_i - \frac a4 \{V_n, V_i\} + \frac 14 C^{\rho\sigma}_\lambda x^\lambda ( \{ F_{\rho i}, V_\sigma \} - \{ V_\rho, \partial_\sigma V_i \} ),\nonumber\\
\widehat V_n   = V_n - \ri a V^j \partial^\star_j - \frac {\ri a}2 \partial_j V^j - \frac a2 V_j V^j + \frac 14 C^{\rho\sigma}_\lambda x^\lambda ( \{ F_{\rho n}, V_\sigma \} - \{ V_\rho, \partial_\sigma V_n \} )  .
\end{gather*}
The action of matter coupled to the gauge f\/ield hence receives corrections \cite{Dimitrijevic:2003,Dimitrijevic:2005c}. The gauge action up to f\/irst order in $a$ is given by
\begin{gather*}
S_g = -\frac 14 \int d^{n+1} x  \left(
F^{\mu\nu} F_{\mu\nu} - \frac 12 C^{\rho\sigma}_\lambda x^\lambda F_{\rho\sigma} F^{\mu\nu} F_{\mu\nu} + 2 C^{\rho\sigma}_\lambda x^\lambda F^{\mu\nu} F_{\mu\rho} F_{\nu\sigma} \right) ,
\end{gather*}
and for matter f\/ields we have
\begin{gather}
S_m = \int d^{n+1} x \bigg(
\bar \psi (\ri \gamma^\mu \mathcal D_\mu - m) \psi - \frac 14 C^{\rho\sigma}_\lambda x^\lambda \bar\psi F_{\rho\sigma} (\ri \gamma^\mu \mathcal D_\mu - m) \psi
\nonumber \\
\phantom{S_m =}{}
 - \frac 12 C^{\rho\sigma}_\lambda \bar \psi\gamma_\rho \mathcal D_\sigma \mathcal D^\lambda \psi - \frac{\ri}{2} C^{\rho\sigma}_\lambda x^\lambda \bar\psi \gamma^\mu F_{\mu\rho} \mathcal D_\sigma \psi - \frac{\ri}{4} C^{\rho\sigma}_\lambda \bar \psi \gamma^\mu F_{\mu\rho} \psi \bigg).
\label{eq:noncanon:kappa:matter-action}
\end{gather}
This action was used for phenomenological considerations in~\cite{Bolokhov:2008}. P.A.~Bolokhov and M.~Pos\-pe\-lov generalized the action \eqref{eq:noncanon:kappa:matter-action} to the case of the Standard Model gauge group $SU(3)\times SU(2) \times U(1)$. Considering nucleon electromagnetic interactions, they could obtain a {\naiv} bound for the non-commutativity scale:
\[
\kappa \sim 1/a > 10^{23}~{\rm GeV}.
\]
The reliability of this bound seems questionable, though, since the calculation relies on some simplifying assumptions.

\subsection{Gauge theory on the fuzzy sphere}
\label{sec:noncanon:fuzzy}

The fuzzy sphere has f\/irst been discussed in \cite{Madore:1990,Madore:1991}~-- for a nice review, see also~\cite{Karabali:2001}. Its generators satisfy linear commutation relations
\begin{gather}
\label{eq:noncanon:fuzzy:algebra}
\co{\hat x_i}{\hat x_j} = \ri \frac{\cth}{r} \sth_{ijk} \hat x_k , \qquad i,j,k\in\{1,2,3\} ,
\end{gather}
where $r^2=\hat x_1^2 + \hat x_2^2 + \hat x_3^2 \in \R$ is the radius of the sphere. The objects
\begin{gather}
\label{eq:noncanon:fuzzy:gen}
\hat R_i = \frac r\cth \hat x_i
\end{gather}
satisfy the $SU(2)$ algebra relations. $\hat R_i$ are chosen to be in an irreducible representation with spin $j$. Therefore, the generators $\hat R_i$ and also $\hat x_i$ are $N \times N$ matrices with $N=2j+1$. The space algebra \eqref{eq:noncanon:fuzzy:algebra} is equipped with a dif\/ferential calculus. Since we are dealing with matrix algebras, all derivations are inner. The dif\/ferentials $\hat \partial_i$ satisfy the same algebra as the coordinates:
\begin{gather*}
\co{\hat \partial_i}{\hat \partial_j} = \frac \ri r \sth_{ijk} \hat \partial_k ,
\end{gather*}
and therefore they can be represented as
\[
\hat \partial_i = - \frac \ri \cth  \hat x_i .
\]
The adjoint action of $\hat R_i$ on a function $\hat f$ generates rotations of $\hat x_i$, hence
\begin{gather*}
\hat L_i \hat f = \co{\hat R_i}{\hat f} ,
\end{gather*}
where $\hat L_i$ denote the generators of angular momentum. The integral over the fuzzy sphere is given by the trace with respect to the matrix space,
\begin{gather*}
\int \hat f = \frac{4\pi r^2}N \text{Tr}\, \hat f .
\end{gather*}
The constant prefactor ensures the correct commutative limit, which is accomplished by kee\-ping~$r$ f\/ixed and taking $\cth\to 0$ (corresponding to $j\to \infty$). The {\nc} Moyal plane is recovered in the limit $r\to \infty$ and keeping $\cth$ f\/ixed (corresponding to $j\to \frac{r^2}\cth$). The {\nc} parameter $\cth$ is f\/ixed by the radius relation:
\[
\cth = \frac{r^2}{\sqrt{j(j+1)}} .
\]
It can be regarded as the elementary area on the sphere, which becomes obvious after a rescaling
\[
\cth' = \frac{r^2}{j(j+1)} = \frac{4\pi r^2}{2\pi N} .
\]

Gauge f\/ields are introduced via the covariant derivatives
\begin{gather*}
\hat D_i = \hat \partial_i -\ri \hat A_i ,
\end{gather*}
where $\hat A_\alpha$ are Hermitian $N\times N$ matrices. The f\/ield strength is given by
\begin{gather*}
\ri \hat F_{ij} = \co{\hat D_i}{\hat D_j} - \frac {\sth_{ijk}}r \hat D_k .
\end{gather*}
Gauge transformations read
\[
\hat D_i' = g \hat D_i g^{-1} ,
\qquad
\hat F_{ij}' = g \hat F_{ij} g^{-1} ,
\]
where $g$ is a $U(N)$ matrix. The restriction of the gauge f\/ield to the sphere is expressed as $\sum_i X_i^2 = r^2$ leading to
\begin{gather}
\label{eq:noncanon:fuzzy:constraint}
\hat x_i \hat A_i + \hat A_i \hat x_i - \cth \hat A_i^2 = 0 .
\end{gather}
Hence, the action for the gauge f\/ield is given by
\begin{gather*}
S_g = \frac{4\pi r^2}N \text{Tr} \, \hat F_{ij} \hat F_{ij}.
\end{gather*}
A complex scalar f\/ield $\hat \Phi$ is coupled to a gauge theory using the minimal coupling:
\begin{gather}
\label{eq:noncanon:fuzzy:min-coupling}
S[\hat \Phi, \hat A] =    \frac{4\pi r^2}{\cth^2 N} \text{Tr} \left( \co{\hat X_i}{\hat\Phi} \co{\hat \Phi}{\hat X_i} + \cth^2 V(\hat \Phi) \right) ,
\end{gather}
where covariant coordinates $\hat X_i = \hat x_i + \cth \hat A_i$ are used. For an earlier reference, see e.g.~\cite{Klimcik:1997}. In the following we will discuss some approaches to gauge theory and their results.

Some topological aspects, such as instantons, monopoles and the axial anomaly have been studied in \cite{Balachandran:1999, Karabali:2001,Ydri:2002nt}. Although conventional lattice regularizations have problems dealing with those aspects, they can be treated on the fuzzy sphere in a natural way.

The {\uim} for $U(1)$ gauge theory on the fuzzy sphere was studied in~\cite{CastroVillarreal:2004}. The quadratic ef\/fective one-loop action was explicitly calculated and a gauge invariant {\uim} was obtained to persist in the limit $N\to \infty$. The authors also predict a f\/irst order phase transition from the one-loop results which has been observed in lattice calculations, see e.g.~\cite{Azuma:2004zq,O'Connor:2006}. The constraint~\eqref{eq:noncanon:fuzzy:constraint} can be interpreted as a scalar excitation tangential to the sphere. Adding a large mass to this scalar mode the {\uim} completely decouples from the gauge sector in the large~$N$ limit.

H.~Steinacker used random matrix techniques to evaluate the path integral for $U(N)$ gauge theory by integrating over eigenvalues \cite{Steinacker:2003}. This allows to compute the path integral explicitly. The starting action is given by
\[
S= \frac 2{g^2N} \text{Tr}   \left( \left(\hat B_i \hat B^i - \frac{N^2-1}4 \right)^2 + (\hat B_i + \ri \sth_{ijk} \hat B^j \hat B^k)(\hat B^i + \ri \sth^{irs} \hat B_r \hat B_s) \right) ,
\]
where the $\hat B$s are covariant coordinates,
\[
\hat B_i = \hat B_{ia} t^a = \hat R_i^{(N)} t^0 + \hat A_{i0} t^0 + \hat A_{ia} t^a ,
\]
where $t^0$ is the identity matrix, $t^a$ denote the Gell-Mann matrices for $SU(N)$, and $\lambda^{(N)}_i \equiv R_i$ has been def\/ined in equation~\eqref{eq:noncanon:fuzzy:gen}. The partition function of the undeformed $U(N)$ Yang--Mills theory on the classical sphere is recovered in the large $N$ limit, as a sum over instanton contributions. The monopole solution could be calculated, but for obtaining $1/N$ corrections the calculations were too involved. The earlier work \cite{Iso:2001} is in the same spirit, where the authors also expand around the classical solution of the fuzzy sphere. They formulate $U(1)$ and $U(N)$ gauge theory and additionally add a Chern--Simons term.

We have seen in \secref{sec:noncanon:kappa:SW} that a Seiberg--Witten map has been calculated for a non-canonical deformation, the $\kappa$-deformation. This has also been done for the case of the fuzzy sphere \cite{Grimstrup:2003dz}. In the limit $r\to \infty$, the canonical expressions are recovered.

The phase structure of the {\nc} $U(1)$ gauge theory has been obtained in~\cite{O'Connor:2006}, using a Monte Carlo simulation. It shows three dif\/ferent phases: A \emph{matrix phase}, which is essentially $SU(N)$ Yang--Mills reduced to a point; a \emph{weak coupling phase} with a constant specif\/ic heat; and a \emph{strong coupling phase} with a non-constant specif\/ic heat. The order parameter is given by the radius of the fuzzy sphere. The dif\/ferent phases meet at a triple point.
Similar non-perturbative structures are obtained on canonically deformed spaces mentioned in the introduction to~\secref{sec:covar}.

Fuzzy spaces have also been discussed in connection with particle phenomenology. In~\cite{Aschieri:2007fb} (see also references therein), gauge theories in higher dimensions are discussed, where the extra dimensions form a fuzzy space. The additional degrees of freedom are interpreted as Kaluza--Klein modes. After dimensional reduction some remarkable features are obtained. The gauge group is broken dynamically, and depending on the parameters of the model, the Standard Model group can be obtained at low energies.

\subsection{Yang--Mills matrix models}
\label{sec:matrixmodels}

In a series of papers~\cite{Yang:2006, Steinacker:2007, Steinacker:2008, Steinacker:2008a,Steinacker:2008ri,Steinacker:2008ya,Klammer:2009dj}, a dif\/ferent interpretation of the origin of the {\uim} in {\nc} gauge models was given by considering matrix models of Yang--Mills type:
\begin{gather}
S_{\rm YM}=-\text{Tr}\co{X^a}{X^b}\co{X^c}{X^d}\eta_{ac}\eta_{bd} ,
\label{eq:matrix_YM-action}
\end{gather}
where $\eta_{ab}$ denotes some $D$ dimensional embedding space. The $X^a$ are Hermitian matrices acting on a Hilbert space $\mathcal{H}$. In the simplest case, these matrices represent generalized ``coordinates'', and if some of them are functions of the others, in the semi-classical limit $X\sim x$ one can interpret these as def\/ining the embedding of a~$2n$-dimensional submanifold $\mathcal{M}^{2n}\in\R^D$ equipped with a~non-trivial induced metric
\[
g_{\m\n}(x)=\pa_\m x^a \pa_\n x^b\eta_{ab} ,
\]
via pull-back of $\eta_{ab}$. This submanifold could then e.g.\ be our ({\nc}) 4-dimensional space-time $\mathcal{M}^4$ endowed with a Poisson structure $\cth^{\m\n}\sim-\ri\co{X^\m}{X^\n}$. In fact, the Poisson structure $\cth^{\m\n}$ (assuming it is non-degenerate) and the induced metric $g_{\m\n}$ combine to an ``ef\/fective'' metric
\begin{gather}
G^{\m\n} =\re^{-\s}\cth^{\m\r}\cth^{\n\s}g_{\r\s} ,\qquad
\re^{-\s} \equiv \frac{\sqrt{\det\cth^{-1}_{\m\n}}}{\sqrt{\det G_{\r\s}}} ,
\label{eq:matrix_effective-metric}
\end{gather}
which is the one that is actually ``felt'' by matter f\/ields. Furthermore, the matrix model action~\eqref{eq:matrix_YM-action} is invariant under the gauge symmetry $X^\m\to gX^\m g^{-1}$, where $g\in U(\infty)$, as well as under global rotation and translation symmetries.

It is remarkable that within the matrix model framework four space-time dimensions, i.e.\ $\m,\n\in\{0,1,2,3\}$,  play a very special role: From the def\/inition of the ef\/fective metric~\eqref{eq:matrix_effective-metric} follows, that if $2n=4$, one has $\det G_{\m\n}=\det g_{\m\n}$. This means that the special class of geometries where $G_{\m\n}=g_{\m\n}$ (which incidentally corresponds to a self-dual symplectic form $\cth^{-1}_{\m\n}$) is a solution of the model. Furthermore, in the 4-dimensional case the Poisson tensor $\cth^{\mu\nu}$ does not enter the Riemannian volume element, which turns out to stabilize f\/lat space.

In order to make things clearer, consider a scalar f\/ield $\ph$ on $\mathcal{M}^4$ in the semi-classical limit where $X^a\sim x^a$ are mere coordinates: In order to preserve gauge invariance, the kinetic term must have the form
\begin{gather*}
\Act[\ph] =-\text{Tr}\co{X^a}{\ph}\co{X^c}{\ph}\eta_{ac}
 \sim \intx \sqrt{\det\cth^{-1}_{\m\n}}\, \cth^{\m\n}\pa_\m x^a\pa_\n\ph\, \cth^{\r\s}\pa_\r x^c\pa_\s\ph  \eta_{ac} \\
\phantom{\Act[\ph]}{}
\sim \intx \sqrt{\det G_{\m\n}} G^{\n\s}\pa_\n\ph\pa_\s\ph ,
\end{gather*}
cf.\ equation~\eqref{eq:noncanon:fuzzy:min-coupling}. This semi-classical ef\/fective action obviously describes a scalar f\/ield on a~4-dimensional space-time with metric $G_{\m\n}$, and if $G_{\m\n}=g_{\m\n}$ it becomes independent of the Poisson tensor $\cth^{\m\n}$ (in this approximation), as claimed above.

In a further step, it is also possible to add $U(N)$ gauge f\/ields $A$ to the matrix model. To show this, we start with the equations of motion of the matrix model action~\eqref{eq:matrix_YM-action}:
\[
\co{X^a}{\co{X^b}{X^c}}\eta_{ab}=0 .
\]
For every solution $X^c$ of this equation, $X^c\otimes\id_N$ is a solution\footnote{One can interpret such a solution as~$N$ coinciding branes.} as well. The f\/luctuations $A_\m$ in the submanifold $\mathcal{M}^4$ around such a background can be parametrized by
\begin{gather*}
Y^a \sim \left(1+\mathcal{A}^\m\pa_\m\right)X^a ,\qquad
\mathcal{A}^\m =-\cth^{\m\n}A_\n(X) ,
\end{gather*}
where the $A_\m$ are some $U(N)$ valued f\/ields\footnote{Notice also the similarity to the covariant coordinates we introduced in \secref{sec:canon:induced_induced}. This is no coincidence: In fact, the ``induced gauge theory'' action~\eqref{eq:induced_induced1loop} we discussed in that section \emph{is} a matrix model one.}. The ef\/fective matrix model action then describes gauge f\/ields in a gravitational background. However, though inseparable, the $U(1)$ and the $SU(N)$ subsectors play very dif\/ferent roles: In fact, the $U(1)$ f\/ields contribute \emph{only} to the gravitational sector, i.e.\ they represent geometrical degrees of freedom. This means, that within the matrix model framework, {\nc} $U(N)$ gauge f\/ield theory describes $SU(N)$ f\/ields coupled to gravity.

Furthermore, there has been a recent proposal, how these $SU(N)$ groups may then be broken down to smaller ones like e.g.\ $SU(3)_c \times SU(2)_L \times U(1)_Q$ (which are required to retrieve the standard model within this framework) by inducing spontaneous symmetry breaking using extra dimensions and fuzzy spheres~\cite{Grosse:2010zq}.

Of course, much more can be said about matrix models. However, for further details we would like to refer to the recent review article in~\cite{Steinacker:2010rh}.

\subsection{Other approaches}
\label{sec:noncanon:other}

We would like to mention two other approaches on non-canonical space-time structures. First, we will discuss the $q$-deformed case and then turn to the recently developed approach on spaces with covariant star products. The former case is related to quantum groups, which have been developed from the study of integrable systems in the framework of quantum inverse scattering. In Sections~\ref{sec:noncanon:twisted} and \ref{sec:noncanon:kappa} we have already encountered Hopf algebras as generalized space-time symmetries. Quantum groups also fall in this category, as they are Hopf algebras with an additional ingredient: the so-called {$\hat R$-matrix}. This matrix is a solution of the Yang--Baxter equation and bridges the gap to statistical physics. The structures are rather involved and therefore not too much is known about quantum f\/ield theory or gauge theory on $q$-deformed spaces.

The latter approach, covariant star products, is especially suited for the discussion of gravitational ef\/fects, but it has also been applied to gauge theory.

\subsubsection[$q$-deformation]{$\boldsymbol{q}$-deformation}
\label{sec:noncanon:other:q}

In this section, we want to discuss the construction of gauge theory on $q$-deformed spaces. These spaces are representations of quantum groups, Hopf algebras which in addition possess a~so-called $\hat R$-matrix. Although we have already introduced some of the notation and def\/initions of Hopf algebras in Sections~\ref{sec:noncanon:twisted} and~\ref{sec:noncanon:kappa}, let us be a bit more careful here, see e.g.~\cite{Majid:1999td}.

A Hopf algebra $\cA$, denoted by $(\cA,m,\eta,\Delta,\epsilon,S)$, consists of an associative algebra $(\cA,m,\eta)$ with a compatible co-algebra structure, given by the structure maps $\Delta$, $\epsilon$ and $S$. In detail, $m:\cA\otimes \cA \to \cA$ denotes the multiplication and $\eta$ the unit map:
\[
\eta: \ \C  \to \cA, \qquad c \mapsto c \id_\cA,
\]
where $\id_\cA\in \cA$ is the unit element. The multiplication is associative. The structure maps of the co-algebra are by def\/inition dual to~$m$ and $\eta$:
\[
\Delta: \ \cA   \longrightarrow \cA\otimes \cA , \qquad
\eta: \ \cA   \longrightarrow \C .
\]
The co-product $\Delta$ satisf\/ies the co-associativity rule
\[
\Delta\circ (\id \otimes \Delta ) = \Delta \circ (\Delta \otimes \id) ,
\]
and for the co-unit $\epsilon$ we have a similar def\/ining relation
\[
(\epsilon \otimes \id)\circ \Delta = (\id \otimes \epsilon)\circ \Delta .
\]
The antipode (``inverse'') $S$ is def\/ined via the relation
\[
m\circ (S\otimes \id) \circ \Delta = \eta \circ \epsilon = m\circ (\id \otimes S) \circ \Delta.
\]
Compatibility between algebra and co-algebra structures means that the co-product $\Delta$ and the co-unit $\epsilon$ are algebra homomorphisms, i.e.
\[
\Delta(ab) = \Delta(a)\Delta(b),\qquad \epsilon(ab) = \epsilon(a)\epsilon(b) ,
\]
with $a,b\in \cA$.
Quantum groups have one additional structure, the $\hat R$-matrix. Let $\hat u^k_m$ be the generators of the Hopf algebra. Then the $\hat R$-matrix deforms the multiplication in the algebra:
\[
\hat R^{ij}_{kl} \hat u^k_m \hat u^l_n = \hat u^i_k \hat u^j_l \hat R^{kl}_{mn} ,
\]
where $\hat R$ itself is a solution of the Yang--Baxter equation:
\begin{gather}
\label{eq:noncanon:others:QYB}
\hat R_{12} \hat R_{23} \hat R_{12} = \hat R_{23} \hat R_{12} \hat R_{23}
 ,
\end{gather}
with $\hat R_{12}{}^{ijk}_{lmn} = \hat R^{ij}_{lm}   \delta^k_n$ and $\hat R_{23}{}^{ijk}_{lmn} = \hat R^{jk}_{mn}   \delta^i_l$.

Quantum spaces with generators $\hat x^i$ are representations of the respective quantum group. The algebra relations of the generators are consistently given by
\[
P_-{}^{ij}_{kl} \hat x^k \hat x^l = 0 ,
\]
where $P_-$ is the $q$-deformed antisymmetric projector, generalizing the commutator, from the projector decomposition of the $\hat R$-matrix of the respective quantum group. Considering the quantum groups $GL_q(N)$ or $SL_q(N)$, we have the following decomposition
\[
\hat R = q P_+ - q^{-1} P_- ,
\]
and in case of $SO_q(N)$,
\[
\hat R = q P_+ - q^{-1}P_- + q^{1-N} P_0 ,
\]
with self-explaining notation. In the commutative limit $q\to 1$, we obtain
\[
\hat R^{ij}_{kl}\to \delta^i_l \delta^j_k .
\]
A covariant (with respect to the action of the quantum group) dif\/ferential calculus also exists and can be def\/ined by the following relations:
\[
d\hat x^i d \hat x^j   = - q^{\pm 1} \hat R^{\pm 1}{}^{ij}_{kl} d\hat x^k d \hat x^l ,\qquad
\hat x^i d\hat x^j   = q^{\pm} \hat R^{\pm 1}{}^{ij}_{kl} d\hat x^k \hat x^l .
\]
Equivalently, we have for partial derivatives ($d=\hat x^i \hat\partial_i$)
\[
  \hat P_-{}^{ij}_{kl} \hat \partial_i \hat \partial_j = 0 , \qquad
 \hat \partial_i \hat x^j = \delta^j_i + q^{\pm 1} \hat R^{\pm 1}{}^{jl}_{ik}\hat x^k \hat \partial_l .
\]
The relation \eqref{eq:noncanon:others:QYB} is also called \emph{braid equation}. There exists a whole graphical apparatus to deal with the braid group. Especially, S.~Majid pushed this mathematical approach, which was also applied to gauge theory~-- see~\cite{Majid:1996} and references therein.

In \cite{Schraml:2002fi}, S.~Schraml computed the Seiberg--Witten map\footnote{The expansion parameter $h$ is def\/ined by $q=e^h$.} up to f\/irst order in $h$ and with respect to a normal ordered star product for a $SL_q(2)$-symmetric quantum space, the so-called \emph{Manin plane}. He considered the $q$-deformed BRST transformation
\begin{gather*}
s \widehat C   = \widehat C \star \widehat C ,\qquad
s \widehat E_i{}^j   = \ri \widehat C \star \widehat E_i{}^j - \ri \widehat E_i{}^k \star (B_k{}^j \widehat C) ,\\
s \widehat A_i   = \hat \partial_i \widehat C + \ri (B_i{}^j \widehat C) \star \widehat A_j - \ri \widehat A_i \star \widehat C,\qquad
s \widehat \psi  = \ri \widehat C \star \widehat \psi,
\end{gather*}
where $\widehat C$ is the ghost f\/ield, $\widehat A_i$ denotes the {\nA} gauge f\/ield, and $\widehat E_i{}^k$ the {\nc} vielbein appearing in the covariant derivatives
\[
\widehat D_i \widehat \psi = \widehat E_i{}^j ( \hat \partial_j - i \widehat A_j ) \widehat \psi .
\]
The operator $B_i{}^k$ is introduced for some technical reasons~\cite{Schraml:2002fi}. To f\/irst order, the gauge equiva\-lence relations yield the following solution:
\begin{gather*}
\widehat C  = C + \frac {\ri h} 2 x^1 x^2 \left( (\partial_2 C) A_1 - A_2 (\partial_1 C) \right) + \mathcal O\big(h^2\big),\\
\widehat A_i   = A_i +h  A^{(1)}_i + \mathcal O\big(h^2\big),\qquad
\widehat E_i{}^j   = \delta_i^j +h  {E^{(1)}}_i{}^j + \mathcal O\big(h^2\big),
\end{gather*}
where
\begin{gather*}
A^{(1)}_1 =  \big(2x^2 \partial_2 + x^1\partial_1\big) A_1 + 2\ri x^2 A_1 A_2 - \frac \ri 2 x^2 A_2 A_1 + \ri x^1 A_1 A_1 \\
\phantom{A^{(1)}_1 =}{}
+ \frac \ri 2 x^1 x^2 ( F_{12} A_1 + \partial_2 A_1 A_1 - A_2 \partial_1 A_1 ) , \\
A^{(1)}_2 =  \big(x^1 \partial_1 + 2x^2 \partial_2 \big)A_2 + \frac \ri 2 x^1 A_2 A_1 + \ri x^2 A_2 A_2 \\
\phantom{A^{(1)}_2 =}{}
  + \frac \ri 2 x^1 x^2 ( - A_2 F_{12} - \partial_2 A_2 A_1 - A_2 \partial_1 A_2 ) ,
\end{gather*}
and
\begin{alignat*}{3}
& {E^{(1)}}_1{}^1   = - \ri \big( 2x^1 A_1 + x^2A_2 \big), \qquad &&
{E^{(1)}}_1{}^2  = - 2 \ri x^2 A_1, & \\
& {E^{(1)}}_2{}^1   = 0 , \qquad &&
{E^{(1)}}_2{}^2  = - \ri h \big(x^1 A_1 + 2 x^2 A_2\big).&
\end{alignat*}
The same approach was also studied in~\cite{Meyer:2003wj}, see also \cite{Mesref:2002}. There, gauge theory is formulated on Euclidean $q$-deformed $2$-dimensional spaces generated by $\hat z$, $\bar{\hat z}$ with relation
\[
\hat z\bar{\hat z} = q^2 \bar{\hat z} \hat z ,
\]
which is covariant under the quantum group $E_q(2)$. In order to formulate an action, one uses the Hermitian star product
\[
(f\star g) (\zeta, \bar \zeta) = m \circ e^{ h( \zeta \partial_\zeta \otimes \bar \zeta \partial_{\bar \zeta} - \bar \zeta \partial_{\bar\zeta} \otimes \zeta \partial_\zeta ) }
\]
and the integration measure $\mu=\frac 1{\zeta \bar\zeta}$, such that
\[
\int d\zeta \, d\bar \zeta \, \mu (f\star g)(\zeta,\bar\zeta) = \int d\zeta \, d\bar \zeta \, \mu (g \star f)(\zeta,\bar\zeta) = \int d\zeta \, d\bar \zeta \,  \mu g(\zeta,\bar\zeta) \cdot f(\zeta,\bar\zeta) .
\]
This property of the integral implies that a variational calculus can be applied, and the gauge invariant action reads
\[
S = \int d\zeta \, d\bar \zeta \, \mu \widehat F_{12} \star \widehat F_{12}  ,
\]
where $\widehat F_{12}$ is the $q$-deformed non-Abelian f\/ield strength.

Non-perturbative methods have been applied e.g.\ in~\cite{Boulatov:1996}. D.V.~Boulatov discussed a~$3$-di\-mensional lattice gauge model with $q$-deformed gauge group $\mathcal U_q(SU(N))$ applying the graphical calculus mentioned above.
He formulated the partition function and discussed some topological invariants. In the continuum limit, the partition function is given by a $3$-fold invariant which coincides with the so-called Turaev--Viro invariant. Furthermore, he conjectured that a continuum limit exists, where both deformed Yang--Mills and Chern--Simons terms are recovered.

Due to the involved structure in the quantum group case, not many results are available, and the conducted work is mainly restricted to the formulation of models and to the discussion of rather general properties. The computation of Feynman rules and explicit perturbative (one-loop) calculations are still missing.

\subsubsection{Gauge theory with covariant star product}
\label{sec:noncanon:other:covariant}

In this section, we will consider a covariant star product with respect to dif\/feomorphism transformations. In~\cite{McCurdy:2009xz} such a star product was constructed for dif\/ferential forms on symplectic manifolds, and generalized to the case of Lie algebra valued dif\/ferential forms in \cite{Chaichian:2009kn}. This approach was also applied to {\nc} gravity, see~\cite{Vassilevich:2009cb}. The starting point is a symplectic structure $\cth^{\nu\mu}$, which is non-degenerate and closed,
\[
\aco{f}{g} = \cth^{\mu\nu} \partial_\mu f \partial_\nu g .
\]
The Poisson bracket of a function $f$ and a form $\alpha$ can be written as
\[
\aco{f}{\alpha} = \cth^{\mu\nu}\partial_\mu f\, \nabla_\nu \alpha ,
\]
the action of the connection $\nabla$ on basis $1$-forms is given by
\[
\nabla_\mu d x^\sigma = - \Gamma^\sigma_{\mu\nu} dx^\nu .
\]
In general, the connection is not torsion-free, therefore two connections $\nabla$ and $\tilde \nabla$ can be def\/ined, acting on $1$-forms as
\[
\nabla_\mu dx^\sigma   = -\Gamma^\sigma_{\mu\nu} dx^\nu , \qquad
\tilde \nabla_\mu dx^\sigma   = - \tilde \Gamma^\sigma_{\mu\nu} dx^\nu = -\Gamma^\sigma_{\nu\mu} dx^\nu .
\]
The star product for two Lie algebra valued dif\/ferential forms $\alpha$ and $\beta$ then reads
\begin{gather}
\label{eq:noncanon:other:covariant-star}
\alpha \star \beta = \alpha   \beta + \sum_{n=1}^\infty \left( \frac{\ri \hbar}{2} \right)^n C_n(\alpha,\beta) = \alpha^a \beta^b T^a T^b + \sum_{n=1}^\infty \left( \frac{\ri \hbar}{2} \right)^n C_n\big(\alpha^a,\beta^b\big) T^a T^b   .
\end{gather}
The bidif\/ferential operators $C_n$ are provided in~\cite{Chaichian:2009kn} up to second order in $\hbar$. The f\/irst order term is given by
\[
C_1(\alpha^a,\beta^b) = \aco{\alpha^a}{\beta^b} = \cth^{\mu\nu} \left( \nabla_\mu \alpha^a \nabla_\nu \beta^b +
(-1)^{|\alpha|} \tilde R^{\sigma\rho}_{\mu\nu} (i_\rho \alpha^a) (i_\sigma \beta^b) \right),
\]
where $|\alpha|$ denotes the degree of the dif\/ferential form $\alpha$, $\tilde R^{\sigma\rho}_{\mu\nu}$ the curvature of the connection $\nabla$, and $i_\rho$ the usual interior product of forms.

This star product is covariant with respect to (coordinate) dif\/feomorphism transformations in the following sense:
\[
(\alpha \star \beta)' = \alpha' \star' \beta' ,
\]
where $\alpha \to \beta'$ is the usual dif\/feomorphism transformation of forms, and $\star'$ is obtained from equation~\eqref{eq:noncanon:other:covariant-star} by transforming the symplectic structure $\cth^{\m\n}$ and the connection.

Due to the problems already described \secref{sec:covar}, the star product does not close in a general Lie algebra, so only Lie algebras such as $U(N)$ can be considered as gauge groups, unless one extends the gauge group to its universal enveloping group or applies Seiberg--Witten maps. The f\/ield strength is introduced as
\[
F = \frac 12 dx^\mu dx^\nu F_{\mu\nu} = dA - \frac \ri 2 \starco{A}{A}.
\]
Furthermore, the following {\nc} action is suggested in~\cite{Chaichian:2009kn}:
\[
S_{\rm NC} = - \frac 1{4g^2} \langle  \hat G^{\mu\rho}\star F_{\rho\nu} \star \hat G^{\nu\sigma} \star F_{\sigma\mu}\rangle ,
\]
where $\langle \cdots \rangle$ denotes the integration \cite{Ho:2001fi}, and $\hat G^{\mu\nu}$ the ``covariantized'' metric of the {\nc} background space, such that under a {\nc} gauge transformation
\[
\delta_{\widehat \lambda} \hat G^{\mu\nu} = \ri \starco{\widehat \lambda}{\hat G^{\mu\nu}}.
\]
Assuming the gauge transformation of the metric, the action is by def\/inition gauge invariant. Furthermore, the integral is cyclic in the semi-classical limit.

\section{Concluding remarks}
\label{sec:conclusion}

In this review we hope to have given an overview of the dif\/ferent current approaches to constructing gauge models on deformed spaces. Supersymmetric models have been omitted since that would have been a review of its own. Our main focus, however, was on the simplest case of a deformed space, namely Euclidean {\moyal} space, and gauge models formulated thereon. But we have also covered a range of various approaches on non-canonical spaces. Especially on those spaces, the generalization of space-time symmetries to Hopf algebraic structures is an essential point and provides some guiding principals. We hope that insights from all the dif\/ferent approaches will lead the way to the construction of a renormalizable model for {\nc} gauge theory.

\subsection*{Acknowledgements}

This work was supported by the ``Fonds zur F\"orderung der Wissenschaftlichen Forschung'' (FWF) under contracts P21610-N16, P20507-N16 and I192-N16.

 \LastPageEnding


\addcontentsline{toc}{section}{References}
\begin{thebibliography}{99}

\footnotesize\itemsep=0pt

\bibitem{Schroedinger:1934}
Schr\"{o}dinger E.,
\"{U}ber die Unanwendbarkeit der Geometrie im Kleinen,
\href{http://dx.doi.org/10.1007/BF01494946}{\textit{Naturwiss.}} \textbf{22} (1934), 518--520.

\bibitem{Heisenberg:1938}
Heisenberg W.,
\"{U}ber die in der Theorie der Elementarteilchen auftretende universelle L\"{a}nge,
\href{http://dx.doi.org/10.1002/andp.19384240105}{\textit{Ann. Physics}} \textbf{32} (1938), 20--33.

\bibitem{Snyder:1946}
 Snyder H.S.,
Quantized space-time,
\href{http://dx.doi.org/10.1103/PhysRev.71.38}{\textit{Phys. Rev.}} \textbf{71} (1947), 38--41.

\bibitem{Snyder:1947}
 Snyder H.S.,
The electromagnetic f\/ield in quantized space-time,
  \href{http://dx.doi.org/10.1103/PhysRev.72.68}{\textit{Phys. Rev.}} \textbf{72} (1947), 68--71.

\bibitem{Doplicher:1994b}
Doplicher S., Fredenhagen K., Roberts J.E.,
The quantum structure of spacetime at the Planck scale and quantum f\/ields,
\href{http://dx.doi.org/10.1007/BF02104515}{\textit{Comm. Math. Phys.}} \textbf{172} (1995), 187--220,
  \href{http://www.arxiv.org/abs/hep-th/0303037}{hep-th/0303037}.

\bibitem{Connes:1994}
Connes A.,
Noncommutative geometry, Academic Press, Inc., San Diego, CA, 1994.

\bibitem{Connes:2006a}
Connes A.,
Noncommutative geometry and the standard model with neutrino mixing,
\href{http://dx.doi.org/10.1088/1126-6708/2006/11/081}{\textit{J.~High Energy Phys.}} \textbf{2006} (2006), no.~11,  081, 19~pages,
\href{http://www.arxiv.org/abs/hep-th/0608226}{hep-th/0608226}.

\bibitem{Barrett:2006a}
 Barrett J.W.,
Lorentzian version of the noncommutative geometry of the Standard Model of particle physics,
\href{http://dx.doi.org/10.1063/1.2408400}{\textit{J. Math. Phys.}} \textbf{48}   (2007), 012303, 7~pages,
  \href{http://www.arxiv.org/abs/hep-th/0608221}{hep-th/0608221}.

\bibitem{Madore:1990}
Madore J.,
The commutative limit of a matrix geometry,
\href{http://dx.doi.org/10.1063/1.529418}{\textit{J.~Math. Phys.}} \textbf{32} (1991), 332--335.

\bibitem{Madore:1991}
Madore J.,
The fuzzy sphere,
\href{http://stacks.iop.org/0264-9381/9/69}{\textit{Classical Quantum Gravity}} \textbf{9} (1992), 69--87.

\bibitem{Lukierski:1991}
Lukierski J., Ruegg H., Nowicki A., Tolstoi V.N.,
$q$-deformation of Poincar\'{e} algebra,
\href{http://dx.doi.org/10.1016/0370-2693(91)90358-W}{\textit{Phys. Lett.~B}} \textbf{264} (1991), 331--338.

\bibitem{Majid:1994}
Majid S., Ruegg H.,
Bicrossproduct structure of $\kappa$-Poincar\'{e} group and non-commutative geometry,
\href{http://dx.doi.org/10.1016/0370-2693(94)90699-8}{\textit{Phys. Lett.~B}} \textbf{334} (1994), 348--354,
\href{http://www.arxiv.org/abs/hep-th/9405107}{hep-th/9405107}.

\bibitem{Wohlgenannt:2003}
Dimitrijevi\'c M., Jonke L., M{\"o}ller L., Tsouchnika E., Wess J., Wohlgenannt M.,
Deformed f\/ield theory on $\kappa$-spacetime,
\href{http://dx.doi.org/10.1140/epjc/s2003-01309-y}{\textit{Eur. Phys. J. C}} \textbf{31} (2003), 129--138,
  \href{http://www.arxiv.org/abs/hep-th/0307149}{hep-th/0307149}.

\bibitem{Reshetikhin:1990}
Reshetikhin N., Takhtadzhyan L., Faddeev L.,
Quantization of Lie groups and Lie algebras,
\textit{Leningrad Math.~J.} \textbf{1} (1990),  193--225.

\bibitem{Wess:1994}
Lorek A., Schmidke W.B., Wess J.,
${\rm SU}_q(2)$ covariant $\hat R$-matrices for reducible representations,
\href{http://dx.doi.org/10.1007/BF00762790}{\textit{Lett. Math. Phys.}} \textbf{31} (1994), 279--288.

\bibitem{Harvey:2001yn}
Harvey J.A.,
Komaba lectures on noncommutative solitons and D-branes,
\href{http://www.arxiv.org/abs/hep-th/0102076}{hep-th/0102076}.

\bibitem{Szabo:2001}
 Szabo R.J.,
Quantum f\/ield theory on noncommutative spaces,
\href{http://dx.doi.org/10.1016/S0370-1573(03)00059-0}{\textit{Phys. Rep.}} \textbf{378} (2003), 207--299,
  \href{http://www.arxiv.org/abs/hep-th/0109162}{hep-th/0109162}.

\bibitem{Paniak:2003yf}
 Paniak L.D., Szabo R.J.,
 Lectures on two-dimensional noncommutative gauge theory. I.~Classical aspects,
  \href{http://www.arxiv.org/abs/hep-th/0302195}{hep-th/0302195}.

\bibitem{Paniak:2003xm}
 Paniak L.D., Szabo R.J.,
 Lectures on two-dimensional noncommutative gauge theory,
in Quantum Field Theory and Noncommutative Geometry, {\it Lecture Notes in Phys.}, Vol.~662, Springer, Berlin, 2005, 205--237,
  \href{http://www.arxiv.org/abs/hep-th/0304268}{hep-th/0304268}.

\bibitem{Bahns:2009}
Bahns D.,
Schwinger functions in noncommutative quantum f\/ield theory,
  \href{http://www.arxiv.org/abs/0908.4537}{arXiv:0908.4537}.

\bibitem{Grosse:2008c}
Grosse H., Lechner G.,
Noncommutative deformations of Wightman quantum f\/ield theories,
\href{http://dx.doi.org/10.1088/1126-6708/2008/09/131}{\textit{J.~High Energy Phys.}} \textbf{2008} (2008), no.~9, 131, 29~pages,
  \href{http://www.arxiv.org/abs/0808.3459}{arXiv:0808.3459}.

\bibitem{Groenewold:1946}
Groenewold H.J.,
On the principles of elementary quantum mechanics,
\href{http://dx.doi.org/10.1016/S0031-8914(46)80059-4}{\textit{Physica}} \textbf{12} (1946), 405--460.

\bibitem{Moyal:1949}
 Moyal J.E.,
Quantum mechanics as a statistical theory,
\href{http://dx.doi.org/10.1017/S0305004100000487}{\textit{Proc. Cambridge Phil. Soc.}} \textbf{45} (1949), 99--124.

\bibitem{Filk:1996}
Filk T.,
Divergencies in a f\/ield theory on quantum space,
  \href{http://dx.doi.org/10.1016/0370-2693(96)00024-X}{\textit{Phys. Lett.~B}} \textbf{376} (1996), 53--58.

\bibitem{Minwalla:1999}
Minwalla S., Van~Raamsdonk M., Seiberg N.,
Noncommutative perturbative dynamics,
\href{http://dx.doi.org/10.1088/1126-6708/2000/02/020}{\textit{J.~High Energy Phys.}} \textbf{2000} (2000), no.~2, 020, 31~pages,
  \href{http://www.arxiv.org/abs/hep-th/9912072}{hep-th/9912072}.

\bibitem{Chepelev:2000}
Chepelev I., Roiban R.,
Convergence theorem for non-commutative Feynman graphs and renormalization,
\href{http://dx.doi.org/10.1088/1126-6708/2001/03/001}{\textit{J.~High Energy Phys.}} \textbf{2001} (2001), no.~3, 001, 72~pages,
  \href{http://www.arxiv.org/abs/hep-th/0008090}{hep-th/0008090}.

\bibitem{Grosse:2003}
Grosse H., Wulkenhaar R.,
Renormalisation of $\phi^4$-theory on noncommutative {$\mathbb R^2$} in the matrix base,
\href{http://dx.doi.org/10.1088/1126-6708/2003/12/019}{\textit{J.~High Energy Phys.}} \textbf{2003} (2003), no.~12,  019, 26~pages,
  \href{http://www.arxiv.org/abs/hep-th/0307017}{hep-th/0307017}.

\bibitem{Grosse:2004b}
Grosse H., Wulkenhaar R.,
Renormalisation of $\phi^4$-theory on noncommutative {$\mathbb R^4$} in the matrix base,
\href{http://dx.doi.org/10.1007/s00220-004-1285-2}{\textit{Comm. Math. Phys.}} \textbf{256} (2005), 305--374,
  \href{http://www.arxiv.org/abs/hep-th/0401128}{hep-th/0401128}.

\bibitem{Langmann:2002}
Langmann E., Szabo R.J.,
Duality in scalar f\/ield theory on noncommutative phase spaces,
\href{http://dx.doi.org/10.1016/S0370-2693(02)01650-7}{\textit{Phys. Lett.~B}} \textbf{533} (2002), 168--177,
  \href{http://www.arxiv.org/abs/hep-th/0202039}{hep-th/0202039}.

\bibitem{Magnen:2007}
Magnen J., Rivasseau V.,
Constructive $\phi^4$ f\/ield theory without tears,
\href{http://dx.doi.org/10.1007/s00023-008-0360-1}{\textit{Ann. Henri Poincar\'e}} \textbf{9} (2008), 403--424,
  \href{http://www.arxiv.org/abs/0706.2457}{arXiv:0706.2457}.

\bibitem{Gurau:2009}
Gurau R., Magnen J., Rivasseau V., Tanasa A.,
A translation-invariant renormalizable non-commutative scalar model,
  \href{http://dx.doi.org/10.1007/s00220-008-0658-3}{\textit{Comm. Math. Phys.}} \textbf{287} (2009), 275--290,
  \href{http://www.arxiv.org/abs/0802.0791}{arXiv:0802.0791}.

\bibitem{Wallet:2007c}
de~Goursac A., Wallet J.-C., Wulkenhaar R.,
Noncommutative induced gauge theory,
\href{http://dx.doi.org/10.1140/epjc/s10052-007-0335-2}{\textit{Eur. Phys. J.~C}} \textbf{51} (2007), 977--987,
  \href{http://www.arxiv.org/abs/hep-th/0703075}{hep-th/0703075}.

\bibitem{Wohlgenannt:2008zz}
Wohlgenannt M.,
Induced gauge theory on a noncommutative space,
\href{http://dx.doi.org/10.1002/prop.200710533}{\textit{Fortschr. Phys.}} \textbf{56} (2008), 547--551.

\bibitem{Blaschke:2007b}
 Blaschke D.N., Grosse H., Schweda M.,
 Non-commutative U(1) gauge theory on $\mathbb R^4_\Theta$ with oscillator term and BRST symmetry,
\href{http://dx.doi.org/10.1209/0295-5075/79/61002}{\textit{Europhys. Lett.}} \textbf{79} (2007),  61002, 3~pages,
  \href{http://www.arxiv.org/abs/0705.4205}{arXiv:0705.4205}.

\bibitem{Blaschke:2009aw}
Blaschke D.N., Grosse H., Kronberger E., Schweda M., Wohlgenannt M.,
Loop calculations for the non-commutative U(1) gauge f\/ield model with oscillator term,
\href{http://dx.doi.org/10.1140/epjc/s10052-010-1295-5}{{\it Eur. Phys. J.~C}} {\bf 67} (2010), 575--582,
  \href{http://www.arxiv.org/abs/0912.3642}{arXiv:0912.3642}.

\bibitem{Buric:2010}
Buric M., Grosse H., Madore J.,
Gauge f\/ields on noncommutative geometries with curvature,
\href{http://dx.doi.org/10.1007/JHEP07(2010)010}{\textit{J.~High Energy Phys.}} \textbf{2010} (2010), no.~7, 010, 18~pages,
\href{http://www.arxiv.org/abs/1003.2284}{arXiv:1003.2284}.

\bibitem{Wohlgenannt:2009}
Buric M., Wohlgenannt M.,
Geometry of the Grosse--Wulkenhaar model,
\href{http://dx.doi.org/10.1007/JHEP03(2010)053}{\textit{J.~High Energy Phys.}} \textbf{2010} (2010), no.~3, 053, 17~pages,
  \href{http://www.arxiv.org/abs/0902.3408}{arXiv:0902.3408}.

\bibitem{Blaschke:2008a}
 Blaschke D.N., Gieres F., Kronberger E., Schweda M., Wohlgenannt M.,
Translation-invariant models for non-commutative gauge f\/ields,
  \href{http://dx.doi.org/10.1088/1751-8113/41/25/252002}{\textit{J.~Phys.~A: Math. Theor.}} \textbf{41} (2008), 252002, 7~pages,
  \href{http://www.arxiv.org/abs/0804.1914}{arXiv:0804.1914}.

\bibitem{Blaschke:2009a}
Blaschke D.N., Rofner A., Schweda M., Sedmik R.I.P.,
One-loop   calculations for a translation invariant non-commutative gauge model,
   \href{http://dx.doi.org/10.1140/epjc/s10052-009-1031-1}{\textit{Eur. Phys. J.~C}} \textbf{62} (2009), 433--443,
  \href{http://www.arxiv.org/abs/0901.1681}{arXiv:0901.1681}.

\bibitem{Blaschke:2009b}
Blaschke D.N., Rofner A., Schweda M., Sedmik R.I.P.,
Improved   localization of a renormalizable non-commutative translation invariant U(1) gauge model,
\href{http://dx.doi.org/10.1209/0295-5075/86/51002}{\textit{Europhys. Lett.}} \textbf{86} (2009), 51002, 6~pages,
  \href{http://www.arxiv.org/abs/0903.4811}{arXiv:0903.4811}.

\bibitem{Blaschke:2009d}
Blaschke D.N., Rofner A., Sedmik R.I.P.,
One-loop calculations and detailed analysis of the localized non-commutative {$1/p^2$ $U(1)$} gauge model,
\href{http://dx.doi.org/10.3842/SIGMA.2010.037}{\textit{SIGMA}} \textbf{6} (2010), 037, 20~pages,
  \href{http://www.arxiv.org/abs/0908.1743}{arXiv:0908.1743}.

\bibitem{Vilar:2009}
 Vilar L.C.Q., Ventura O.S., Tedesco D.G., Lemes V.E.R.,
 On   the renormalizability of noncommutative U(1) gauge theory~-- an algebraic approach,
 \href{http://dx.doi.org/10.1088/1751-8113/43/13/135401}{\textit{J.~Phys.~A: Math. Theor.}} \textbf{43} (2010), 135401, 13~pages,
  \href{http://www.arxiv.org/abs/0902.2956}{arXiv:0902.2956}.

\bibitem{Blaschke:2009e}
 Blaschke D.N., Rofner A., Sedmik R.I.P., Wohlgenannt M.,
 On non-commutative $U_\star(1)$ gauge models and renormalizability,
  \href{http://www.arxiv.org/abs/0912.2634}{arXiv:0912.2634}.

\bibitem{Gribov:1978}
 Gribov V.N.,
 Quantization of non-Abelian gauge theories,
  \href{http://dx.doi.org/10.1016/0550-3213(78)90175-X}{\textit{Nuclear Phys.~B}} \textbf{139} (1978), 1--19.

\bibitem{Zwanziger:1989}
Zwanziger D.,
Local and renormalizable action from the Gribov horizon,
\href{http://dx.doi.org/10.1016/0550-3213(89)90122-3}{\textit{Nuclear Phys.~B}} \textbf{323} (1989), 513--544.

\bibitem{Zwanziger:1993}
Zwanziger D.,
Renormalizability of the critical limit of lattice gauge theory by BRS invariance,
\href{http://dx.doi.org/10.1016/0550-3213(93)90506-K}{\textit{Nuclear Phys.~B}} \textbf{399} (1993), 477--513.

\bibitem{Baulieu:2009}
Baulieu L., Sorella S.P.,
Soft breaking of BRST invariance for introducing non-perturbative infrared ef\/fects in a local and renormalizable way, \href{http://dx.doi.org/10.1016/j.physletb.2008.11.036}{\textit{Phys. Lett.~B}} \textbf{671} (2009), 481--485,
  \href{http://www.arxiv.org/abs/0808.1356}{arXiv:0808.1356}.

\bibitem{Seiberg:1999}
Seiberg N.,  Witten E.,
String theory and noncommutative geometry,
\href{http://dx.doi.org/10.1088/1126-6708/1999/09/032}{\textit{J.~High Energy Phys.}} \textbf{1999} (1999), no.~9, 032, 93~pages,
\href{http://www.arxiv.org/abs/hep-th/9908142}{hep-th/9908142}.

\bibitem{Jurco:2001a}
Jur\v{c}o B., M{\"o}ller L., Schraml S., Schupp P., Wess J.,
Construction of non-Abelian gauge theories on noncommutative spaces,
\href{http://dx.doi.org/10.1007/s100520100731}{\textit{Eur. Phys. J.~C}} \textbf{21} (2001), 383--388,
  \href{http://www.arxiv.org/abs/hep-th/0104153}{hep-th/0104153}.

\bibitem{Barnich:2003wq}
Barnich G., Brandt F., Grigoriev M.,
Local BRST cohomology and Seiberg--Witten maps in noncommutative Yang--Mills theory,
\href{http://dx.doi.org/10.1016/j.nuclphysb.2003.10.043}{\textit{Nuclear Phys.~B}} \textbf{677} (2004), 503--534,
  \href{http://www.arxiv.org/abs/hep-th/0308092}{hep-th/0308092}.

\bibitem{Wohlgenannt:2001}
Calmet X., Jur\v{c}o B., Schupp P., Wess J., Wohlgenannt M.,
The standard model on non-commutative space-time,
\href{http://dx.doi.org/10.1007/s100520100873}{\textit{Eur. Phys. J.~C}} \textbf{23} (2002), 363--376,
  \href{http://www.arxiv.org/abs/hep-ph/0111115}{hep-ph/0111115}.

\bibitem{Melic:2005am}
Meli\'c B., Passek-Kumeri\v{c}ki K., Trampeti\'c J., Schupp P., Wohlgenannt M.,
The standard model on non-commutative space-time: strong interactions included,
\href{http://dx.doi.org/10.1140/epjc/s2005-02301-3}{\textit{Eur. Phys. J.~C}} \textbf{42} (2005), 499--504,
  \href{http://www.arxiv.org/abs/hep-ph/0503064}{hep-ph/0503064}.

\bibitem{Wohlgenannt:2005b}
Meli\'c B., Passek-Kumeri\v{c}ki K., Trampeti\'c J., Schupp P., Wohlgenannt M.,
The standard model on non-commutative space-time: electroweak  currents and Higgs sector,
\href{http://dx.doi.org/10.1140/epjc/s2005-02318-6}{\textit{Eur. Phys. J.~C}} \textbf{42} (2005), 483--497,
  \href{http://www.arxiv.org/abs/hep-ph/0502249}{hep-ph/0502249}.

\bibitem{Buric:2006wm}
Buri\'c M., Radovanovi\'c V., Trampeti\'c J.,
The one-loop renormalization of the gauge sector in the $\theta$-expanded  noncommutative standard model,
\href{http://dx.doi.org/10.1088/1126-6708/2007/03/030}{\textit{J.~High Energy Phys.}} \textbf{2007} (2007), no.~3, 030, 17~pages,
  \href{http://www.arxiv.org/abs/hep-th/0609073}{hep-th/0609073}.

\bibitem{Buric:2002}
Buri\'c M., Radovanovi\'c V.,
The one-loop ef\/fective action for quantum electrodynamics on noncommutative space,
\href{http://dx.doi.org/10.1088/1126-6708/2002/10/074}{\textit{J.~High Energy Phys.}} \textbf{2002} (2002), no.~10, 074, 17~pages,
  \href{http://www.arxiv.org/abs/hep-th/0208204}{hep-th/0208204}.

\bibitem{Bichl:2001d}
 Bichl A.A., Grimstrup J.M., Grosse H., Popp L., Schweda M., Wulkenhaar R.,
Renormalization of the noncommutative photon self-energy to all orders via Seiberg--Witten map,
\href{http://dx.doi.org/10.1088/1126-6708/2001/06/013}{\textit{J.~High Energy Phys.}} \textbf{2001} (2001), no.~6, 013, 15~pages,
  \href{http://www.arxiv.org/abs/hep-th/0104097}{hep-th/0104097}.

\bibitem{Wulkenhaar:2001}
Wulkenhaar R.,
Non-renormalizability of $\theta$-expanded non-commutative QED,
\href{http://dx.doi.org/10.1088/1126-6708/2002/03/024}{\textit{J.~High Energy Phys.}} \textbf{2002} (2002), no.~3, 024, 35~pages,
  \href{http://www.arxiv.org/abs/hep-th/0112248}{hep-th/0112248}.

\bibitem{Martin:2007}
 Mart\'{\i}n C.P., Tamarit C.,
 Renormalisability of the matter determinants in noncommutative gauge theory in the enveloping-algebra formalism, \href{http://dx.doi.org/10.1016/j.physletb.2007.08.033}{\textit{Phys. Lett.~B}} \textbf{658} (2008), 170--173,
  \href{http://www.arxiv.org/abs/0706.4052}{arXiv:0706.4052}.

\bibitem{Tamarit:2008}
Tamarit C., Trampeti\'c J.,
Noncommutative fermions and quarkonia decays,
\href{http://dx.doi.org/10.1103/PhysRevD.79.025020}{\textit{Phys. Rev.~D}} \textbf{79} (2009), 025020, 15~pages,
  \href{http://www.arxiv.org/abs/0812.1731}{arXiv:0812.1731}.

\bibitem{Martin:2009}
 Mart\'{\i}n C.P., Tamarit C.,
Renormalisability of noncommutative GUT inspired f\/ield theories with anomaly safe groups,
\href{http://dx.doi.org/10.1088/1126-6708/2009/12/042}{\textit{J.~High Energy Phys.}} \textbf{2009} (2009), no.~12, 042, 18~pages,
  \href{http://www.arxiv.org/abs/0910.2677}{arXiv:0910.2677}.

\bibitem{Trampetic:2002eb}
Trampeti\'c J.,
Rare and forbidden decays,
\textit{Acta Phys. Polon.~B} \textbf{33} (2002), 4317--4372,
  \href{http://www.arxiv.org/abs/hep-ph/0212309}{hep-ph/0212309}.

\bibitem{Melic:2005su}
Meli\'c B., Passek-Kumeri\v{c}ki K., Trampeti\'c J.,
$K \to \pi \gamma$ decay and space-time noncommutativity,
\href{http://dx.doi.org/10.1103/PhysRevD.72.057502}{\textit{Phys. Rev.~D}} \textbf{72} (2005), 057502, 3~pages
  \href{http://www.arxiv.org/abs/hep-ph/0507231}{hep-ph/0507231}.

\bibitem{Buric:2007qx}
Buri\'c M., Latas D., Radovanovi\'c V., Trampeti\'c J.,
Nonzero $Z \to \gamma \gamma$ decays in the renormalizable gauge sector of the  noncommutative standard model, \href{http://dx.doi.org/10.1103/PhysRevD.75.097701}{\textit{Phys. Rev.~D}} \textbf{75} (2007), 097701, 4~pages.

\bibitem{Schupp:2002up}
Schupp P., Trampeti\'c J., Wess J., Raf\/felt G.,
The photon neutrino interaction in non-commutative gauge f\/ield theory and astrophysical bounds,
\href{http://dx.doi.org/10.1140/epjc/s2004-01874-5}{\textit{Eur. Phys. J.~C}} \textbf{36} (2004), 405--410,
  \href{http://www.arxiv.org/abs/hep-ph/0212292}{hep-ph/0212292}.

\bibitem{Horvat:2009cm}
Horvat R., Trampeti\'c J.,
Constraining spacetime noncommutativity with primordial nucleosynthesis,
\href{http://dx.doi.org/10.1103/PhysRevD.79.087701}{\textit{Phys. Rev.~D}} \textbf{79} (2009), 087701, 4~pages,
  \href{http://www.arxiv.org/abs/0901.4253}{arXiv:0901.4253}.

\bibitem{Schupp:2008}
Schupp P.,  You J.,
UV/IR mixing in noncommutative QED def\/ined by Seiberg--Witten map,
\href{http://dx.doi.org/10.1088/1126-6708/2008/08/107}{\textit{J.~High Energy Phys.}} \textbf{2008} (2008), no.~8, 107, 10~pages,
  \href{http://www.arxiv.org/abs/0807.4886}{arXiv:0807.4886}.

\bibitem{Tureanu:2010}
Raasakka M., Tureanu A.,
On UV/IR mixing via Seiberg--Witten map for noncommutative QED,
\href{http://dx.doi.org/10.1103/PhysRevD.81.125004}{\textit{Phys. Rev.~D}} {\bf 81} (2010), 125004, 8~pages,
  \href{http://www.arxiv.org/abs/1002.4531}{arXiv:1002.4531}.

\bibitem{Slavnov:2003}
 Slavnov A.A.,
 Consistent non-commutative quantum gauge theories?,
\href{http://dx.doi.org/10.1016/S0370-2693(03)00726-3}{\textit{Phys. Lett.~B}} \textbf{565} (2003), 246--252,
  \href{http://www.arxiv.org/abs/hep-th/0304141}{hep-th/0304141}.

\bibitem{Slavnov:2004}
 Slavnov A.A.,
A gauge-invariant U(1)-model on a noncommutative plane in an axial gauge,
 \href{http://dx.doi.org/10.1023/B:TAMP.0000039828.45005.51}{\textit{Theoret. and Math. Phys.}} \textbf{140} (2004), 1222--1228.

\bibitem{Blaschke:2005c}
 Blaschke D.N., Hohenegger S., Schweda M.,
 Divergences in non-commutative gauge theories with the Slavnov term,
\href{http://dx.doi.org/10.1088/1126-6708/2005/11/041}{\textit{J.~High Energy Phys.}} \textbf{2005} (2005), no.~11, 041, 29~pages,
  \href{http://www.arxiv.org/abs/hep-th/0510100}{hep-th/0510100}.

\bibitem{Blaschke:2006a}
 Blaschke D.N., Gieres F., Piguet O., Schweda M.,
 A vector supersymmetry in noncommutative U(1) gauge theory with the Slavnov term,
  \href{http://dx.doi.org/10.1088/1126-6708/2006/05/059}{\textit{J.~High Energy Phys.}} \textbf{2006} (2006), no.~5, 059, 16~pages,
  \href{http://www.arxiv.org/abs/hep-th/0604154}{hep-th/0604154}.

\bibitem{Blaschke:2007a}
Blaschke D.N., Hohenegger S.,
A generalization of Slavnov-extended non-commutative gauge theories,
\href{http://dx.doi.org/10.1088/1126-6708/2007/08/032}{\textit{J.~High Energy Phys.}} \textbf{2007} (2007), no.~8, 032, 21~pages,
  \href{http://www.arxiv.org/abs/0705.3007}{arXiv:0705.3007}.

\bibitem{Dimitrijevic:2003}
Dimitrijevi\'c M., Meyer F., M{\"o}ller L., Wess J.,
Gauge theories on the $\kappa$-Minkowski spacetime,
\href{http://dx.doi.org/10.1140/epjc/s2004-01887-0}{\textit{Eur. Phys.~J.~C}} \textbf{36} (2004), 117--126,
  \href{http://www.arxiv.org/abs/hep-th/0310116}{hep-th/0310116}.

\bibitem{Dimitrijevic:2005c}
Dimitrijevi\'c M., Jonke L., M\"oller L.,
U(1) gauge f\/ield theory on $\kappa$-Minkowski space,
\href{http://dx.doi.org/10.1088/1126-6708/2005/09/068}{\textit{J.~High Energy Phys.}} \textbf{2005} (2005), no.~9,  068, 15~pages,
  \href{http://www.arxiv.org/abs/hep-th/0504129}{hep-th/0504129}.

\bibitem{Bolokhov:2008}
 Bolokhov P.A., Pospelov M.,
 Low-energy constraints on $\kappa$-Minkowski extension of the Standard Model,
 \href{http://dx.doi.org/10.1016/j.physletb.2009.04.086}{\textit{Phys. Lett.~B}}  \textbf{677} (2009), 160--163,
  \href{http://www.arxiv.org/abs/0807.1522}{arXiv:0807.1522}.

\bibitem{Harikumar:2010}
Harikumar E.,
Maxwell's equations on the $\kappa$-Minkowski spacetime and electric-magnetic duality,
\href{http://dx.doi.org/10.1209/0295-5075/90/21001}{\textit{Europhys. Lett.}} \textbf{90} (2010), 21001, 6~pages, \href{http://www.arxiv.org/abs/1002.3202}{arXiv:1002.3202}.

\bibitem{Klimcik:1997}
Klim\v{c}\'{\i}k C.,
Gauge theories on the noncommutative sphere,
\href{http://dx.doi.org/10.1007/s002200050501}{\textit{Comm. Math. Phys.}} \textbf{199} (1998) 257--279,
 \mbox{\href{http://www.arxiv.org/abs/hep-th/9710153}{hep-th/9710153}}.

\bibitem{Balachandran:1999}
 Balachandran A.P., Vaidya S.,
 Instantons and chiral anomaly in fuzzy physics,
 \href{http://dx.doi.org/10.1142/S0217751X01003212}{\textit{Internat. J. Modern Phys.~A}} \textbf{16} (2001), 17--39,
  \href{http://www.arxiv.org/abs/hep-th/9910129}{hep-th/9910129}.

\bibitem{Karabali:2001}
Karabali D., Nair  V.P., Polychronakos A.P.,
Spectrum of  Schr\"odinger f\/ield in a noncommutative magnetic monopole,
\href{http://dx.doi.org/10.1016/S0550-3213(02)00062-7}{\textit{Nuclear   Phys.~B}} \textbf{627} (2002), 565--579,
  \href{http://www.arxiv.org/abs/hep-th/0111249}{hep-th/0111249}.

\bibitem{Iso:2001}
Iso S., Kimura Y., Tanaka K., Wakatsuki K.,
Noncommutative gauge theory on fuzzy sphere from matrix model,
\href{http://dx.doi.org/10.1016/S0550-3213(01)00173-0}{\textit{Nuclear Phys.~B}} \textbf{604} (2001), 121--147,
  \href{http://www.arxiv.org/abs/hep-th/0101102}{hep-th/0101102}.

\bibitem{Steinacker:2003}
Steinacker H.,
Quantized gauge theory on the fuzzy sphere as random matrix model,
\href{http://dx.doi.org/10.1016/j.nuclphysb.2003.12.005}{\textit{Nuclear Phys.~B}} \textbf{679} (2004), 66--98,
  \href{http://www.arxiv.org/abs/hep-th/0307075}{hep-th/0307075}.

\bibitem{CastroVillarreal:2004}
Castro-Villarreal~P., Delgadillo-Blando R., Ydri B.,
A gauge-invariant UV-IR mixing and the corresponding phase transition for U(1) f\/ields on the fuzzy sphere, \href{http://dx.doi.org/10.1016/j.nuclphysb.2004.10.032}{\textit{Nuclear Phys.~B}} \textbf{704} (2005), 111--153,
  \href{http://www.arxiv.org/abs/hep-th/0405201}{hep-th/0405201}.

\bibitem{O'Connor:2006}
O'Connor D., Ydri B.,
Monte Carlo simulation of a NC gauge theory on the fuzzy sphere,
\href{http://dx.doi.org/10.1088/1126-6708/2006/11/016}{\textit{J.~High Energy Phys.}} \textbf{2006} (2006), no.~11, 016, 37~pages,
  \href{http://www.arxiv.org/abs/hep-lat/0606013}{hep-lat/0606013}.

\bibitem{Alexanian:2001qj}
Alexanian G., Balachandran A.P., Immirzi G., Ydri B.,
Fuzzy $\mathbb{C}P^2$,
\href{http://dx.doi.org/10.1016/S0393-0440(01)00070-5}{\textit{J. Geom. Phys.}} \textbf{42} (2002), 28--53,
  \href{http://www.arxiv.org/abs/hep-th/0103023}{hep-th/0103023}.

\bibitem{Grosse:2004c}
Grosse H., Steinacker H.,
Finite gauge theory on fuzzy $\mathbb{C}P^2$,
\href{http://dx.doi.org/10.1016/j.nuclphysb.2004.11.058}{\textit{Nuclear Phys.~B}} \textbf{707} (2005), 145--198,
  \mbox{\href{http://www.arxiv.org/abs/hep-th/0407089}{hep-th/0407089}}.

\bibitem{Boulatov:1996}
 Boulatov D.V.,
 Quantum deformation of lattice gauge theory,
  \href{http://dx.doi.org/10.1007/s002200050111}{\textit{Comm. Math. Phys.}} \textbf{186} (1997), 295--322,
  \href{http://www.arxiv.org/abs/hep-th/9604117}{hep-th/9604117}.

\bibitem{Majid:1996}
Majid S.,
Diagrammatics of braided group gauge theory,
\href{http://dx.doi.org/10.1142/S021821659900047X}{\textit{J.~Knot Theory Ramifications}} \textbf{8} (1999), 731--771,
  \href{http://www.arxiv.org/abs/q-alg/9603018}{q-alg/9603018}.

\bibitem{Mesref:2002}
Mesref L.,
A map between $q$-deformed noncommutative and ordinary gauge theories,
\href{http://dx.doi.org/10.1088/1367-2630/5/1/307}{\textit{New J. Phys.}} \textbf{5} (2003), 7.1--7.9,
  \href{http://www.arxiv.org/abs/hep-th/0209005}{hep-th/0209005}.

\bibitem{Schraml:2002fi}
Schraml S.,
Non-Abelian gauge theory on $q$-quantum spaces,
  \href{http://www.arxiv.org/abs/hep-th/0208173}{hep-th/0208173}.

\bibitem{Meyer:2003wj}
Meyer F., Steinacker H.,
Gauge f\/ield theory on the $E_q(2)$-covariant plane,
\href{http://dx.doi.org/10.1142/S0217751X04019512}{\textit{Internat. J. Modern Phys.~A}} \textbf{19} (2004), 3349--3375,
  \href{http://www.arxiv.org/abs/hep-th/0309053}{hep-th/0309053}.

\bibitem{Ishibashi:1996xs}
Ishibashi N., Kawai H., Kitazawa Y., Tsuchiya A.,
A large-$N$ reduced model as superstring,
\href{http://dx.doi.org/10.1016/S0550-3213(97)00290-3}{\textit{Nuclear Phys.~B}} \textbf{498} (1997), 467--491,
  \href{http://www.arxiv.org/abs/hep-th/9612115}{hep-th/9612115}.

\bibitem{Jack:2001cr}
Jack I., Jones D.R.T.,
Ultra-violet f\/initeness in noncommutative supersymmetric theories,
\href{http://dx.doi.org/10.1088/1367-2630/3/1/319}{\textit{New J. Phys.}} \textbf{3} (2001), 19.1--19.8,
  \href{http://www.arxiv.org/abs/hep-th/0109195}{hep-th/0109195}.

\bibitem{Matusis:2000jf}
Matusis A., Susskind L., Toumbas N.,
The IR/UV connection in the non-commutative gauge theories,
\href{http://dx.doi.org/10.1088/1126-6708/2000/12/002}{\textit{J.~High Energy Phys.}} \textbf{2000} (2000), no.~12, 002, 18~pages,
  \href{http://www.arxiv.org/abs/hep-th/0002075}{hep-th/0002075}.

\bibitem{SheikhJabbari:1999iw}
 Sheikh-Jabbari M.M.,
One loop renormalizability of supersymmetric Yang-Mills theories on noncommutative two-torus,
\href{http://dx.doi.org/10.1088/1126-6708/1999/06/015}{\textit{J.~High Energy Phys.}} \textbf{1999} (1999), no.~6,  015, 16~pages,
  \href{http://www.arxiv.org/abs/hep-th/9903107}{hep-th/9903107}.

\bibitem{Rivelles:2000}
 Girotti H.O., Gomes M., Rivelles  V.O., da~Silva A.J.,
 A~consistent noncommutative f\/ield theory: the Wess--Zumino model,
\href{http://dx.doi.org/10.1016/S0550-3213(00)00483-1}{\textit{Nuclear Phys.~B}} \textbf{587} (2000), 299--310,
  \href{http://www.arxiv.org/abs/hep-th/0005272}{hep-th/0005272}.

\bibitem{Schweda:2000b}
 Bichl A.A., Grimstrup J., Grosse H., Popp L., Schweda M., Wulkenhaar R.,
The superf\/ield formalism applied to the noncommutative Wess--Zumino model,
\href{http://dx.doi.org/10.1088/1126-6708/2000/10/046}{\textit{J.~High Energy Phys.}} \textbf{2000} (2000), no.~10, 046, 19~pages,
  \mbox{\href{http://www.arxiv.org/abs/hep-th/0007050}{hep-th/0007050}}.

\bibitem{Zanon:2000}
Zanon D.,
Noncommutative $N = 1,2$ super ${\rm U}(N)$ Yang--Mills: UV/IR mixing and ef\/fective action results at one loop,
\href{http://dx.doi.org/10.1016/S0370-2693(01)00194-0}{\textit{Phys. Lett.~B}}  \textbf{502} (2001), 265--273,
  \href{http://www.arxiv.org/abs/hep-th/0012009}{hep-th/0012009}.

\bibitem{Ferrara:2000mm}
Ferrara S., Lled\'o M.A.,
Some aspects of deformations of supersymmetric f\/ield theories,
\href{http://dx.doi.org/10.1088/1126-6708/2000/05/008}{\textit{J.~High Energy Phys.}} \textbf{2000} (2000), no.~5, 008, 22~pages,
  \href{http://www.arxiv.org/abs/hep-th/0002084}{hep-th/0002084}.

\bibitem{Schweda:2002c}
 Bichl A.A., Ertl M., Gerhold A., Grimstrup J.M., Popp L., Putz V., Schweda M., Grosse H., Wulkenhaar R., Non-commutative U(1) super-Yang--Mills theory: perturbative self-energy corrections,
 \href{http://dx.doi.org/10.1142/S0217751X04018221}{\textit{Internat. J. Modern Phys.~A}} \textbf{19} (2004), 4231--4249,
  \href{http://www.arxiv.org/abs/hep-th/0203141}{hep-th/0203141}.

\bibitem{Ooguri:2003qp}
Ooguri H., Vafa C.,
The $C$-deformation of gluino and non-planar diagrams,
\textit{Adv. Theor. Math. Phys.} \textbf{7} (2003), 53--85,
  \href{http://www.arxiv.org/abs/hep-th/0302109}{hep-th/0302109}.

\bibitem{Araki:2003se}
Araki T., Ito K., Ohtsuka A.,
Supersymmetric gauge theories on noncommutative superspace,
\href{http://dx.doi.org/10.1016/j.physletb.2003.08.038}{\textit{Phys. Lett.~B}} \textbf{573} (2003), 209--216,
  \href{http://www.arxiv.org/abs/hep-th/0307076}{hep-th/0307076}.

\bibitem{Ooguri:2003tt}
Ooguri H., Vafa C.,
Gravity induced $C$-deformation,
\textit{Adv. Theor. Math. Phys.} \textbf{7} (2004), 405--417,
  \mbox{\href{http://www.arxiv.org/abs/hep-th/0303063}{hep-th/0303063}}.

\bibitem{Ferrari:2004ex}
 Ferrari A.F., Girotti H.O., Gomes M., Petrov A.Yu., Ribeiro A.A.,   Rivelles V.O., da~Silva A.J.,
Towards a~consistent noncommutative supersymmetric Yang--Mills theory: superf\/ield covariant analysis,
  \href{http://dx.doi.org/10.1103/PhysRevD.70.085012}{\textit{Phys. Rev.~D}} \textbf{70} (2004), 085012, 11~pages,
  \href{http://www.arxiv.org/abs/hep-th/0407040}{hep-th/0407040}.

\bibitem{Martin:2009mu}
 Mart\'{\i}n C.P., Tamarit C.,
 Noncommutative $N=1$ super Yang--Mills, the Seiberg--Witten map and UV divergences,
 \href{http://dx.doi.org/10.1088/1126-6708/2009/11/092}{\textit{J.~High Energy Phys.}} \textbf{2009} (2009), no.~11, 092, 28~pages, \href{http://www.arxiv.org/abs/0907.2437}{arXiv:0907.2437}.

\bibitem{Szabo:2006}
 Szabo R.J.,
 Symmetry, gravity and noncommutativity,
 \href{http://dx.doi.org/10.1088/0264-9381/23/22/R01}{\textit{Classical Quantum Gravity}} \textbf{23} (2006), R199--R242,
  \href{http://www.arxiv.org/abs/hep-th/0606233}{hep-th/0606233}.

\bibitem{Rivelles:2002}
 Rivelles V.O.,
 Noncommutative f\/ield theories and gravity,
  \href{http://dx.doi.org/10.1016/S0370-2693(03)00271-5}{\textit{Phys. Lett.~B}} \textbf{558} (2003), 191--196,
  \mbox{\href{http://www.arxiv.org/abs/hep-th/0212262}{hep-th/0212262}}.

\bibitem{Yang:2004vd}
 Yang H.S.,
 Exact Seiberg--Witten map and induced gravity from noncommutativity,
 \href{http://dx.doi.org/10.1142/S0217732306021682}{\textit{Modern Phys. Lett.~A}} \textbf{21} (2006), 2637--2647,
  \href{http://www.arxiv.org/abs/hep-th/0402002}{hep-th/0402002}.

\bibitem{Banerjee:2004rs}
Banerjee R., Yang H.S.,
Exact Seiberg--Witten map, induced gravity and topological invariants in non-commutative f\/ield theories,
\href{http://dx.doi.org/10.1016/j.nuclphysb.2004.12.003}{\textit{Nuclear Phys.~B}} \textbf{708} (2005), 434--450,
  \href{http://www.arxiv.org/abs/hep-th/0404064}{hep-th/0404064}.

\bibitem{Yang:2006}
 Yang H.S.,
 Instantons and emergent geometry,
 \href{http://dx.doi.org/10.1209/0295-5075/88/31002}{\textit{Europhys. Lett.}} \textbf{88} (2009), 31002, 6~pages,
  \href{http://www.arxiv.org/abs/hep-th/0608013}{hep-th/0608013}.

\bibitem{Steinacker:2007}
Steinacker H.,
Emergent gravity from noncommutative gauge theory,
\href{http://dx.doi.org/10.1088/1126-6708/2007/12/049}{\textit{J.~High Energy Phys.}} \textbf{2007} (2007), no.~12, 049, 36~pages,
  \href{http://www.arxiv.org/abs/0708.2426}{arXiv:0708.2426}.

\bibitem{Steinacker:2008}
Grosse H., Steinacker H., Wohlgenannt M.,
Emergent gravity, matrix models and UV/IR mixing,
\href{http://dx.doi.org/10.1088/1126-6708/2008/04/023}{\textit{J.~High Energy Phys.}} \textbf{2008} (2008), no.~4, 023, 30~pages,
  \href{http://www.arxiv.org/abs/0802.0973}{arXiv:0802.0973}.

\bibitem{Steinacker:2008a}
Klammer D., Steinacker H.,
Fermions and emergent noncommutative gravity,
\href{http://dx.doi.org/10.1088/1126-6708/2008/08/074}{\textit{J.~High Energy Phys.}} \textbf{2008} (2008), no.~8, 074, 27~pages,
  \href{http://www.arxiv.org/abs/0805.1157}{arXiv:0805.1157}.

\bibitem{Steinacker:2008ya}
Steinacker H.,
Covariant f\/ield equations, gauge f\/ields and conservation laws from Yang--Mills matrix models,
\href{http://dx.doi.org/10.1088/1126-6708/2009/02/044}{\textit{J.~High Energy Phys.}} \textbf{2009} (2009), no.~2, 044, 30~pages,
  \href{http://www.arxiv.org/abs/0812.3761}{arXiv:0812.3761}.

\bibitem{Freidel:2005}
Freidel L., Livine E.R.,
Ponzano--Regge model revisited. III.~Feynman diagrams and ef\/fective f\/ield theory,
\href{http://dx.doi.org/10.1088/0264-9381/23/6/012}{\textit{Classical Quantum Gravity}} \textbf{23} (2006), 2021--2061,
  \href{http://www.arxiv.org/abs/hep-th/0502106}{hep-th/0502106}.

\bibitem{Freidel:2005a}
Freidel L., Livine E.R.,
3D quantum gravity and ef\/fective noncommutative quantum f\/ield theory,
\href{http://dx.doi.org/10.1103/PhysRevLett.96.221301}{\textit{Phys. Rev. Lett.}} \textbf{96} (2006), 221301, 4~pages,
  \href{http://www.arxiv.org/abs/hep-th/0512113}{hep-th/0512113}.

\bibitem{Vassilevich:2007}
 Gitman D.M., Vassilevich D.V.,
 Space-time noncommutativity with a bifermionic parameter,
 \href{http://dx.doi.org/10.1142/S0217732308024481}{\textit{Modern Phys. Lett.~A}} \textbf{23} (2008), 887--893,
  \href{http://www.arxiv.org/abs/hep-th/0701110}{hep-th/0701110}.

\bibitem{Terashima:2000}
Terashima S.,
A note on superf\/ields and noncommutative geometry,
\href{http://dx.doi.org/10.1016/S0370-2693(00)00486-X}{\textit{Phys. Lett.~B}} \textbf{482} (2000), 276--282,
  \href{http://www.arxiv.org/abs/hep-th/0002119}{hep-th/0002119}.

\bibitem{Matsubara:2000gr}
Matsubara K.,
Restrictions on gauge groups in noncommutative gauge theory,
\href{http://dx.doi.org/10.1016/S0370-2693(00)00549-9}{\textit{Phys. Lett.~B}} \textbf{482} (2000), 417--419,
  \href{http://www.arxiv.org/abs/hep-th/0003294}{hep-th/0003294}.

\bibitem{Bars:2001}
Bars I., Sheikh-Jabbari M.M., Vasiliev M.A.,
Noncommutative ${\rm o}_*(N)$ and ${\rm usp}_*(2N)$ algebras and the corresponding gauge f\/ield theories,
\href{http://dx.doi.org/10.1103/PhysRevD.64.086004}{\textit{Phys. Rev.~D}} \textbf{64} (2001), 086004, 13~pages,
  \href{http://www.arxiv.org/abs/hep-th/0103209}{hep-th/0103209}.

\bibitem{Chaichian:2001mu}
Chaichian M., Pre\v{s}najder P., Sheikh-Jabbari M.M., Tureanu A.,
Noncommutative gauge f\/ield theories: a~no-go theorem,
\href{http://dx.doi.org/10.1016/S0370-2693(01)01478-2}{\textit{Phys. Lett.~B}} \textbf{526} (2002), 132--136,
  \href{http://www.arxiv.org/abs/hep-th/0107037}{hep-th/0107037}.

\bibitem{Bonora:2000td}
Bonora L., Schnabl M., Sheikh-Jabbari M.M., Tomasiello A.,
Noncommutative ${\rm SO}(n)$ and ${\rm Sp}(n)$ gauge theories,
\href{http://dx.doi.org/10.1016/S0550-3213(00)00527-7}{\textit{Nuclear Phys.~B}} \textbf{589} (2000), 461--474,
  \href{http://www.arxiv.org/abs/hep-th/0006091}{hep-th/0006091}.

\bibitem{Bietenholz:2006cz}
Bietenholz W., Nishimura J., Susaki Y., Volkholz J.,
A~non-perturbative study of 4d U(1) non-commutative gauge theory~-- the fate of one-loop instability,
\href{http://dx.doi.org/10.1088/1126-6708/2006/10/042}{\textit{J.~High Energy Phys.}} \textbf{2006} (2006), no.~10, 042, 31~pages,
  \href{http://www.arxiv.org/abs/hep-th/0608072}{hep-th/0608072}.

\bibitem{Nishimura:2007ix}
Nishimura J., Bietenholz W., Susaki Y., Volkholz J.,
A non-perturbative study of non-commutative U(1) gauge theory,
\href{http://dx.doi.org/10.1143/PTPS.171.178}{{\it Progr. Theor. Phys. Suppl.}} (2007),  no.~171, 178--183,
  \href{http://www.arxiv.org/abs/0706.3244}{arXiv:0706.3244}.

\bibitem{Belov:2000}
 Aref'eva I.Y., Belov D.M., Koshelev A.S., Rytchkov O.A.,
UV/IR mixing for noncommutative complex scalar f\/ield theory interacting with gauge f\/ields,
 \href{http://dx.doi.org/10.1016/S0920-5632(01)01531-6}{\textit{Nuclear Phys. Proc. Suppl.}} \textbf{102} (2001), 11--17,
  \mbox{\href{http://www.arxiv.org/abs/hep-th/0003176}{hep-th/0003176}}.

\bibitem{Micu:2000}
Micu A., Sheikh~Jabbari M.M.,
Noncommutative $\Phi^4$ theory at two loops,
\href{http://dx.doi.org/10.1088/1126-6708/2001/01/025}{\textit{J.~High Energy Phys.}} \textbf{2001} (2001), no.~1, 025, 45~pages,
  \href{http://www.arxiv.org/abs/hep-th/0008057}{hep-th/0008057}.

\bibitem{Schweda:2002b}
 Grimstrup J.M., Grosse H., Popp L., Putz V., Schweda M., Wickenhauser M., Wulkenhaar R.,
 IR-singularities in noncommutative perturbative dynamics?,
 \href{http://dx.doi.org/10.1209/epl/i2003-10285-9}{\textit{Europhys. Lett.}} \textbf{67} (2004), 186--190,
  \mbox{\href{http://www.arxiv.org/abs/hep-th/0202093}{hep-th/0202093}}.

\bibitem{Martin:1999aq}
 Mart\'{\i}n C.P., S\'anchez-Ruiz D.,
One-loop UV divergent structure of ${\rm U}(1)$ Yang--Mills theory on noncommutative $\mathbb R^4$,
 \href{http://dx.doi.org/10.1103/PhysRevLett.83.476}{\textit{Phys. Rev. Lett.}} \textbf{83} (1999), 476--479,
  \href{http://www.arxiv.org/abs/hep-th/9903077}{hep-th/9903077}.

\bibitem{Hayakawa:1999}
Hayakawa M.,
Perturbative analysis on infrared aspects of noncommutative QED on~$\mathbb{R}^4$,
\href{http://dx.doi.org/10.1016/S0370-2693(00)00242-2}{\textit{Phys. Lett.~B}} \textbf{478} (2000), 394--400,
  \href{http://www.arxiv.org/abs/hep-th/9912094}{hep-th/9912094}.

\bibitem{Hayakawa:1999b}
Hayakawa M.,
Perturbative analysis on infrared and ultraviolet aspects of noncommutative QED on $\mathbb{R}^4$,
  \href{http://www.arxiv.org/abs/hep-th/9912167}{hep-th/9912167}.

\bibitem{Grosse:2000}
Grosse H., Krajewski T., Wulkenhaar R.,
Renormalization of noncommutative Yang--Mills theories: a simple example,
  \href{http://www.arxiv.org/abs/hep-th/0001182}{hep-th/0001182}.

\bibitem{Martin:2000}
 Mart\'{\i}n C.P., S\'anchez-Ruiz D.,
 The BRS invariance of noncommutative ${\rm U}(N)$ Yang--Mills theory at the one-loop level,
 \href{http://dx.doi.org/10.1016/S0550-3213(01)00015-3}{\textit{Nuclear Phys.~B}} \textbf{598} (2001), 348--370,
  \href{http://www.arxiv.org/abs/hep-th/0012024}{hep-th/0012024}.

\bibitem{Liao:2001}
Liao Y.,
One loop renormalization of spontaneously broken U(2) gauge theory on noncommutative spacetime,
\href{http://dx.doi.org/10.1088/1126-6708/2001/11/067}{\textit{J.~High Energy Phys.}} \textbf{2001} (2001), no.~11,  067, 39~pages,
\href{http://www.arxiv.org/abs/hep-th/0110112}{hep-th/0110112}.

\bibitem{Nakajima:2002}
Nakajima T.,
UV/IR mixing and anomalies in noncommutative gauge theories,
  \href{http://www.arxiv.org/abs/hep-th/0205058}{hep-th/0205058}.

\bibitem{Ruiz:2000}
Ruiz Ruiz F.,
 Gauge-f\/ixing independence of IR divergences in non-commutative U(1), perturbative tachyonic instabilities and
  supersymmetry,
\href{http://dx.doi.org/10.1016/S0370-2693(01)00145-9}{\textit{Phys. Lett.~B}} \textbf{502} (2001), 274--278,
  \href{http://www.arxiv.org/abs/hep-th/0012171}{hep-th/0012171}.

\bibitem{Blaschke:2005b}
Attems M., Blaschke D.N., Ortner M., Schweda M., Stricker S., Weiretmayr M.,
Gauge independence of IR singularities in non-commutative QFT~-- and interpolating gauges,
\href{http://dx.doi.org/10.1088/1126-6708/2005/07/071}{\textit{J.~High Energy Phys.}} \textbf{2005} (2005), no.~7, 071, 10~pages,
  \href{http://www.arxiv.org/abs/hep-th/0506117}{hep-th/0506117}.

\bibitem{VanRaamsdonk:2001jd}
Van~Raamsdonk M.,
The meaning of infrared singularities in noncommutative gauge theories,
\href{http://dx.doi.org/10.1088/1126-6708/2001/11/006}{\textit{J.~High Energy Phys.}} \textbf{2001} (2001), no.~11, 006, 17~pages,
  \href{http://www.arxiv.org/abs/hep-th/0110093}{hep-th/0110093}.

\bibitem{Armoni:2000xr}
Armoni A.,
Comments on perturbative dynamics of non-commutative Yang--Mills theory,
\href{http://dx.doi.org/10.1016/S0550-3213(00)00557-5}{\textit{Nuclear Phys.~B}} \textbf{593} (2001), 229--242,
  \href{http://www.arxiv.org/abs/hep-th/0005208}{hep-th/0005208}.

\bibitem{Armoni:2001}
Armoni A., Lopez E.,
UV/IR mixing via closed strings and tachyonic instabilities,
\href{http://dx.doi.org/10.1016/S0550-3213(02)00290-0}{\textit{Nuclear Phys.~B}} \textbf{632} (2002), 240--256,
  \href{http://www.arxiv.org/abs/hep-th/0110113}{hep-th/0110113}.

\bibitem{Martin:2000bk}
 Mart\'{\i}n C.P., Ruiz Ruiz F.,
 Paramagnetic dominance, the sign of the beta function and UV/IR mixing in non-commutative U(1),
 \href{http://dx.doi.org/10.1016/S0550-3213(00)00726-4}{\textit{Nuclear Phys.~B}} \textbf{597} (2001), 197--227,
  \href{http://www.arxiv.org/abs/hep-th/0007131}{hep-th/0007131}.

\bibitem{Szabo:1998}
Landi G., Lizzi F., Szabo R.J.,
String geometry and the noncommutative torus,
\href{http://dx.doi.org/10.1007/s002200050839}{\textit{Comm. Math. Phys.}} \textbf{206} (1999), 603--637,
  \href{http://www.arxiv.org/abs/hep-th/9806099}{hep-th/9806099}.

\bibitem{Szabo:1999}
Landi G., Lizzi  F., Szabo R.J.,
From large $N$ matrices to the noncommutative torus,
\href{http://dx.doi.org/10.1007/s002200000356}{\textit{Comm. Math. Phys.}} \textbf{217} (2001), 181--201,
  \href{http://www.arxiv.org/abs/hep-th/9912130}{hep-th/9912130}.

\bibitem{Wulkenhaar:1999}
Krajewski T., Wulkenhaar R.,
Perturbative quantum gauge f\/ields on the noncommutative torus,
\href{http://dx.doi.org/10.1142/S0217751X00000495}{\textit{Internat. J. Modern Phys.~A}} \textbf{15} (2000), 1011--1029,
  \href{http://www.arxiv.org/abs/hep-th/9903187}{hep-th/9903187}.

\bibitem{Chaichian:2001py}
Chaichian M., Pre\v{s}najder P., Sheikh-Jabbari M.M., Tureanu A.,
Non-commutative standard model: model building,
\href{http://dx.doi.org/10.1140/epjc/s2003-01204-7}{\textit{Eur. Phys. J.~C}} \textbf{29} (2003), 413--432,
  \href{http://www.arxiv.org/abs/hep-th/0107055}{hep-th/0107055}.

\bibitem{Jurco:2000fs}
Jur\v{c}o B., Schupp P., Wess J.,
Noncommutative gauge theory for Poisson manifolds,
\href{http://dx.doi.org/10.1016/S0550-3213(00)00363-1}{\textit{Nuclear Phys.~B}} \textbf{584} (2000), 784--794,
  \href{http://www.arxiv.org/abs/hep-th/0005005}{hep-th/0005005}.

\bibitem{Wohlgenannt:2003a}
Wohlgenannt M.,
Introduction to a non-commutative version of the standard model,
  \href{http://www.arxiv.org/abs/hep-th/0302070}{hep-th/0302070}.

\bibitem{Aschieri:2002mc}
Aschieri P., Jur\v{c}o B., Schupp P., Wess J.,
Non-commutative GUTs, standard model and $C$, $P$, $T$,
\href{http://dx.doi.org/10.1016/S0550-3213(02)00937-9}{\textit{Nuclear Phys.~B}} \textbf{651} (2003), 45--70,
  \href{http://www.arxiv.org/abs/hep-th/0205214}{hep-th/0205214}.

\bibitem{Moller:2004qq}
M{\"o}ller L.,
Second order of the expansions of action functionals of the noncommutative standard model,
\href{http://dx.doi.org/10.1088/1126-6708/2004/10/063}{\textit{J.~High Energy Phys.}} \textbf{2004} (2004), no.~10, 063, 20~pages,
  \href{http://www.arxiv.org/abs/hep-th/0409085}{hep-th/0409085}.

\bibitem{Alboteanu:2007bp}
Alboteanuv A., Ohl T., Ruckl R.,
The noncommutative standard model at $O(\theta^2)$,
\href{http://dx.doi.org/10.1103/PhysRevD.76.105018}{\textit{Phys. Rev.~D}} \textbf{76} (2007), 105018, 10~pages,
  \href{http://www.arxiv.org/abs/0707.3595}{arXiv:0707.3595}.

\bibitem{Trampetic:2007hx}
Trampetic J., Wohlgenannt M.,
Comment on the 2nd order Seiberg--Witten maps,
\href{http://dx.doi.org/10.1103/PhysRevD.76.127703}{\textit{Phys. Rev.~D}} \textbf{76} (2007), 127703, 4~pages,
  \href{http://www.arxiv.org/abs/0710.2182}{arXiv:0710.2182}.

\bibitem{Bichl:2001b}
 Bichl A.A., Grimstrup J.M., Popp L., Schweda M., Wulkenhaar R.,
Deformed QED via Seiberg--Witten map,
  \href{http://www.arxiv.org/abs/hep-th/0102103}{hep-th/0102103}.

\bibitem{Carlson:2002wj}
Carlson C.E., Carone C.D., Zobin N.,
Noncommutative gauge theory without Lorentz violation,
\href{http://dx.doi.org/10.1103/PhysRevD.66.075001}{\textit{Phys. Rev.~D}} \textbf{66} (2002), 075001, 8~pages,
  \href{http://www.arxiv.org/abs/hep-th/0206035}{hep-th/0206035}.

\bibitem{Martin:2006gw}
 Mart\'{\i}n C.P., S\'anchez-Ruiz D., Tamarit C.,
 The noncommutative U(1)  Higgs--Kibble model in the enveloping-algebra formalism and its renormalizability, \href{http://dx.doi.org/10.1088/1126-6708/2007/02/065}{\textit{J.~High Energy Phys.}} \textbf{2007} (2007), no.~2, 065, 23~pages,
  \href{http://www.arxiv.org/abs/hep-th/0612188}{hep-th/0612188}.

\bibitem{Buric:2005xe}
Buri\'c M., Latas D., Radovanovi\'c V.,
Renormalizability of noncommutative ${\rm SU}(N)$ gauge theory,
\href{http://dx.doi.org/10.1088/1126-6708/2006/02/046}{\textit{J.~High Energy Phys.}} \textbf{2006} (2006), no.~2, 046, 13~pages,
  \href{http://www.arxiv.org/abs/hep-th/0510133}{hep-th/0510133}.

\bibitem{Latas:2007eu}
Latas D., Radovanovi\'c V., Trampeti\'c J.,
Non-commutative $SU(N)$ gauge theories and asymptotic freedom,
\href{http://dx.doi.org/10.1103/PhysRevD.76.085006}{\textit{Phys. Rev.~D}} \textbf{76} (2007), 085006, 7~pafes,
  \href{http://www.arxiv.org/abs/hep-th/0703018}{hep-th/0703018}.

\bibitem{Buric:2007ix}
Buri\'c M., Latas D., Radovanovi\'c V., Trampeti\'c J.,
Absence of the 4$\psi$ divergence in noncommutative chiral models,
\href{http://dx.doi.org/10.1103/PhysRevD.77.045031}{\textit{Phys. Rev.~D}} \textbf{77} (2008), 045031, 7~pages,
  \href{http://www.arxiv.org/abs/0711.0887}{arXiv:0711.0887}.

\bibitem{Martin:2009sg}
 Mart\'{\i}n C.P., C.~Tamarit,
 Noncommutative GUT inspired theories and the UV f\/initeness of the fermionic four point functions,
\href{http://dx.doi.org/10.1103/PhysRevD.80.065023}{\textit{Phys. Rev.~D}} \textbf{80} (2009), 065023, 6~pages,
  \href{http://www.arxiv.org/abs/0907.2464}{arXiv:0907.2464}.

\bibitem{Banerjee:2001un}
Banerjee R., Ghosh S.,
Seiberg--Witten map and the axial anomaly in non-commutative f\/ield theory,
\href{=http://dx.doi.org/10.1016/S0370-2693(02)01566-6}{\textit{Phys. Lett.~B}} \textbf{533} (2002), 162--167,
  \href{http://www.arxiv.org/abs/hep-th/0110177}{hep-th/0110177}.

\bibitem{Martin:2005jy}
 Mart\'{\i}n C.P., Tamarit C.,
 Noncommutative ${\rm SU}(N)$ theories, the axial anomaly, Fujikawa's method and the Atiyah--Singer index, \href{http://dx.doi.org/10.1016/j.physletb.2005.06.028}{\textit{Phys. Lett.~B}} \textbf{620} (2005), 187--194,
  \href{http://www.arxiv.org/abs/hep-th/0504171}{hep-th/0504171}.

\bibitem{Brandt:2003fx}
Brandt F., Mart\'{\i}n  C.P., Ruiz Ruiz F.,
Anomaly freedom in Seiberg--Witten noncommutative gauge theories,
\href{http://dx.doi.org/10.1088/1126-6708/2003/07/068}{\textit{J.~High Energy Phys.}} \textbf{2003} (2003), no.~7, 068, 26~pages,
  \href{http://www.arxiv.org/abs/hep-th/0307292}{hep-th/0307292}.

\bibitem{Hewett:2000zp}
 Hewett J.L., Petriello F.J., Rizzo T.G.,
 Signals for non-commutative interactions at linear colliders,
\href{http://dx.doi.org/10.1103/PhysRevD.64.075012}{\textit{Phys. Rev.~D}} \textbf{64} (2001), 075012, 23~pages,
  \href{http://www.arxiv.org/abs/hep-ph/0010354}{hep-ph/0010354}.

\bibitem{Carlson:2001sw}
 Carlson C.E., Carone C.D., Lebed R.F.,
 Bounding noncommutative QCD,
 \href{http://dx.doi.org/10.1016/S0370-2693(01)01045-0}{\textit{Phys. Lett.~B}} \textbf{518} (2001), 201--206,
  \href{http://www.arxiv.org/abs/hep-ph/0107291}{hep-ph/0107291}.

\bibitem{Kersting:2001zz}
Kersting N.,
$(g-2)_\mu$ from noncommutative geometry,
\href{http://dx.doi.org/10.1016/S0370-2693(01)01518-0}{\textit{Phys.   Lett.~B}} \textbf{527} (2002), 115--118,
  \href{http://www.arxiv.org/abs/hep-ph/0109224}{hep-ph/0109224}.

\bibitem{Carone:2004wt}
 Carone C.D.,
 Phenomenology of noncommutative f\/ield theories,
  \href{http://dx.doi.org/10.1088/1742-6596/37/1/017}{\textit{J. Phys. Conf. Ser.}} \textbf{37} (2006), 96--106,
  \href{http://www.arxiv.org/abs/hep-ph/0409348}{hep-ph/0409348}.

\bibitem{Minkowski:2003jg}
Minkowski P., Schupp P., Trampeti\'c J.,
Neutrino dipole moments and charge radii in non-commutative space-time,
\href{http://dx.doi.org/10.1140/epjc/s2004-01969-y}{\textit{Eur. Phys. J.~C}}  \textbf{37} (2004), 123--128,
  \href{http://www.arxiv.org/abs/hep-th/0302175}{hep-th/0302175}.

\bibitem{Schweda:1996}
Emery S., Kr\"uger M., Rant J., Schweda~M., Sommer T.,
Two-dimensional BF model quantized in the axial gauge,
\textit{Nuovo Cimento~A} \textbf{111} (1998), 1321--1335,
  \href{http://www.arxiv.org/abs/hep-th/9609240}{hep-th/9609240}.

\bibitem{Schweda:1998}
 Del~Cima O.M., Landsteiner K., Schweda M.,
 Twisted $N = 4$ SUSY algebra in topological models of Schwarz type,
\href{http://dx.doi.org/10.1016/S0370-2693(98)01065-X}{\textit{Phys. Lett.~B}} \textbf{439} (1998), 289--300,
  \href{http://www.arxiv.org/abs/hep-th/9806137}{hep-th/9806137}.

\bibitem{Schweda:1999}
 Del~Cima O.M., Grimstrup J.M., Schweda M.,
 On the f\/initeness of a new topological model in $D = 3$,
\href{http://dx.doi.org/10.1016/S0370-2693(99)00954-5}{\textit{Phys. Lett.~B}} \textbf{463} (1999), 48--56,
  \href{http://www.arxiv.org/abs/hep-th/9906146}{hep-th/9906146}.

\bibitem{Gieres:2000}
Gieres F., Grimstrup J.M., Pisar T., Schweda M.,
Vector supersymmetry in topological f\/ield theories,
\href{http://dx.doi.org/10.1088/1126-6708/2000/06/018}{\textit{J.~High Energy Phys.}} \textbf{2000} (2000), no.~6, 018, 22~pages,
  \href{http://www.arxiv.org/abs/hep-th/0002167}{hep-th/0002167}.

\bibitem{Batalin:1981}
Batalin I.A., Vilkovisky G.A.,
Gauge algebra and quantization,
\href{http://dx.doi.org/10.1016/0370-2693(81)90205-7}{\textit{Phys. Lett.~B}} \textbf{102} (1981), 27--31.

\bibitem{Batalin:1983}
 Batalin I.A., Vilkovisky G.A.,
 Quantization of gauge theories with linearly dependent generators,
 \href{http://dx.doi.org/10.1103/PhysRevD.28.2567}{\textit{Phys. Rev.~D}} \textbf{28} (1983), 2567--2582.

\bibitem{Piguet:1995}
Piguet O., Sorella S.P.,
Algebraic renormalization. Perturbative renormalization, symmetries and anomalies,
{\it Lecture Notes in Physics. New Series m: Monographs}, Vol.~28, Springer-Verlag, Berlin, 1995.

\bibitem{Szabo:2008}
Fischer A., Szabo R.J.,
Duality covariant quantum f\/ield theory on noncommutative Minkowski space,
\href{http://dx.doi.org/10.1088/1126-6708/2009/02/031}{\textit{J.~High Energy Phys.}} \textbf{2009} (2009), no.~2, 031, 36~pages,
  \href{http://www.arxiv.org/abs/0810.1195}{arXiv:0810.1195}.

\bibitem{Rivasseau:2005a}
Rivasseau V., Vignes-Tourneret F., Wulkenhaar R.,
Renormalization of noncommutative $\phi^4$-theory by multi-scale analysis,
\href{http://dx.doi.org/10.1007/s00220-005-1440-4}{\textit{Comm. Math. Phys.}} \textbf{262} (2006), 565--594,
  \href{http://www.arxiv.org/abs/hep-th/0501036}{hep-th/0501036}.

\bibitem{Disertori:2006uy}
Disertori M., Rivasseau V.,
Two- and three-loop beta function of non-commutative $\Phi_4^4$ theory,
\href{http://dx.doi.org/10.1140/epjc/s10052-007-0211-0}{\textit{Eur. Phys.~J.~C}} \textbf{50} (2007), 661--671,
  \href{http://www.arxiv.org/abs/hep-th/0610224}{hep-th/0610224}.

\bibitem{Landau:1955}
 Landau L.D.,
 On the quantum theory of f\/ields,
in  Niels Bohr and the Development of Physics,  McGraw-Hill Book Co., New York, N.Y., 1955, 52--69.


\bibitem{Grosse:2004a}
Grosse H., Wulkenhaar R.,
The $\beta$-function in duality-covariant non-commutative $\phi^4$-theory,
\href{http://dx.doi.org/10.1140/epjc/s2004-01853-x}{\textit{Eur. Phys.~J.~C}} \textbf{35} (2004), 277--282,
  \href{http://www.arxiv.org/abs/hep-th/0402093}{hep-th/0402093}.

\bibitem{Rivasseau:2006b}
Disertori M., Gurau R., Magnen J., Rivasseau V.,
Vanishing of beta function of non-commutative $\Phi_4^4$ theory to all orders,
\href{http://dx.doi.org/10.1016/j.physletb.2007.04.007}{\textit{Phys.   Lett.~B}} \textbf{649} (2007), 95--102,
  \href{http://www.arxiv.org/abs/hep-th/0612251}{hep-th/0612251}.

\bibitem{Blaschke:2008b}
 Blaschke D.N., Gieres F., Kronberger E., Reis T., Schweda M., Sedmik R.I.P.,
Quantum corrections for translation-invariant renormalizable non-commutative $\phi^4$ theory,
\href{http://dx.doi.org/10.1088/1126-6708/2008/11/074}{\textit{J.~High Energy Phys.}} \textbf{2008} (2008), no.~11, 074, 16~pages,
  \href{http://www.arxiv.org/abs/0807.3270}{arXiv:0807.3270}.

\bibitem{Madore:2000aq}
Madore J.,
An introduction to noncommutative dif\/ferential geometry and its physical applications, 2nd ed.,
 {\it London Mathematical Society Lecture Note Series}, Vol.~257, Cambridge University Press, Cambridge, 1999.

\bibitem{Rivasseau:1991ub}
Rivasseau V.,
From perturbative to constructive renormalization,
{\it Princeton Series in Physics}, Princeton University Press, Princeton, NJ, 1991.

\bibitem{Sorella:2005}
 Sobreiro R.F., Sorella S.P.,
 Introduction to the Gribov  ambiguities in Euclidean Yang--Mills theories,
  \mbox{\href{http://www.arxiv.org/abs/hep-th/0504095}{hep-th/0504095}}.

\bibitem{Kummer-memorial}
Grumiller D., Rebhan A., Vassilevich D. (Editors),
 Fundamental interactions~-- a memorial volume for Wolfgang Kummer, World
  Scientif\/ic,  Singapore, 2009.

\bibitem{vanBaal:2000zc}
van Baal P.,
QCD in a f\/inite volume,
  \href{http://www.arxiv.org/abs/hep-ph/0008206}{hep-ph/0008206}.

\bibitem{Dudal:2009xh}
Dudal D., Sorella S.P., Vandersickel N., Verschelde H.,
Gribov no-pole condition, Zwanziger horizon function, Kugo--Ojima conf\/inement criterion, boundary conditions, BRST breaking and all that,
\href{http://dx.doi.org/10.1103/PhysRevD.79.121701}{\textit{Phys.   Rev.~D}} \textbf{79} (2009), 121701, 5~pages,
  \href{http://www.arxiv.org/abs/0904.0641}{arXiv:0904.0641}.

\bibitem{Dudal:2008}
Dudal D., Gracey J., Sorella S.P., Vandersickel N., Verschelde H.,
A ref\/inement of the Gribov--Zwanziger approach in the Landau gauge: infrared propagators in harmony with the lattice results,
  \href{http://dx.doi.org/10.1103/PhysRevD.78.065047}{\textit{Phys. Rev.~D}} \textbf{78} (2008), 065047, 30~pages,
  \href{http://www.arxiv.org/abs/0806.4348}{arXiv:0806.4348}.

\bibitem{Ryder:1996}
 Ryder L.H.,
 Quantum f\/ield theory, 2nd ed., Cambridge University Press,  Cambridge, 1996.

\bibitem{Schweda-book:1998}
Boresch A., Emery S., Moritsch O., Schweda M., Sommer T., Zerrouki H.,
Applications of noncovariant gauges in the algebraic renormalization procedure, World Scientif\/ic Publishing Co., Inc., River Edge, NJ, 1998.

\bibitem{Sedmik:2009}
 Sedmik R.I.P.,
 On the development of non-commutative translation-invariant quantum gauge f\/ield mo\-dels,
 PhD Thesis, Vienna University of Technology, 2009, available at \url{http://media.obvsg.at/AC07806220}.

\bibitem{Putz:2003}
Aigner F., Hillbrand M., Knapp J., Milovanovic G., Putz V., Sch\"{o}fbeck R., Schweda M.,
Technical remarks and comments on the UV/IR-mixing problem of a noncommutative scalar quantum f\/ield theory, \href{http://dx.doi.org/10.1023/B:CJOP.0000038525.68786.c9}{\textit{Czechoslovak J. Phys.}} \textbf{54} (2004), 711--719,
  \href{http://www.arxiv.org/abs/hep-th/0302038}{hep-th/0302038}.


\bibitem{Denk:2003}
Denk S., Schweda M.,
Time ordered perturbation theory for non-local interactions: applications to NCQFT,
\href{http://dx.doi.org/10.1088/1126-6708/2003/09/032}{\textit{J.~High Energy Phys.}} \textbf{2003} (2003), no.~9, 032, 22~pages,
  \href{http://www.arxiv.org/abs/hep-th/0306101}{hep-th/0306101}.

\bibitem{Bahns:2003}
Bahns D., Doplicher S., Fredenhagen K., Piacitelli G.,
Ultraviolet f\/inite quantum f\/ield theory on quantum spacetime,
\href{http://dx.doi.org/10.1007/s00220-003-0857-x}{\textit{Comm. Math. Phys.}} \textbf{237} (2003), 221--241,
  \href{http://www.arxiv.org/abs/hep-th/0301100}{hep-th/0301100}.

\bibitem{Blaschke:2009c}
Blaschke D.N., Kronberger E., Rofner A., Schweda M., Sedmik R.I.P., Wohlgenannt M.,
On the problem of renormalizability in non-commutative gauge f\/ield models~-- a critical review,
\href{http://dx.doi.org/10.1002/prop.200900102}{\textit{Fortschr. Phys.}} \textbf{58} (2010), 364--372,
  \href{http://www.arxiv.org/abs/0908.0467}{arXiv:0908.0467}.

\bibitem{Rofner:2009}
Rofner A.,
Investigations on the renormalizability of a non-commutative $U(1)$ gauge theory,
PhD Thesis, Vienna University of Technology, 2009,
 available at  \url{http://media.obvsg.at/AC07806222}.

\bibitem{workinprogress}
Blaschke D.N., Kronberger E., Sedmik R., Wohlgenannt M.,
On the renormalizability of a non-commutative gauge model with soft-breaking, work in progress.

\bibitem{Denk:2004}
Denk S.,
Perturbative aspects of non-local and non-commutative quantum  field theories,
 PhD Thesis, Vienna University of Technology, 2004,
 available at \url{http://media.obvsg.at/AC04351727-2001}.

\bibitem{Liao:2002}
Liao Y., Sibold K.,
Time-ordered perturbation theory on non-commutative spacetime. II.~Unitarity,
\href{http://dx.doi.org/10.1007/s10052-002-1018-7}{\textit{Eur. Phys.~J.~C}} \textbf{25} (2002), 479--486,
  \href{http://www.arxiv.org/abs/hep-th/0206011}{hep-th/0206011}.

\bibitem{Fischer:2002}
Bozkaya H., Fischer P., Grosse H., Pitschmann M., Putz V., Schweda M., Wulkenhaar R.,
Space/time noncommutative f\/ield theories and causality,
\href{http://dx.doi.org/10.1140/epjc/s2003-01210-9}{\textit{Eur. Phys. J.~C}} \textbf{29} (2003), 133--141,
  \href{http://www.arxiv.org/abs/hep-th/0209253}{hep-th/0209253}.

\bibitem{Ohl:2003}
Ohl T., R\"uckl R., Zeiner J.,
Unitarity of time-like noncommutative gauge theories: rhe violation of Ward identities in time-ordered perturbation
  theory,
\href{http://dx.doi.org/10.1016/j.nuclphysb.2003.10.022}{\textit{Nuclear Phys.~B}} \textbf{676} (2004), 229--242,
  \href{http://www.arxiv.org/abs/hep-th/0309021}{hep-th/0309021}.

\bibitem{Bahns:2002}
Bahns D., Doplicher S., Fredenhagen K., Piacitelli G.,
On the  unitarity problem in space/time noncommutative theories,
\href{http://dx.doi.org/10.1016/S0370-2693(02)01563-0}{\textit{Phys. Lett.~B}} \textbf{533} (2002), 178--181,
  \href{http://www.arxiv.org/abs/hep-th/0201222}{arXiv:hep-th/0201222}.

\bibitem{Bahns:2004}
Bahns D., Doplicher S., Fredenhagen K., Piacitelli G.,
Field theory on noncommutative spacetimes: quasiplanar Wick products,
\href{http://dx.doi.org/10.1103/PhysRevD.71.025022}{\textit{Phys. Rev.~D}} \textbf{71} (2005), 025022, 12~pages,
  \href{http://www.arxiv.org/abs/hep-th/0408204}{hep-th/0408204}.

\bibitem{Kulish:1999}
 Kulish P.P., Mudrov A.I.,
 Twist-related geometries on $q$-Minkowski space,
 \textit{Proc. Steklov Inst. Math.}  (1999), no.~3 (226),  86--97,
  \href{http://www.arxiv.org/abs/math.QA/9901019}{math.QA/9901019}.

\bibitem{Oeckl:2000eg}
Oeckl R.,
Untwisting noncommutative $\mathbb R^d$ and the equivalence of quantum f\/ield theories,
\href{http://dx.doi.org/10.1016/S0550-3213(00)00281-9}{\textit{Nuclear Phys.~B}} \textbf{581} (2000), 559--574,
  \href{http://www.arxiv.org/abs/hep-th/0003018}{hep-th/0003018}.

\bibitem{Wess:2009zza}
Wess J.,
Deformed gauge theories,
in Noncommutative Spacetimes,
\href{http://dx.doi.org/10.1007/978-3-540-89793-4_2}{\textit{Lecture Notes in Physics}}, Vol.~774, Springer, Berlin, 2009, 23--37.

\bibitem{Aschieri:2006ye}
Aschieri P., Dimitrijevi\'c M., Meyer F., Schraml S., Wess J.,
Twisted gauge theories,
\href{http://dx.doi.org/10.1007/s11005-006-0108-0}{\textit{Lett. Math. Phys.}} \textbf{78} (2006), 61--71,
  \href{http://www.arxiv.org/abs/hep-th/0603024}{hep-th/0603024}.

\bibitem{Vassilevich:2006tc}
 Vassilevich D.V.,
 Twist to close,
 \href{=http://dx.doi.org/10.1142/S0217732306020755}{\textit{Modern Phys. Lett.~A}}  \textbf{21} (2006), 1279--1283,
  \href{http://www.arxiv.org/abs/hep-th/0602185}{hep-th/0602185}.

\bibitem{Banerjee:2006jy}
Banerjee R., Samanta S.,
Gauge generators, transformations and identities on a noncommutative space,
\href{http://dx.doi.org/10.1140/epjc/s10052-007-0280-0}{\textit{Eur. Phys. J.~C}} \textbf{51} (2007), 207--215,
  \href{http://www.arxiv.org/abs/hep-th/0608214}{hep-th/0608214}.

\bibitem{Balachandran:2007kv}
 Balachandran A.P., Pinzul A., Qureshi  B.A., Vaidya S.,
 Twisted gauge and gravity theories on the Groenewold--Moyal plane,
\href{http://dx.doi.org/10.1103/PhysRevD.76.105025}{\textit{Phys. Rev.~D}} \textbf{76} (2007), 105025, 10~pages,
  \href{http://www.arxiv.org/abs/0708.0069}{arXiv:0708.0069}.

\bibitem{Akofor:2008ae}
Akofor E., Balachandran A.P., Joseph A.,
Quantum f\/ields on the Groenewold--Moyal plane,
\href{http://dx.doi.org/10.1142/S0217751X08040317}{\textit{Internat. J. Modern Phys.~A}} \textbf{23} (2008), 1637--1677,
  \href{http://www.arxiv.org/abs/0803.4351}{arXiv:0803.4351}.

\bibitem{Lukierski:2001}
Lukierski J.,
From noncommutative space-time to quantum relativistic symmetries with fundamental mass parameter,
  \href{http://www.arxiv.org/abs/hep-th/0112252}{hep-th/0112252}.

\bibitem{Dimitrijevic:2004}
Dimitrijevi\'c M., M{\"o}ller L.,  Tsouchnika E.,
Derivatives, forms and vector f\/ields on the $\kappa$-deformed Euclidean space,
\href{http://dx.doi.org/10.1088/0305-4470/37/41/010}{\textit{J.~Phys.~A: Math. Gen.}} \textbf{37} (2004), 9749--9770,
  \href{http://www.arxiv.org/abs/hep-th/0404224}{hep-th/0404224}.

\bibitem{Meljanac:2007xb}
Meljanac S., Samsarov A., Stoji\'c M.,  Gupta K.S.,
$\kappa$-Minkowski space-time and the star product realizations,
\href{http://dx.doi.org/10.1140/epjc/s10052-007-0450-0}{\textit{Eur. Phys. J.~C}} \textbf{53} (2008), 295--309,
  \href{http://www.arxiv.org/abs/0705.2471}{arXiv:0705.2471}.

\bibitem{Ydri:2002nt}
Ydri B.,
Noncommutative chiral anomaly and the Dirac--Ginsparg--Wilson operator,
\href{http://dx.doi.org/10.1088/1126-6708/2003/08/046}{\textit{J.~High Energy Phys.}} \textbf{2003} (2003), no.~8, 046, 18~pages,
  \href{http://www.arxiv.org/abs/hep-th/0211209}{hep-th/0211209}.

\bibitem{Azuma:2004zq}
Azuma T., Bal S., Nagao~K., Nishimura J.,
Nonperturbative studies of fuzzy spheres in a matrix model with the Chern--Simons term,
  \href{http://dx.doi.org/10.1088/1126-6708/2004/05/005}{\textit{J.~High Energy Phys.}} \textbf{2004} (2004), no.~5, 005,36~pages
  \href{http://www.arxiv.org/abs/hep-th/0401038}{hep-th/0401038}.

\bibitem{Grimstrup:2003dz}
 Grimstrup J.M.,
 Noncommutative coordinate transformations and the Seiberg--Witten map,
 \textit{Acta Phys. Polon.~B} \textbf{34} (2003), 4855--4866.

\bibitem{Aschieri:2007fb}
Aschieri P., Steinacker H., Madore J., Manousselis P., Zoupanos G.,
Fuzzy extra dimensions: dimensional reduction, dynamical generation and renormalizability,
  \href{http://www.arxiv.org/abs/0704.2880}{arXiv:0704.2880}.

\bibitem{Steinacker:2008ri}
Steinacker H.,
Emergent gravity and noncommutative branes from Yang--Mills matrix models,
\href{http://dx.doi.org/10.1016/j.nuclphysb.2008.10.014}{\textit{Nuclear Phys.~B}} \textbf{810} (2009), 1--39,
  \href{http://www.arxiv.org/abs/0806.2032}{arXiv:0806.2032}.

\bibitem{Klammer:2009dj}
Klammer D., Steinacker H.,
Fermions and noncommutative emergent gravity. II.~Curved branes in extra dimensions,
\href{http://dx.doi.org/10.1007/JHEP02(2010)074}{\textit{J.~High Energy Phys.}} \textbf{2010} (2010), no.~2, 074, 32~pages,
  \href{http://www.arxiv.org/abs/0909.5298}{arXiv:0909.5298}.

\bibitem{Grosse:2010zq}
Grosse H., Lizzi F., Steinacker H.,
Noncommutative gauge theory and symmetry breaking in matrix models,
\href{http://dx.doi.org/10.1103/PhysRevD.81.085034}{\textit{Phys. Rev.~D}} \textbf{81} (2010), 085034, 12~pages,
  \href{http://www.arxiv.org/abs/1001.2703}{arXiv:1001.2703}.

\bibitem{Steinacker:2010rh}
Steinacker H.,
Emergent geometry and gravity from matrix models: an introduction,
\href{http://dx.doi.org/10.1088/0264-9381/27/13/133001}{\textit{Classical Quantum Gravity}} \textbf{27} (2010), 133001, 46~pages,
  \href{http://www.arxiv.org/abs/1003.4134}{arXiv:1003.4134}.

\bibitem{Majid:1999td}
Majid S.,
Quantum groups and noncommutative geometry,
\href{http://dx.doi.org/10.1063/1.533331}{\textit{J.~Math. Phys.}} \textbf{41} (2000), 3892--3942,
  \mbox{\href{http://www.arxiv.org/abs/hep-th/0006167}{hep-th/0006167}}.

\bibitem{McCurdy:2009xz}
McCurdy S., Zumino B.,
Covariant star product for exterior dif\/ferential forms on symplectic manifolds,
\href{http://dx.doi.org/10.1063/1.3327559}{\textit{AIP Conf. Proc.}} \textbf{1200} (2010), 204--214,
  \href{http://www.arxiv.org/abs/0910.0459}{arXiv:0910.0459}.

\bibitem{Chaichian:2009kn}
Chaichian M., Tureanu A., Zet G.,
Gauge f\/ield theories with covariant star-product,
\href{http://dx.doi.org/10.1088/1126-6708/2009/07/084}{\textit{J.~High Energy Phys.}} \textbf{2009} (2009), no.~7, 084, 13~pages,
  \href{http://www.arxiv.org/abs/0905.0608}{arXiv:0905.0608}.

\bibitem{Vassilevich:2009cb}
 Vassilevich D.V.,
 Dif\/feomorphism covariant star products and noncommutative gravity,
 \href{http://dx.doi.org/10.1088/0264-9381/26/14/145010}{\textit{Classical Quantum Gravity}} \textbf{26} (2009), 145010, 8~pages,
  \href{http://www.arxiv.org/abs/0904.3079}{arXiv:0904.3079}.

\bibitem{Ho:2001fi}
 Ho P.-M., Miao S.-P.,
Noncommutative dif\/ferential calculus for a D-brane in a nonconstant $B$ f\/ield background with $H=0$,
 \href{http://dx.doi.org/10.1103/PhysRevD.64.126002}{\textit{Phys. Rev.~D}} \textbf{64} (2001),  126002, 8~pages,
  \href{http://www.arxiv.org/abs/hep-th/0105191}{hep-th/0105191}.

\end{thebibliography}
\end{document}